\documentclass[12pt,prd, aps, showpacs, superscriptaddress,floatfix,nofloatinbib]{revtex4-1}

\usepackage{graphicx}
\usepackage{hyperref}

\begin{document}

\bibliographystyle{h-physrev5}

\title{SU(2) chiral perturbation theory low-energy constants from\\
2+1 flavor staggered lattice simulations}


\newcommand{\ur}{Universit\"at Regensburg, Universit\"atsstr.~31, D-93053 Regensburg, Germany}
\newcommand{\buw}{Bergische Universit\"at Wuppertal, Gau\ss{}str.~20, D-42119 Wuppertal, Germany}
\newcommand{\jsc}{J\"ulich Supercomputing Centre, Forschungszentrum J\"ulich, D-52425 J\"ulich, Germany}
\newcommand{\budapest}{Institute for Theoretical Physics, E\"otv\"os University, H-1117 Budapest, Hungary}

%
\author{Szabolcs~Bors\'anyi}\affiliation{\buw}
\author{Stephan~D\"urr}\email[E-mail: ]{durr\,(AT)\,itp.unibe.ch}\affiliation{\buw}\affiliation{\jsc}
\author{Zolt\'an~Fodor}\affiliation{\buw}\affiliation{\jsc}\affiliation{\budapest}
\author{Stefan~Krieg}\affiliation{\buw}\affiliation{\jsc}
\author{Andreas~Sch\"afer}\affiliation{\ur}
\author{Enno~E.~Scholz}\email[E-mail: ]{enno.scholz\,(AT)\,physik.uni-regensburg.de}\affiliation{\ur}
\author{K\'alm\'an~K.~Szab\'o}\affiliation{\buw}


\date{May 4, 2012 and July 24, 2013}

\begin{abstract}
\vfill
\noindent
We extract the next-to-leading-order low-energy constants $\bar\ell_3$ and $\bar\ell_4$ of SU(2)
chiral perturbation theory, based on precise lattice data for the pion mass and
decay constant on ensembles generated by the Wuppertal-Budapest Collaboration
for QCD thermodynamics.
These ensembles feature 2+1 flavors of two-fold stout-smeared dynamical staggered
fermions combined with Symanzik glue, with pion masses varying from 135 to
435\,MeV, lattice scales between 0.7 and 2.0\,GeV, while $m_s$ is kept
fixed at its physical value.
Moderate taste splittings and the scale being set through the pion decay
constant allow us to restrict ourselves to the taste pseudoscalar state and to
use formulas from continuum chiral perturbation theory.
Finally, by dropping the data points near 135\,MeV from the fits, we can
explore the range of pion masses that is needed in SU(2) chiral perturbation theory to reliably
extrapolate to the physical point.
\vfill
$\phantom{.}$
\end{abstract}

\maketitle

\newcommand{\pad}{\partial}
\newcommand{\til}{\tilde}
\renewcommand{\dag}{^\dagger}
\newcommand{\<}{\langle}
\renewcommand{\>}{\rangle}
\newcommand{\Mpi}{M_\pi}
\newcommand{\Fpi}{F_\pi}
\newcommand{\Mka}{M_K}
\newcommand{\Fka}{F_K}
\newcommand{\Met}{M_\et}
\newcommand{\Fet}{F_\et}
\newcommand{\Mss}{M_{\bar{s}s}}
\newcommand{\Fss}{F_{\bar{s}s}}
\newcommand{\Mcc}{M_{\bar{c}c}}
\newcommand{\Fcc}{F_{\bar{c}c}}
\newcommand{\fpi}{f_\pi}

\newcommand{\al}{\alpha}
\newcommand{\be}{\beta}
\newcommand{\ga}{\gamma}
\newcommand{\de}{\delta}
\newcommand{\ep}{\epsilon}
\newcommand{\ve}{\varepsilon}
\newcommand{\ze}{\zeta}
\newcommand{\et}{\eta}
\renewcommand{\th}{\theta}
\newcommand{\vt}{\vartheta}
\newcommand{\io}{\iota}
\newcommand{\ka}{\kappa}
\newcommand{\la}{\lambda}
\newcommand{\rh}{\rho}
\renewcommand{\vr}{\varrho}
\newcommand{\si}{\sigma}
\newcommand{\ta}{\tau}
\newcommand{\ph}{\phi}
\newcommand{\vp}{\varphi}
\newcommand{\ch}{\chi}
\newcommand{\ps}{\psi}
\newcommand{\om}{\omega}

\newcommand{\bdm}{\begin{displaymath}}
\newcommand{\edm}{\end{displaymath}}
\newcommand{\bea}{\begin{eqnarray}}
\newcommand{\eea}{\end{eqnarray}}
\newcommand{\beq}{\begin{equation}}
\newcommand{\eeq}{\end{equation}}

\newcommand{\mr}{\mathrm}
\newcommand{\mb}{\mathbf}
\newcommand{\ri}{\mr{i}}
\newcommand{\Nf}{N_{\!f}}
\newcommand{\Nc}{N_{ c }}
\newcommand{\Nt}{N_{ t }}
\newcommand{\MeV}{\,\mr{MeV}}
\newcommand{\GeV}{\,\mr{GeV}}
\newcommand{\fm}{\,\mr{fm}}
\newcommand{\MSbar}{\overline{\mr{MS}}}


\section{Introduction}
\label{sec:Introduction}

One of the most fascinating aspects of QCD \cite{Fritzsch:1973pi}, the theory
of strong interactions, is the nonanalytic behavior of its Green's functions
close to the chiral limit, that is with two or three quark masses taken small,
$m_q\ll\Lambda^2/B$, where $\Lambda\!\sim\!1\GeV$ is a typical hadronic scale
and $B$ is a condensate parameter which we will define (and determine) below.

The structure of these nonanalytic contributions can be worked out in the
effective field theory approach which is known as
chiral perturbation theory (ChPT) \cite{Gasser:1983yg,Gasser:1984gg}.
In this setup physical quantities are expanded in powers of $p/\Lambda$,
where $p$ is an external momentum, and the quark mass is treated as
$m_q\!\simeq\!O(p^2)$.
Depending on whether this is done for the two ($u,d$) or three ($u,d,s$)
lightest flavors, the framework is known as SU(2) or SU(3) ChPT.

In either case, at the leading order (LO) of the chiral expansion there are two
low-energy constants, defined from the pion decay constant and the scalar condensate as
\beq
F=\lim_{m_u,m_d\to0}\Fpi
\;,\qquad
B=\lim_{m_u,m_d\to0}\{-\<0|\bar{q}q|0\>/\Fpi^2\}
\label{def_FB}
\eeq
where $q$ denotes one of the light flavors (i.e.\ $q\!=\!u$ or $q\!=\!d$) in
the SU(2) case, and similarly (with also $m_s$ sent to $0$) in the SU(3) case.
At the next-to-leading order (NLO), i.e.\ at $O(p^4)$ in the chiral counting, seven
new low-energy constants show up in the SU(2) framework \cite{Gasser:1983yg}, or ten low-energy plus
two high-energy constants for SU(3) \cite{Gasser:1984gg}.
These low-energy constants parametrize the above-mentioned chiral logarithms
in the Green's functions of QCD.
Their numerical values can be determined either from experiment or from an
{\it ab initio} solution of QCD in the relevant (small coupling and light quark
mass) regime, as is provided by lattice QCD
\cite{Wilson:1974sk,Creutz:1980zw,Durr:2008zz}.
Since the chiral logarithms show up as rather subtle effects, meaningful results can
only be obtained from lattice data which have excellent statistical precision
and explore, at the same time, a wide enough range of lattice spacings, quark
masses, and box volumes such that all sources of systematic error can be
controlled and eventually removed.

In this paper we provide such a determination of the SU(2) low-energy constants
$\bar\ell_3$ and $\bar\ell_4$.
Their numerical values are extracted from the quark-mass dependence of $\Mpi$
and $\Fpi$, respectively, and complemented by numerical values of the leading-order low-energy constants $F$ and $B$.
We use staggered fermion simulations with $\Nf=2\!+\!1$ dynamical flavors, that
is, two degenerate light quarks of variable mass $m_l$ and an active strange
quark whose mass $m_s$ is pinned down at its physical value.
As a result of this our values of the low-energy constants are supposed to
coincide with those in the real word.
The inverse lattice spacings cover the range $0.7\GeV\leq a^{-1}\leq 2.0\GeV$
(see Sec.\,\ref{sec:scale} and Table \ref{tab:robust} for details).
A preliminary account of our work (based on a smaller dataset) was given in Ref.\,\cite{Scholz:2011rk}.

The remainder of this paper is structured as follows.
In Sec.\,\ref{sec:Lattice} we specify the gauge and fermion actions used and list the
ensembles which go into the determination of the chiral low-energy constants.
Furthermore, details are given how we calculate the pion mass and decay
constant, and how we correct the latter for the effect of the finite spatial
volume of the box (which is always a small correction, since our data satisfy
$3.3\!\leq\!\Mpi L\!\leq\!6.8$).
In Sec.\,\ref{sec:scale} we specify the procedure through which we set, for each bare
coupling $\be\!=\!6/g_0^2$, the lattice spacing $a$ and the physical values of
the bare quark masses $m_l\!=\!(m_u\!+\!m_d)/2$ and $m_s$.
Section \ref{sec:NLOChPT} contains the core part of the present investigation, an analysis of
our data with SU(2) ChPT at NLO, with details of how we select adequate
mass windows and determine the systematic uncertainty of the fitted
low-energy constants.
Section \ref{sec:NNLOChPT} contains a similar though less mature analysis at next-to-next-to-leading order (NNLO), where again
the main goal is to determine the NLO coefficients, with and without the
help of some priors on the remaining NLO and NNLO low-energy constants.
This helps to give a reliable estimate of the theoretical uncertainty of the
NLO results obtained.
A summary and a comparison with the findings of other recent lattice
studies of SU(2) NLO low-energy constants is presented in Sec.\,\ref{sec:Conclusions}.


\section{Lattice data}
\label{sec:Lattice}

In this section we specify the lattice actions used, list the ensembles which
go into the determination of the SU(2) low-energy constants, give details of
how we extract the pion mass and decay constant on a given ensemble, and
describe the procedure by which we remove the (small) impact of the finite
spatial box size $L$ on the data.

\subsection{Lattice action and ensemble generation}
\label{subsec:ensembles}

The lattices are generated with a tree-level Symanzik improved gauge action
\cite{Luscher:1985zq} and $2\!+\!1$ flavors of staggered quarks with two levels
of stout-smearing \cite{Morningstar:2003gk}.
The action is specified in full detail in Ref.\,\cite{Aoki:2005vt}.
The algorithm used is a combination of HMC and RHMC with standard improvements
(see, e.g., Ref.\,\cite{Durr:2008rw} for an overview).
Some ensembles were generated for scale setting purposes in previous finite
temperature studies \cite{Aoki:2005vt,Aoki:2006br,Aoki:2006we,Aoki:2009sc,
Borsanyi:2010bp,Borsanyi:2010cj}, and some were generated specifically for the
present investigation.
The taste splitting $M_{PX}^2-M_{PP}^2$ [where the first subscript indicates
that the state is a pseudoscalar in spinor space and the second one refers to
its taste, with $PP$ indicating the Goldstone state that couples to the
operator $(\ga_5\otimes\xi_5$)] is in good approximation independent of the
quark mass \cite{Bazavov:2009bb}.
Building on this information, the masses of the taste partners occurring in the
present investigation can be reconstructed from the Goldstone masses given
below and the splittings presented in Refs.\,\cite{Borsanyi:2010bp,Borsanyi:2010cj}.

We adopt a mass independent scale setting, that is the lattice spacing $a$
depends only on the coupling $\be$ in the gauge action, not on the quark masses
$m_l,m_s$.
With this choice it is straightforward to adjust, for each $\be$, the strange
quark mass roughly to its physical value by tuning the ratio $(2\Mka^2-\Mpi^2)/M_\ph^2$
to its physical value.
In the numerator the FLAG values \cite{Colangelo:2010et} of the pseudoscalar
meson masses are used which correct for isospin breaking and QED effects
(cf.\ Sec.\,\ref{sec:scale} below).
In the denominator the PDG value \cite{Nakamura:2010zzi} of the vector meson
mass is used, despite the fact that our $M_\ph$ involves only the connected
contribution (the difference is believed to be small, cf.\ Ref.\,\cite{Durr:2008rw}).
Starting from the symmetric point $m_l=m_s \simeq m_s^\mr{phys}$, one can lower the
light quark mass $m_l$, at fixed $m_s$, until the ratio $\Mpi^2/\fpi^2$
assumes its physical value.
This is one possible definition of the physical point in which we effectively
set the lattice spacing through $f_\pi$ with details given in
Sec.\,\ref{sec:scale} (other definitions differ from this one just in the choice
of which ratios are affected by cutoff effects and which are not).

From a more practical point of view it suffices to say that we simulate, for
each $\be$, a number of $(m_l,m_s)$ combinations, where $m_s$ is held fixed and
is close to whichever definition of the physical strange quark mass that one
may adopt, while $m_l$ varies between roughly the physical light quark mass and
four to ten times this value (depending on $\be$).
The precise value of $m_l^\mr{phys}$ is determined, {\it a posteriori}, by means of
an interpolation, as described in Section \ref{sec:scale} below.
A summary of our ensembles and their bare parameters is given in Table~\ref{tab:ensembles}.

%

\begin{table}
\begin{center}
\hfill
\raisebox{-\depth}{%
\begin{tabular}{lcllr}
\hline\hline
$\beta$ & $L^3\times T$ & $am_l$ & $am_s$ & \#conf \\\hline\hline 
3.45 & $24^3\times32$ & 0.0057619 & 0.1573 &  158 \\
     & $16^3\times32$ & 0.0172857 &        &  226 \\
     & $12^3\times28$ & 0.0288095 &        & 1839 \\
     & $12^3\times28$ & 0.0403333 &        & 1612 \\
     & $12^3\times28$ & 0.0518571 &        & 1504 \\ \hline
3.55 & $24^3\times32$ & 0.00374878 & 0.1023417 &  301 \\
     & $16^3\times32$ & 0.01312073 &           &   85 \\
     & $16^3\times32$ & 0.01874390 &           &  207 \\
     & $12^3\times28$ & 0.02624146 &           & 1865 \\
     & $12^3\times28$ & 0.03373902 &           & 1702 \\ \hline
3.67 & $32^3\times48$ & 0.00231904 & 0.06330976 & 166 \\
     & $24^3\times32$ & 0.00927616 &            & 135 \\
     & $16^3\times32$ & 0.01391424 &            & 502 \\
     & $16^3\times32$ & 0.01739280 &            & 467 \\
     & $14^3\times32$ & 0.02203088 &            & 460 \\
\hline\hline
\end{tabular}}
\hfill
\raisebox{-\depth}{%
\begin{tabular}{lcllr}
\hline\hline
$\beta$ & $L^3\times T$ & $am_l$ & $am_s$ & \#conf \\\hline\hline 
3.75 & $48^3\times96$ & 0.00172000 & 0.048 &  180 \\
     & $40^3\times64$ & 0.00240000 &       &  380 \\
     & $32^3\times64$ & 0.00342857 &       &  200 \\
     & $40^3\times64$ & 0.00480000 &       &  379 \\
     & $32^3\times64$ & 0.00685000 &       &  323 \\ \hline
3.792 & $48^3\times64$ & 0.00160714 & 0.045 &  429 \\
      & $40^3\times64$ & 0.00225000 &       &  510 \\
      & $40^3\times64$ & 0.00321429 &       &  202 \\
      & $40^3\times64$ & 0.00450000 &       &  668 \\
      & $32^3\times64$ & 0.00674300 &       &  371 \\ \hline 
3.85 & $48^3\times64$ & 0.00144606 & 0.0394774 & 326 \\
     & $40^3\times64$ & 0.00197387 &           & 466 \\
     & $40^3\times64$ & 0.00281981 &           & 385 \\
     & $48^3\times64$ & 0.00394774 &           & 400 \\
     & $32^3\times48$ & 0.00578424 &           &  49 \\
     & $32^3\times48$ & 0.00867636 &           &  59 \\
     & $24^3\times48$ & 0.01156848 &           & 143 \\
     & $24^3\times48$ & 0.01446060 &           & 165 \\
\hline\hline
\end{tabular}}
\hfill$\phantom{.}$
%
\caption{Overview of the staggered $2\!+\!1$ flavor ensembles used in this work.}
\label{tab:ensembles}
\end{center}
\end{table}

\subsection{Calculating meson masses and decay constants}
\label{subsec:meson}

A specific advantage of staggered fermions (or of any other discretization
with some form of chiral symmetry) is that the decay constant $f\!=\!\sqrt{2}F$
of a pseudoscalar meson (in the following we will distinguish the two
normalizations by using either the upper-case or the lower-case symbol) to the
zero component of an axial current can be extracted without recurrence to any
lattice-to-continuum matching factor.

We start from the two-point function $C_{PP}(t)$ between two pointlike
pseudoscalar density operators (at least one of which is projected to zero
spatial momentum) which, for an intermediate window of the Euclidean time $t$,
takes the form
\beq
C_{PP}(t)=A_{PP}\;\Big[\exp(-Mt)+\exp(-M(T\!-\!t))\Big]
\label{coshform}
\eeq
with $T$ the lattice extent in the fourth direction.
The mass $M$ corresponds to the mass of the lightest asymptotic state with
the right quantum numbers (here $\pi$ or $K$), whereas the amplitude $A_{PP}$
is proportional to the squared matrix element, i.e.\
$A_{PP}\propto\<0|P|xy\>^2/M$, where $|xy\>$ denotes the pseudoscalar state
put together from flavors $x,y$ (here $x,y=l,s$).

In practice it means that we determine, in a first step, the mass and the
amplitude from the $PP$ correlator.
We do this either via the effective mass and amplitude method where the determination of
\bea
M_\mr{eff}(t)&=&
\frac{1}{2}\log
\left(\frac%
{C(t\!-\!1)+\sqrt{C(t\!-\!1)^2-C(T/2)^2}}
{C(t\!+\!1)+\sqrt{C(t\!+\!1)^2-C(T/2)^2}}
\right)
\label{def_Meff}
\\
A_\mr{eff}(t)&=&\frac%
{C(t)}
{\exp[-M_\mr{eff}(t)\,t]+\exp[-M_\mr{eff}(t)\,(T\!-\!t)]}
\label{def_Aeff}
\eea
is followed by a fit to a constant over some time region
$t\in[t_\mr{min}:t_\mr{max}]$, or using a direct fit of the correlator to
the functional form (\ref{coshform}).
In either case the data are symmetrized about $T/2$, and
$t_\mr{max}\!\leq\!T/2$.
The decay constant is defined as $f_{xy}=\<0|A_4|xy\>/M_{xy}$, and via the PCAC
relation this is transformed into
\beq
f_{xy}=(m_x\!+\!m_y)\frac{\<0|P|xy\>}{\sqrt{2}M_{xy}^2}
\eeq
where $m_{x,y}$ denotes the quark mass of the flavor $x$ or $y$.
Putting things together, it follows that the decay constant may be obtained
from the amplitude and the mass as
\beq
f_{xy}\propto(m_x\!+\!m_y)\sqrt{\frac{A_{xy}}{M_{xy}^3}}
\eeq
where the missing prefactors (e.g.\ $L^3$) reflect normalization conventions
which depend on the geometry, but not on the quark masses.

\begin{figure}
\begin{center}
\includegraphics[width=.49\textwidth]{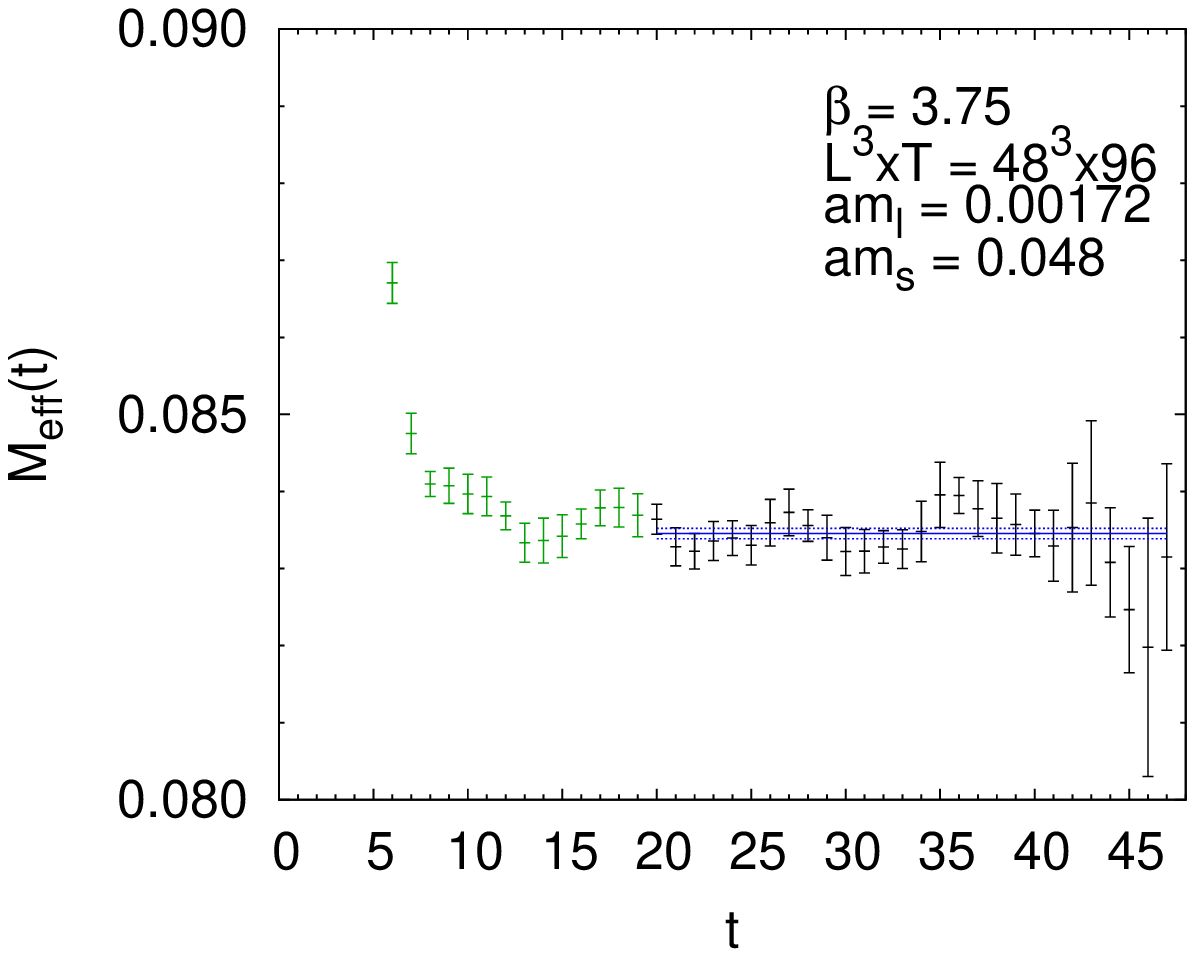}%
\includegraphics[width=.49\textwidth]{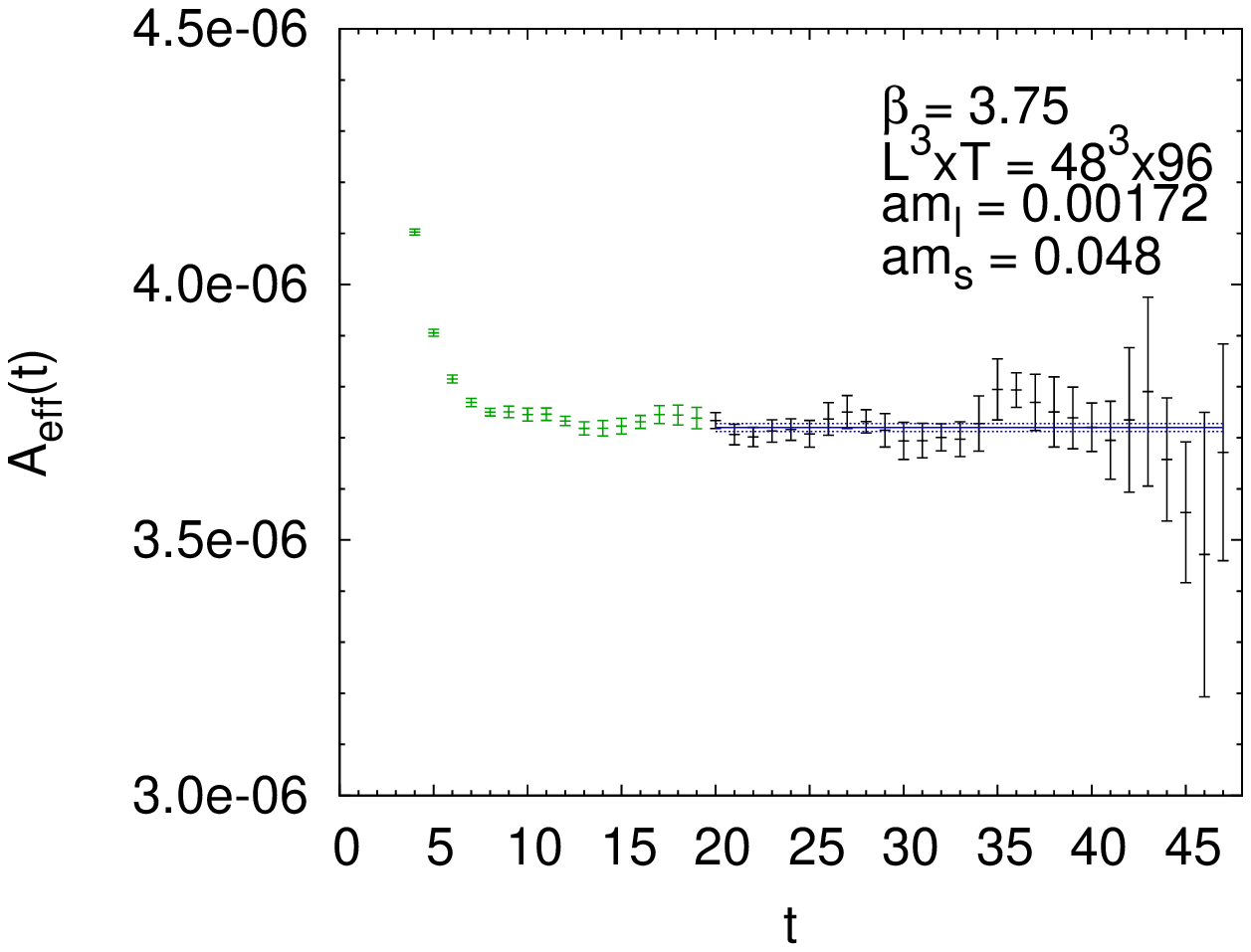}
\caption{Effective masses (\textit{left panel}) and amplitudes (\textit{right panel}) from the pion correlator at $\beta=3.75$, $am_l=0.00172$, $am_s = 0.048$, including the plateau-fits (\textit{blue lines}).}
\label{fig:plateaux}
\end{center}
\end{figure}

For the interim step, i.e.\ the determination of $M_\mr{eff},A_\mr{eff}$ from
the correlators, a typical plateau is shown in Fig.\,\ref{fig:plateaux}.
We looked for a potential zig-zag of the data close to the mid-time point
$T/2$.
This, if present, is commonly attributed to a back-propagating parity partner
\cite{Bazavov:2009bb} and reflects an effect which is specific to the staggered
discretization.
An advantage of the symmetric definition of the effective mass (\ref{def_Meff})
is that $M_\mr{eff}(t)$ for an odd value of $t$ uses only data from the
original correlator at even $t$ and vice versa.
Accordingly, we can compare the results of ($i$) a plateau average of
$M_\mr{eff}(t)$ for odd $t$, ($ii$) a plateau average of
$M_\mr{eff}(t)$ for even $t$, and ($iii$) the result of a direct fit to the
Ansatz (\ref{coshform}) (which does not distinguish between even and odd
time-slices).
We have carefully analyzed the impact of these options and found them
completely insignificant compared to both the statistical uncertainty and
(even more so) the theoretical uncertainty inherent in the precise choice of
the masses and lattice spacings included in the chiral fit.
The latter represent relevant options that will be discussed in detail in
Sections \ref{sec:NLOChPT} and \ref{sec:NNLOChPT} below.

The statistical errors are determined via a jackknife procedure (using an
extension known as superjackknife \cite{DelDebbio:2007pz,Bratt:2010jn} which
allows one to deal with ensembles of unequal size).
Typically the data are blocked in sets of $1,2,5,10$ configurations [where a
configuration corresponds to $O(10)$ trajectories], and we determine at which
level the jackknife error saturates.

We stress that all the fitting is performed within the jackknife procedure.
A common issue in many lattice calculations is that the covariance matrix (in
Euclidean time direction) of the local masses---e.g.\ the $28\times28$ matrix
$C$ that corresponds to $t\in[20,...,47]$ of Fig.\,\ref{fig:plateaux}---may not
be invertible, at least not on all jackknife samples.
This precludes a clean-cut definition of the $\chi^2$ of such a fit to the
primary data; one often truncates $C$ to its diagonal and near-diagonal parts
or uses pseudoinverses based on the singular value decomposition of $C$
to come up with a modified $\chi^2$.
We find that the ``fit within jackknife'' approach yields very robust values
of the statistical error of the fitted mass plateaus, regardless of which
effective $C^{-1}$ is used.
In the end we opted for using uncorrelated fits to the primary data to avoid
an uninvertible correlation matrix in some occasions.
The way in which the correlations among the secondary data $a^{-1}, M_{xy}, f_{xy}$
are treated will be discussed in Sec.\,\ref{subsec:NLOformulae} below.

%
%

\subsection{Finite-volume corrections}
\label{subsec:FV}

Pseudoscalar masses and decay constants experience a systematic shift due to
the finite spatial box length $L$.
Approximate three-loop and two-loop expressions have been given for the ratios
$\Mpi(L)/\Mpi(\infty)$ and $\fpi(L)/\fpi(\infty)$, respectively, in
Ref.\,\cite{Colangelo:2005gd}. For the range of quark masses used in this
work they are supposed to give a reliable estimate of these (small) shifts.
The numerical values thus obtained vary between 0.1 and 2.7 per mil for the
pion mass, and between 0.2 and 7.5 per mil for the pion decay constant.
For each ensemble we thus calculate these ratios and apply them to our data.
In the following, only the finite-volume corrected data are used.

\subsection{Other systematic uncertainties}
\label{subsec:other}

There is a number of systematic uncertainties which we cannot estimate
from our data set, since they are not systematically probed or varied.
These include the effect that a slight mistuning of the dynamical strange
quark mass has on the chiral low-energy constants, the effect of the
quenching of the charm quark, and the way in which we correct for the fact
that isospin is broken (by both electromagnetism and $m_u \neq m_d$)
in nature but not in our simulations.
The size of such effects can only be assessed by means of theoretical
arguments; see, e.g., the discussion in the FLAG report \cite{Colangelo:2010et}.
For instance a slight mistuning of the dynamical strange quark mass can be
believed to be tiny, since the FLAG compilation could not even detect a
statistically significant difference between SU(2) low-energy constants
determined from $\Nf=2$ and $\Nf=2+1$ simulations.
   In consequence a similar statement holds with respect to the quenching
   of the charm quark.
   Estimates of the impact of 
   isospin breakings on the definition of the physical values of $\Mpi$ and
   $\fpi$ are found in Ref.\,\cite{Colangelo:2010et}.
We find that all theoretically accounted sources of systematic
uncertainty are negligible compared to the systematic uncertainties that
emerge from the chiral fits (see Secs.\,\ref{sec:NLOChPT} and \ref{sec:NNLOChPT} below).

\section{Determining the lattice scale and physical quark mass}
\label{sec:scale}

As indicated in the previous section, we wish to determine for each $\be$ the
lattice spacing $a$ and the physical value of $am_l$.
Since the simulation points for any $\be$ are at a fixed value of $am_s$ (which
is tuned to its physical value, see Sec.\,\ref{subsec:ensembles}),
it is clear that the observables to be used shall include only the light but
not the strange flavor, and the obvious choice is thus $\fpi=\sqrt{2}\Fpi$
and $M_\pi$.

On a more technical level we proceed by means of a two-step procedure.
First, we extrapolate the ratio $(aM_{ll})^2/(af_{ll})^2=(a\Mpi)^2/(a\fpi)^2$
of the squared pion mass and decay constant to its physical value, $1.06846$.
Here we use the isospin averaged and electromagnetically corrected pion mass
$\Mpi^\mr{phys}=134.8\MeV$ from FLAG \cite{Colangelo:2010et}, and
the PDG value of the pion decay constant $\fpi^\mr{phys}=130.41\MeV$
\cite{Nakamura:2010zzi}.
In this step the purpose of the square is to reduce the amount of curvature,
and we interpolate the data by means of a low-order polynomial and rational
Ansatz (typically with three parameters applied to the five lightest data points,
i.e., with two degrees of freedom).
We stress that the point where this ratio assumes the desired value, $1.06846$,
is always very close to the lightest simulated quark mass.
In view of this it should not come as a surprise that the values of
$am_l^\mr{phys}$ that stem from the polynomial and the rational fit are
always very close to each other (on the scale set by the statistical error).
We use the average of the two as our central value; the difference should be
seen as indicative of the systematic uncertainty of $am_l^\mr{phys}$ from
this set of ensembles.

\begin{figure}
\begin{center}
\includegraphics[width=.47\textwidth]{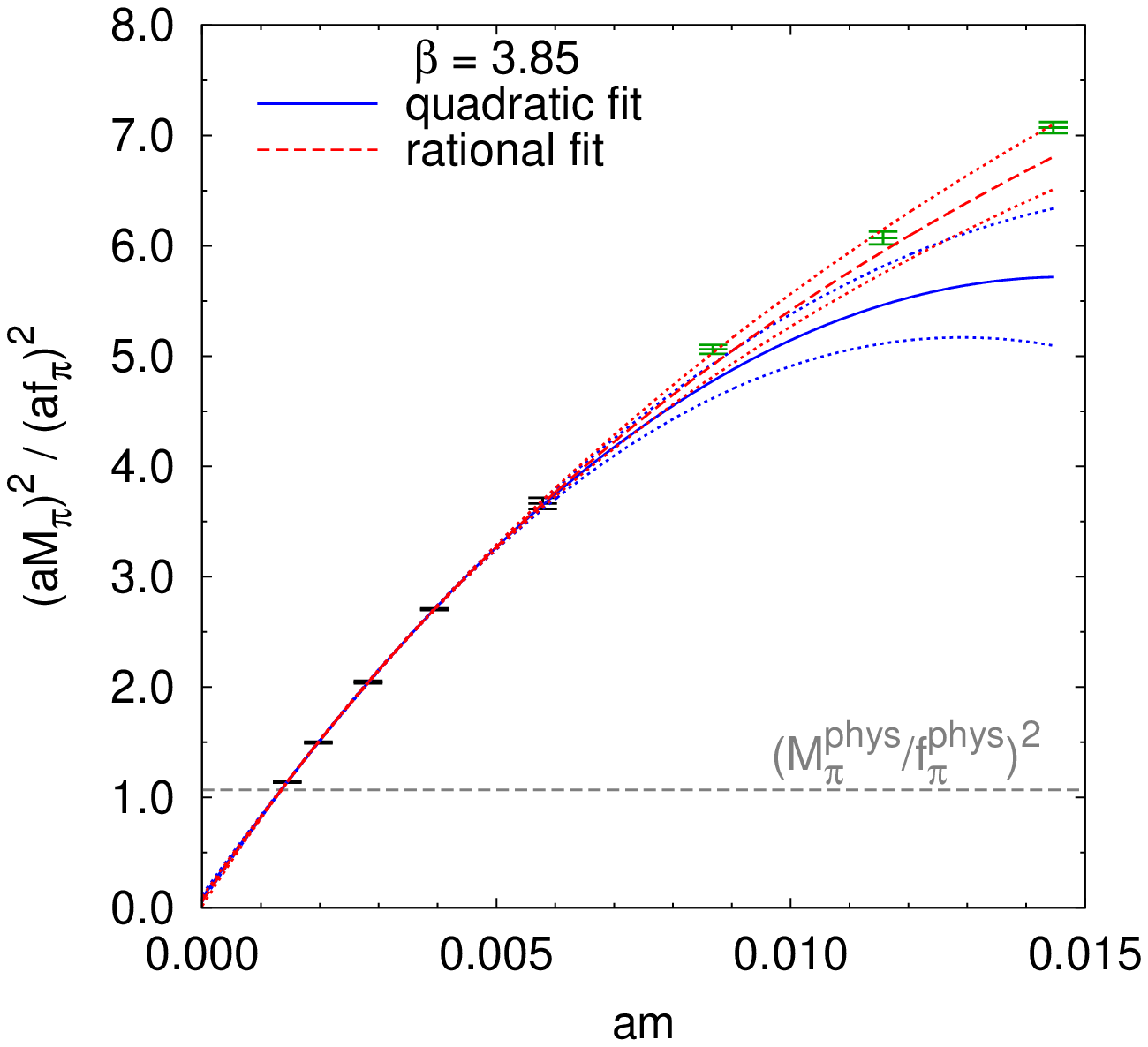}
\includegraphics[width=.47\textwidth]{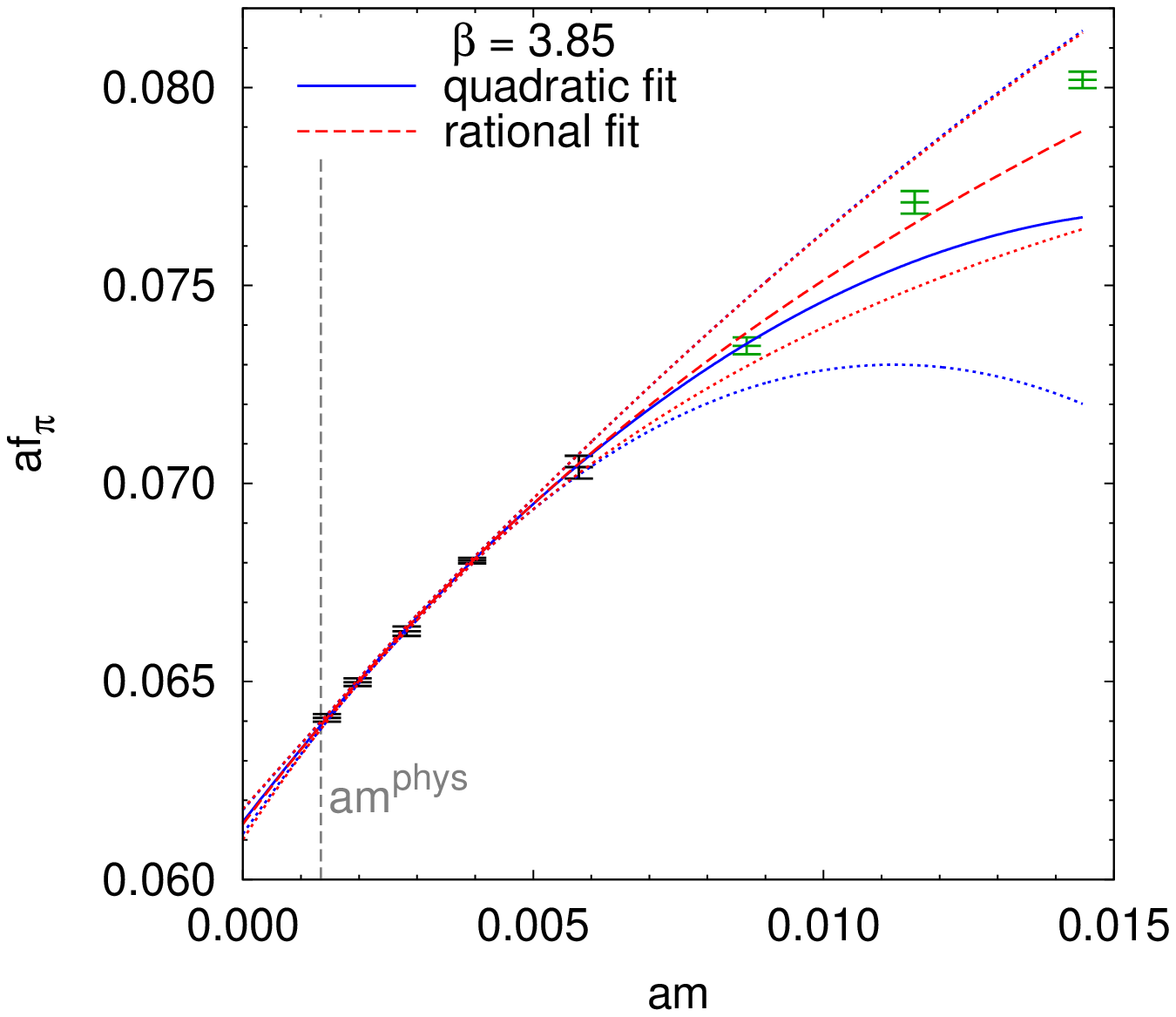}
\caption{Fits to the ratio $(aM_\pi)^2/(af_\pi)^2$ (\textit{left}) and the decay constant $af_\pi$ (\textit{right}) at $\beta=3.85$ to obtain the physical light quark mass $am^{\rm phys}$ and the lattice scale $1/a$. Shown are quadratic (\textit{solid blue lines}) and rational (\textit{dashed red lines}) fits which include the \textit{black data points}; the \textit{green data points} were excluded from the fits. Error bands are indicated by \textit{dotted lines}. The \textit{dashed gray lines} mark the physical value of the ratio and the physical light quark mass extracted therefrom.}
\label{fig:robust}
\end{center}
\end{figure}

In the second step we consider $a\fpi$ as a function of $am_l$.
Again, we interpolate the data with the same polynomial and rational Ansatz,
and determine the ordinate value at the abscissa point that was specified in
the previous step.
This value $a\fpi$ is then identified with the product of the lattice
spacing $a$ and the PDG value $\fpi^\mr{phys}=130.41\MeV$
\cite{Nakamura:2010zzi}; this yields the lattice spacing $a$ in fm for the
lattice theory at that particular value of the coupling $\be$.
A typical example of this two-step procedure is shown in
Fig.\,\ref{fig:robust}; in the relevant range (close to the lightest mass
point) the difference between the two Ans\"atze is invisibly small.
Furthermore, in Table~\ref{tab:robust} the results for the physical light quark
masses and lattice spacings obtained by this method are displayed.

As a final comment, let us remark that already in these two steps one could, in
principle, use ChPT.
We rather prefer to stay with the simple yet robust procedure as sketched
above.
This ensures that the fact that some chiral fits go wild (when an inadequate
fitting window is used, cf.\ the discussion in Secs.\,\ref{sec:NLOChPT} and
\ref{sec:NNLOChPT}) is not linked to a potential mishap in the
physical mass and scale determination.
In other words, we take the lattice spacing and the physical light quark mass
from an ``ideal'' simulation where $\Mpi/\fpi$ is exactly tuned to its physical
value, and use this knowledge regardless of how many data points enter the
chiral fits described below.
Needless to comment that for those NLO and NNLO fits which work fine (and which
include the lightest data points), their intrinsic physical mass and scale
determination were always found to be in very good agreement with the result of
the procedure described in the previous two paragraphs.

\begin{table}
\begin{center}
\begin{tabular}{cccc}
\hline\hline
$\beta$ & $am^{\rm phys}$ & $1/a\,/\,{\rm GeV}$ & $a\,/\,{\rm fm}$ \\\hline
3.45  & $5.771(18)\cdot10^{-3}$  & 0.69468(67) & 0.28406(27) \\
3.55  & $3.612(18)\cdot10^{-3}$  & 0.9165(12)  & 0.21531(28) \\
3.67  & $2.191(15)\cdot10^{-3}$  & 1.3063(25)  & 0.15105(29) \\
3.75  & $1.6889(55)\cdot10^{-3}$ & 1.6288(15)  & 0.12115(11) \\
3.792 & $1.5355(70)\cdot10^{-3}$ & 1.7935(22)  & 0.11002(14) \\
3.85  & $1.3430(98)\cdot10^{-3}$ & 2.0410(35)  & 0.09668(16) \\
\hline\hline
\end{tabular}
\caption{Physical light quark mass $am^{\rm phys}$, lattice scale $1/a$ and spacing $a$ at different $\beta$ obtained by the method described in Sec.~\ref{sec:scale}.}
\label{tab:robust}
\end{center}
\end{table}

\section{Fits to NLO-SU(2) ChPT}
\label{sec:NLOChPT}

In this section we will describe how we fit the quark-mass dependence of the meson decay constant and its squared mass to the prediction of NLO ChPT and in this way obtain the low-energy constants (LECs) appearing in these ChPT formulas. Since in the simulations considered here the strange quark mass was fixed to its physical value, we will only deal with the light quark-mass dependence described by SU(2) ChPT. The LECs extracted in this way therefore contain the correct contribution of the effects due to the strange quark present in nature. Further, we are restricting ourselves to the case of a degenerate light quark mass, which we will denote by $m$ (or $am$ in lattice units) throughout the remainder of this paper, i.e., from now on we drop the subscript $l$ on the light quark-mass parameter. At the physical point, this mass corresponds to the average mass of the two light quarks observed in nature: $m^{\rm phys}\:=\:(m_u^{\rm phys}+m_d^{\rm phys})/2$.

In the following we will try to fit our data to continuum ChPT and not consider
variants of ChPT, which take into account lattice discretization effects and/or
taste violations present in the staggered formulation; see, e.g.,
Refs.\,\cite{Aubin:2003mg,Aubin:2003uc,Bazavov:2009bb}.
Whether a continuum Ansatz is suitable to describe our data is not {\it a priori}
clear and needs to be tested.
It is valid if, within the statistical precision of our data, no cutoff
dependence is seen in the observables considered.
As will be shown below, this is indeed the case for our $\Mpi$ and $\fpi$ data
in the relevant region.
This can be traced back to the combined effect of (a) the two levels of stout
smearing in the action (see Sec.\,\ref{sec:Lattice}) and (b) the specific
choice of our scaling trajectory, i.e., the scale and the physical quark mass being
set through $\fpi$ and $\Mpi$ (see Sec.\,\ref{sec:scale}).
While (b) ensures that discretization effects on both $\Mpi$ and $\fpi$ vanish
at the physical point, (a) keeps them small in its vicinity due to the
suppression of taste violations (see, e.g., Refs.\,\cite{Blum:1996uf,Orginos:1999cr}
and \cite{Borsanyi:2010bp}).
As we will see below, the combination of these two effects leads to a
suppression of discretization effects within our statistical precision over a
sufficiently wide region of quark masses and lattice spacings.

\subsection{Methods and fit formulas}
\label{subsec:NLOformulae}

ChPT up to NLO predicts for the decay constant $f_\pi$ and the squared mass $M_\pi^2$ of a meson consisting of two mass-degenerate quarks of mass $m$ the following functional form \cite{Gasser:1983yg,Gasser:1984gg,Colangelo:2010et}
%
\begin{eqnarray}
\label{eq:NLO:Msq}
M_{\pi}^2 &=& \chi\left[1\:+\:\frac{\chi}{16\pi^2f^2}\log\frac{\chi}{\Lambda_3^2}\right]\,,\\
\label{eq:NLO:f}
f_{\pi} &=& f\,\left[1\:-\:\frac{\chi}{8\pi^2f^2}\log\frac{\chi}{\Lambda_4^2}\right]\,,\\
\label{eq:NLO:chi}
\chi &=& 2Bm\,.
\end{eqnarray}
%
At this order four LECs appear: the decay constant $f$ in the SU(2)-chiral limit ($m\to0$), the condensate parameter $B$, and two low-energy scales $\Lambda_3$ and $\Lambda_4$. The condensate parameter $B$ depends like the quark mass $m$ on the renormalization scheme, but the combination $\chi$ is renormalization scheme independent, and it is this combination which will be used exclusively in this work. The low-energy scales are related to the LECs $\bar{\ell}_3$ and $\bar{\ell}_4$ at the scale of the physical pion mass $M_\pi^{\rm phys}$ via
%
\begin{equation}
\label{eq:lbar34:def}
\bar{\ell}_i \;=\; \log\frac{\Lambda_i^2}{(M_\pi^{\rm phys})^2}\,,\quad i=3,4\,.
\end{equation}

Since we used the physical values $f_\pi^{\rm phys}$ and $(M_\pi^{\rm phys})^2$ to set the scale and determine the physical light quark mass $m^{\rm phys}$ for each set of lattice ensembles with a given gauge coupling $\beta$, our Ansatz should reproduce the physical point. Note also, that each set of lattice ensembles contained one simulated point in close vicinity of the physical point. Therefore, we could impose the following constraints to the chiral formulas [Eqs.~(\ref{eq:NLO:Msq}) and (\ref{eq:NLO:f})]:
%
\begin{equation}
\label{eq:NLO:constraints}
\left.M_{\pi}^2\right|_{m=m^{\rm phys}}\:=\:(M_\pi^{\rm phys})^2\,,\;\;\;\left.f_{\pi}\right|_{m=m^{\rm phys}}\:=\:f_\pi^{\rm phys}\,.
\end{equation}
%
As it is easily derived, these two constraints result in the relations
%
\begin{eqnarray}
\label{eq:NLO:Lam3constr}
\log\frac{\chi^{\rm phys}}{\Lambda_3^2}&=&\frac{16\pi^2f^2}{(\chi^{\rm phys})^2}\Big((M_\pi^{\rm phys})^2-\chi^{\rm phys}\Big)\,,\\
\label{eq:NLO:Lam4constr}
\log\frac{\chi^{\rm phys}}{\Lambda_4^2}&=&\frac{8\pi^2f^2}{f_\pi^{\rm phys}}\Big(f-f_\pi^{\rm phys}\Big)
\end{eqnarray}
%
between the LECs, where $\chi^{\rm phys}\,=\,2Bm^{\rm phys}$. Using them to eliminate, e.g., the low-energy scales from the NLO ChPT formulas, the constrained formulas can be written as
\begin{eqnarray}
\label{eq:NLO:Msqconstr}
M_{\pi}^2 &=& \chi\left[1\:+\:\frac{\chi}{16\pi^2f^2}\log\frac{\chi}{\chi^{\rm phys}}\:+\:\frac{\chi}{\chi^{\rm phys}}\,\frac{(M_\pi^{\rm phys})^2-\chi^{\rm phys}}{\chi^{\rm phys}}\right]\,,\\
\label{eq:NLO:fconstr}
f_{\pi} &=& f\left[1\:-\:\frac{\chi}{8\pi^2f^2}\log\frac{\chi}{\chi^{\rm phys}}\:-\:\frac{\chi}{\chi^{\rm phys}}\,\frac{f-f_\pi^{\rm phys}}{f}\right]\,.
\end{eqnarray}
Note that $\chi/\chi^{\rm phys}\,=\,m/m^{\rm phys}$. These formulas now depend on two LECs, $B$ and $f$, and the two physical input values $M_\pi^{\rm phys}$ and $f_\pi^{\rm phys}$. In our fitting procedure only $\chi^{\rm phys}$ and $f$ will be treated as free parameters. For that reason, we also like to refer to these fits as parameter-reduced fits. [Of course, by treating $M_\pi^{\rm phys}$ and $f_\pi^{\rm phys}$ as free parameters as well, one would recover the unconstrained Eqs.~(\ref{eq:NLO:Msq}) and (\ref{eq:NLO:f}), respectively.] 

In the following we want to perform combined (i.e., fitting $M_{\pi}^2$ and $f_{\pi}$ simultaneously) global fits to our lattice data at the different gauge couplings available to us. For this reason we make use of the lattice scale and physical light quark mass determined beforehand (see Sec.\,\ref{sec:scale}). The meson mass and decay constant in lattice units at a given gauge coupling $\beta$ are converted into physical units by
\[ M_{\pi}^2\;=\;(1/a)_\beta^2\,(aM_{\pi})_\beta^2\,,\;\;\;f_{\pi}\;=\;(1/a)_\beta\,(af_{\pi})_\beta\,,\]
respectively. Furthermore, we rewrite the combination $\chi=2Bm$ as
\[\chi\;=\;(2Bm^{\rm phys})\,\frac{(am)_\beta}{(am^{\rm phys})_\beta}\;=\;\chi^{\rm phys}\,\frac{(am)_\beta}{(am^{\rm phys})_\beta}\]
and will determine only the renormalization-independent combination $\chi^{\rm phys}=2Bm^{\rm phys}$ in our fits. The factor $m^\mr{phys}$ will be removed, based on external data, in Sec.\,\ref{sec:Conclusions}.

It follows that with this setup, there is no correlation between secondary quantities (the pion mass $\Mpi$ and decay constant $\fpi$) from ensembles with different $\beta$ (cf.\ Table \ref{tab:ensembles}), but there is a substantial correlation among $\Mpi$ and $\fpi$ on any individual ensemble (because they are extracted from the same pseudoscalar correlator), and there is a weak correlation among all $\Mpi$ and $\fpi$ with the same value of $\beta$ (because the scale setting attributes a joint $(1/a)_\be$ and $(am^\mr{phys})_\be$ to all ensembles with a common $\be$).
As explained at the end of Sec.\,\ref{subsec:meson} we perform all fitting within the jackknife procedure.
In the present context this means that the statistical uncertainty from the scale setting is propagated into the uncertainty of the fitted NLO ChPT low-energy constants.
Similarly to what was reported at the end of Sec.\,\ref{subsec:meson} for primary quantities, we now observe a genuine robustness of the fitted parameters to the details of the pseudoinverse that is formed from the covariance matrix among the secondary observables.
The respective correlated $\chi^2$ and $p$-values of the fit do, however, show a clear sensitivity to the details of the pseudoinverse.
Therefore we decided to always quote results obtained with uncorrelated fits, but with the statistical uncertainty (of both the fitted parameters and the uncorrelated $\chi^2$ values) determined through the outer jackknife procedure (using the superjackknife extension
described in Refs.\,\cite{DelDebbio:2007pz,Bratt:2010jn}).

\subsection{Combined global fits}
\label{subsec:NLOfits}

\subsubsection{NLO ChPT without constraints}
\label{subsubsec:NLOfits_free}

\begin{figure}
\begin{center}
\includegraphics[width=.7\textwidth]{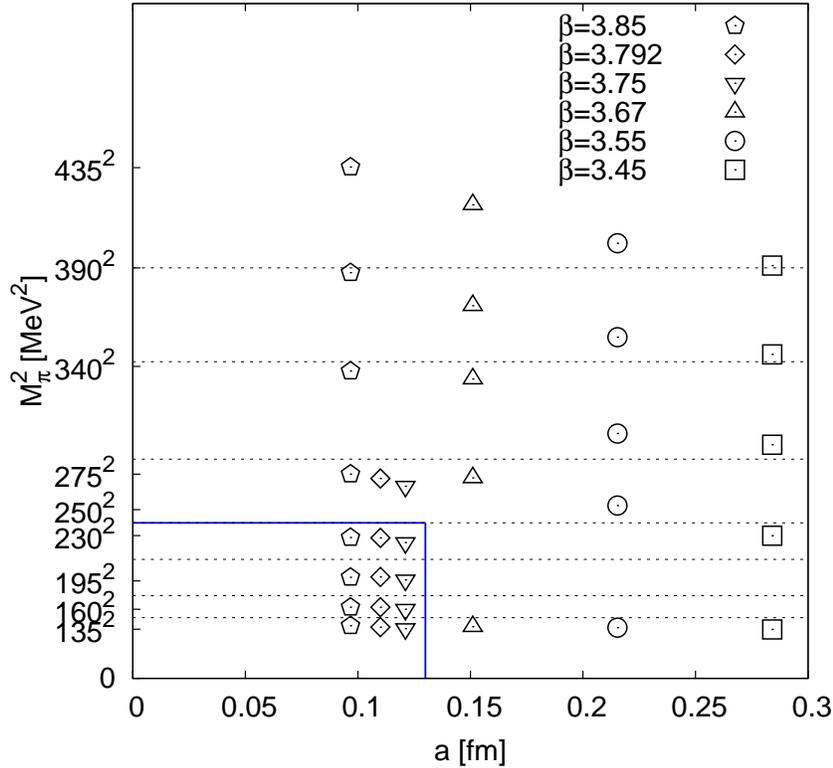}
\caption{Landscape plot of the simulated pion masses $M_\pi^2$ and lattice spacings $a$. The \textit{horizontal dashed lines} indicate the various mass cuts applied in our chiral fits. The \textit{full blue line} indicates our preferred fit range $(a<0.12\fm, \Mpi<240\MeV)$, see text for details.}
\label{fig:landscape}
\end{center}
\end{figure}

We begin the discussion of the chiral fits with the results of applying the unconstrained fit formulas, Eqs.~(\ref{eq:NLO:Msq}) and (\ref{eq:NLO:f}), to our data. The combined fit has four free parameters: $\chi^{\rm phys}=2Bm^{\rm phys}$, $f$, $\Lambda_3$, and $\Lambda_4$. {\it A priori}, it is not clear whether all simulated lattice spacings will lie in the scaling region, especially for the very coarse lattices with lattice spacings of up to 0.28 fm (corresponding to a lattice scale of $1/a\approx0.7\,{\rm GeV}$) this is questionable. Also, the range of quark masses or equivalently meson masses to which NLO-SU(2) ChPT is applicable will have to be determined. Eventually, the fit quality, which we measure by the standard $\chi^2/\textrm{d.o.f.}$, will be used to decide on these issues. In Fig.\,\ref{fig:landscape} we provide a landscape of the simulated meson masses at the various lattice spacings. The horizontal dashed lines indicate the various mass cuts we applied in our global combined fits. One should keep in mind that for the combined fit, each point in the landscape plot represents two data points: one for the meson mass and one for the meson decay constant.

\begin{figure}
\begin{center}
\includegraphics[width=.47\textwidth]{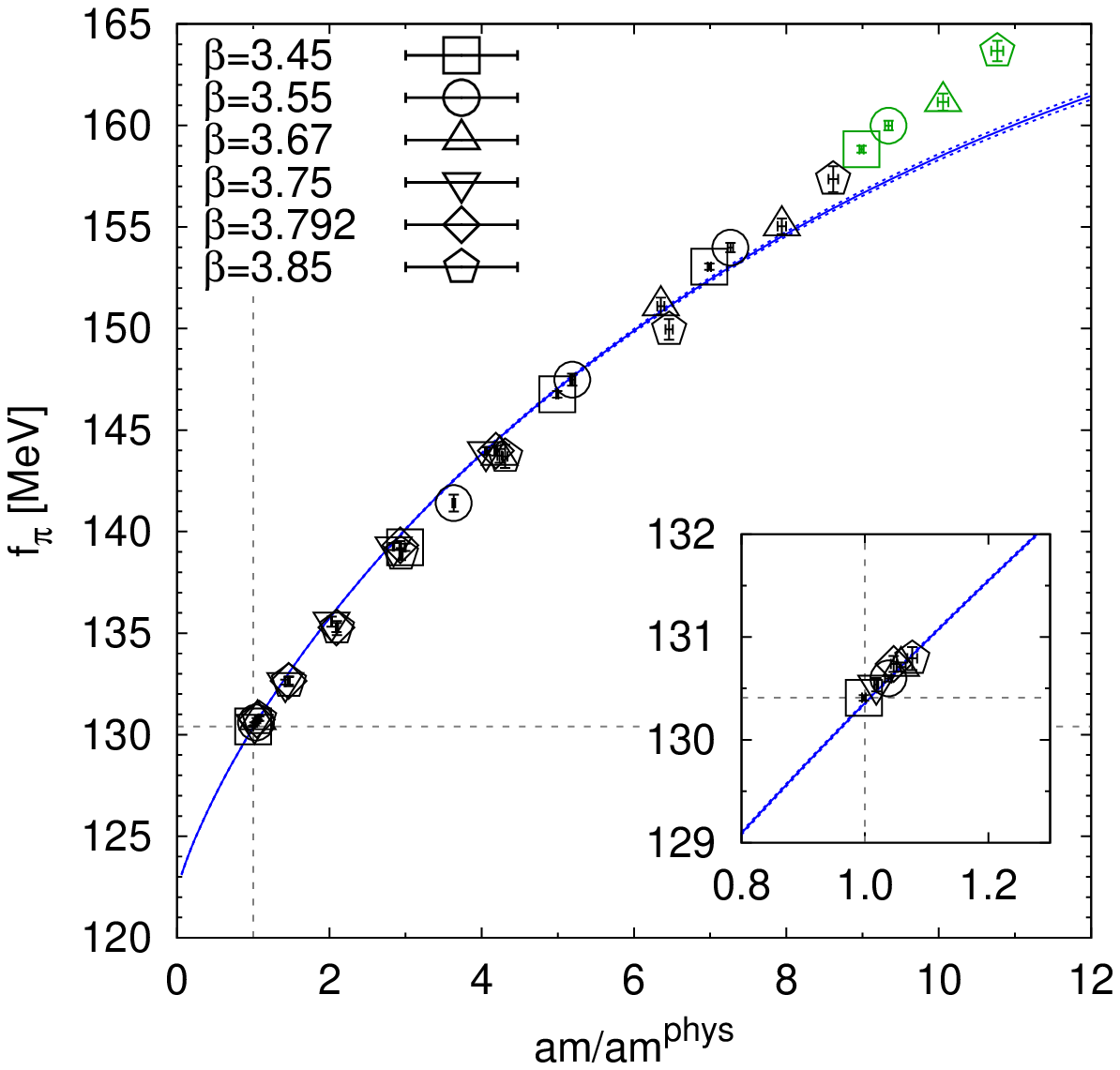}%
\includegraphics[width=.47\textwidth]{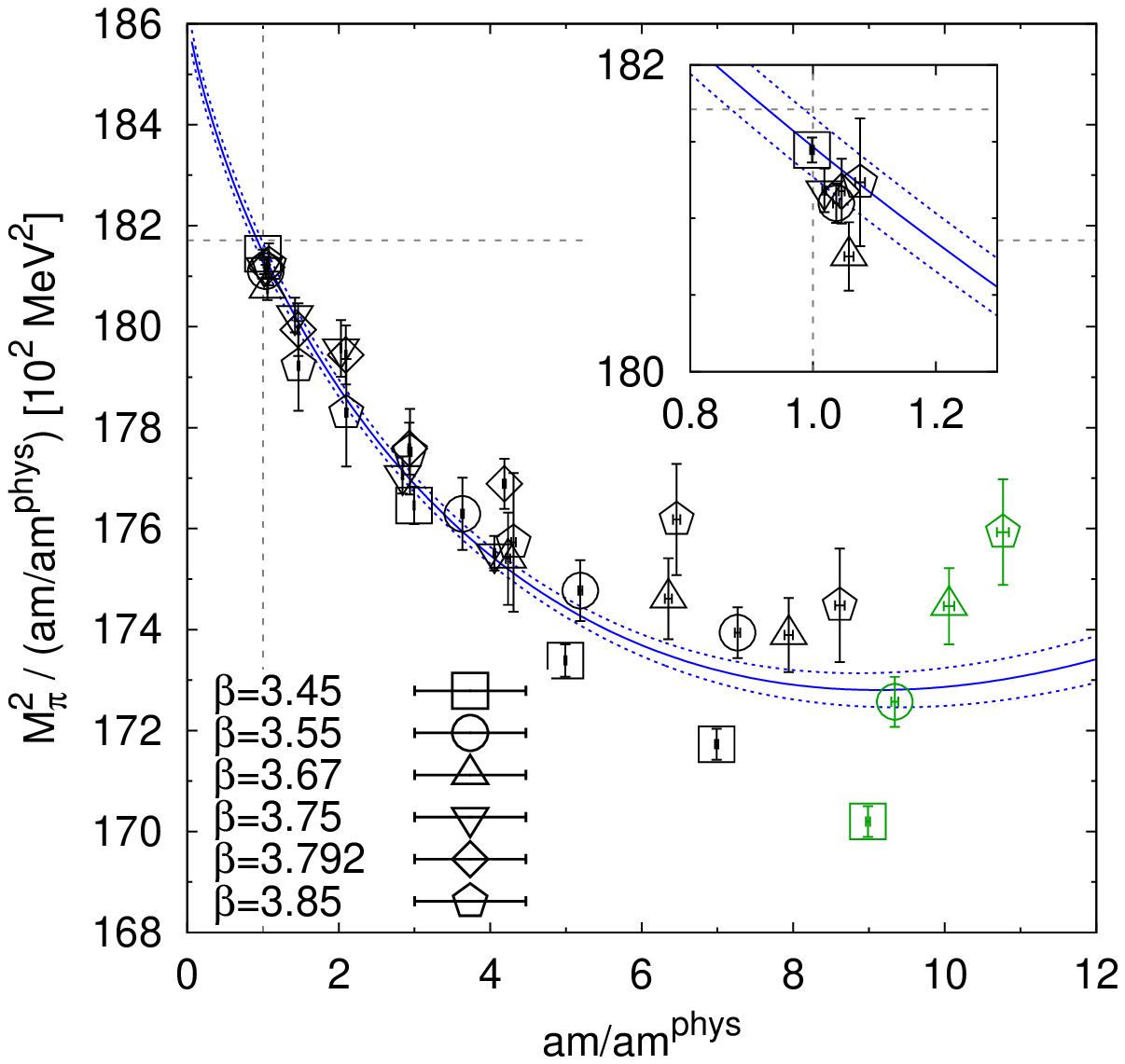}
\caption{Combined fit for all lattice scales, $135\,{\rm MeV}\:\leq\:M_{\pi}\:\leq\:390\,{\rm MeV}$. \textit{Left panel:} meson decay constant, \textit{right panel:} squared meson mass divided by quark-mass ratio. Points marked by \textit{black symbols} are included in the fit, those marked by \textit{green symbols} are excluded. The physical values are indicated by \textit{dashed gray lines}.} 
\label{fig:NLO:allB_135-390}
\end{center}
\end{figure}

\begin{figure}
\begin{center}
\includegraphics[width=.45\textwidth]{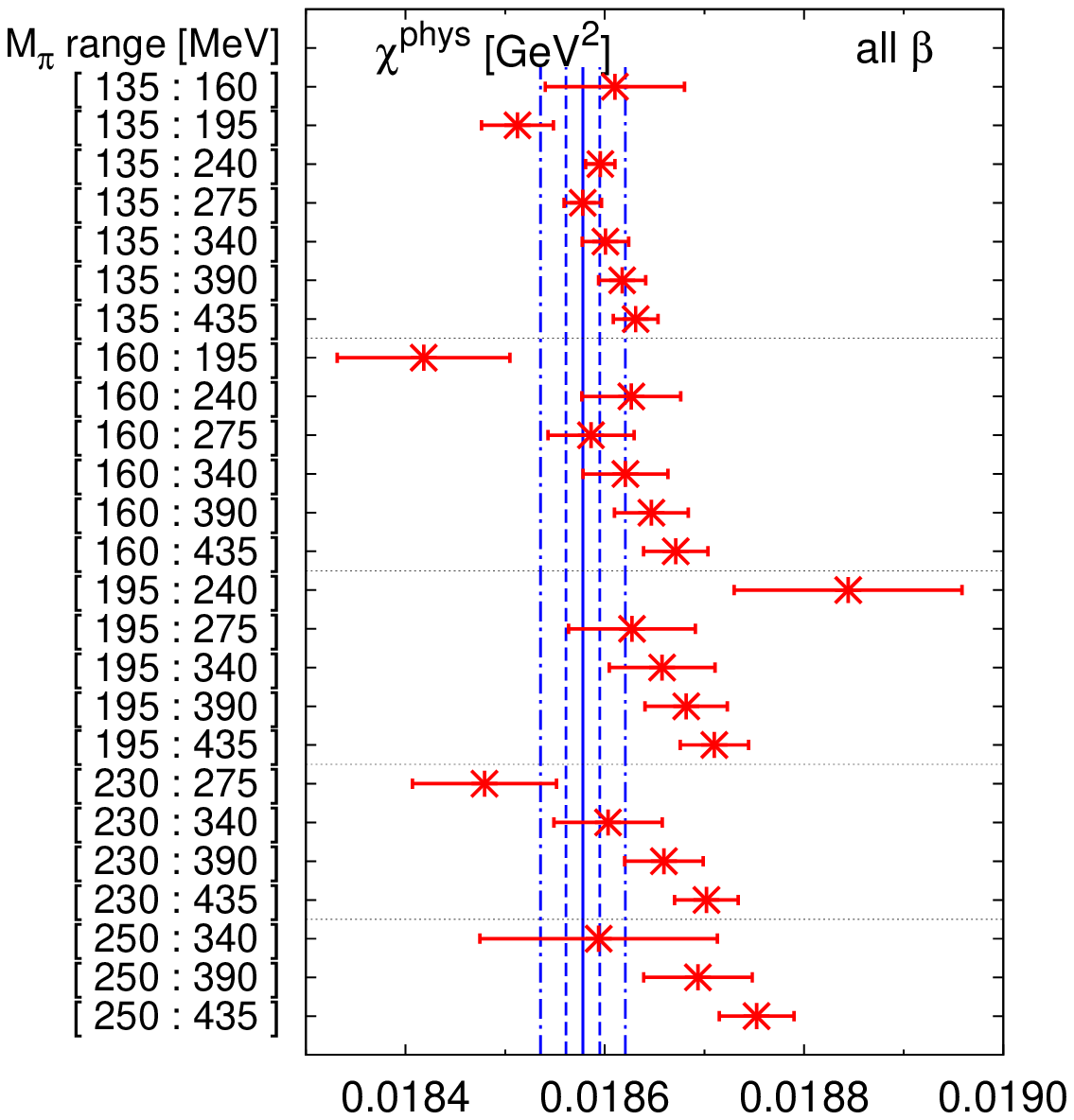}%
\includegraphics[width=.45\textwidth]{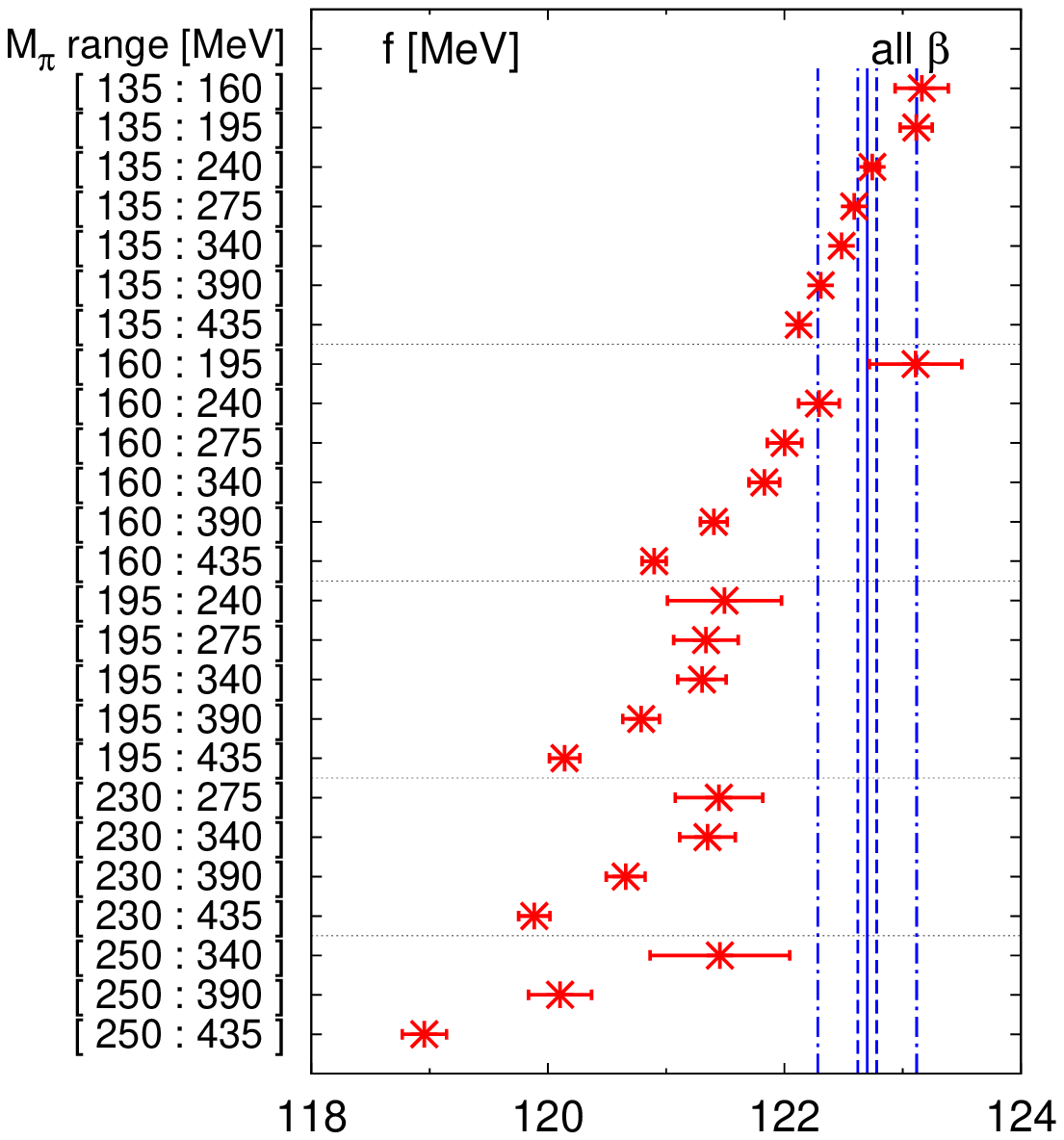}\\
\includegraphics[width=.45\textwidth]{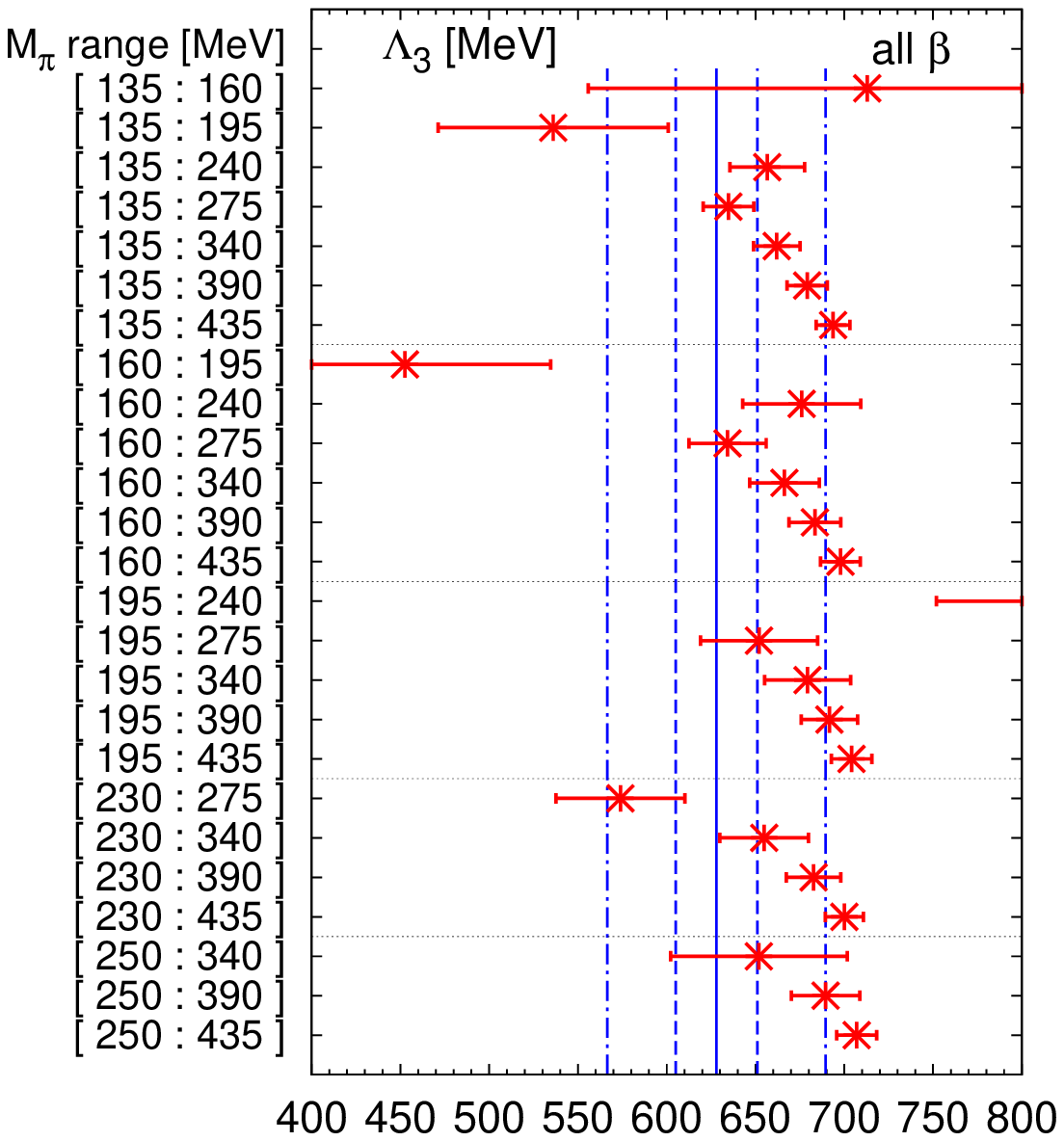}%
\includegraphics[width=.45\textwidth]{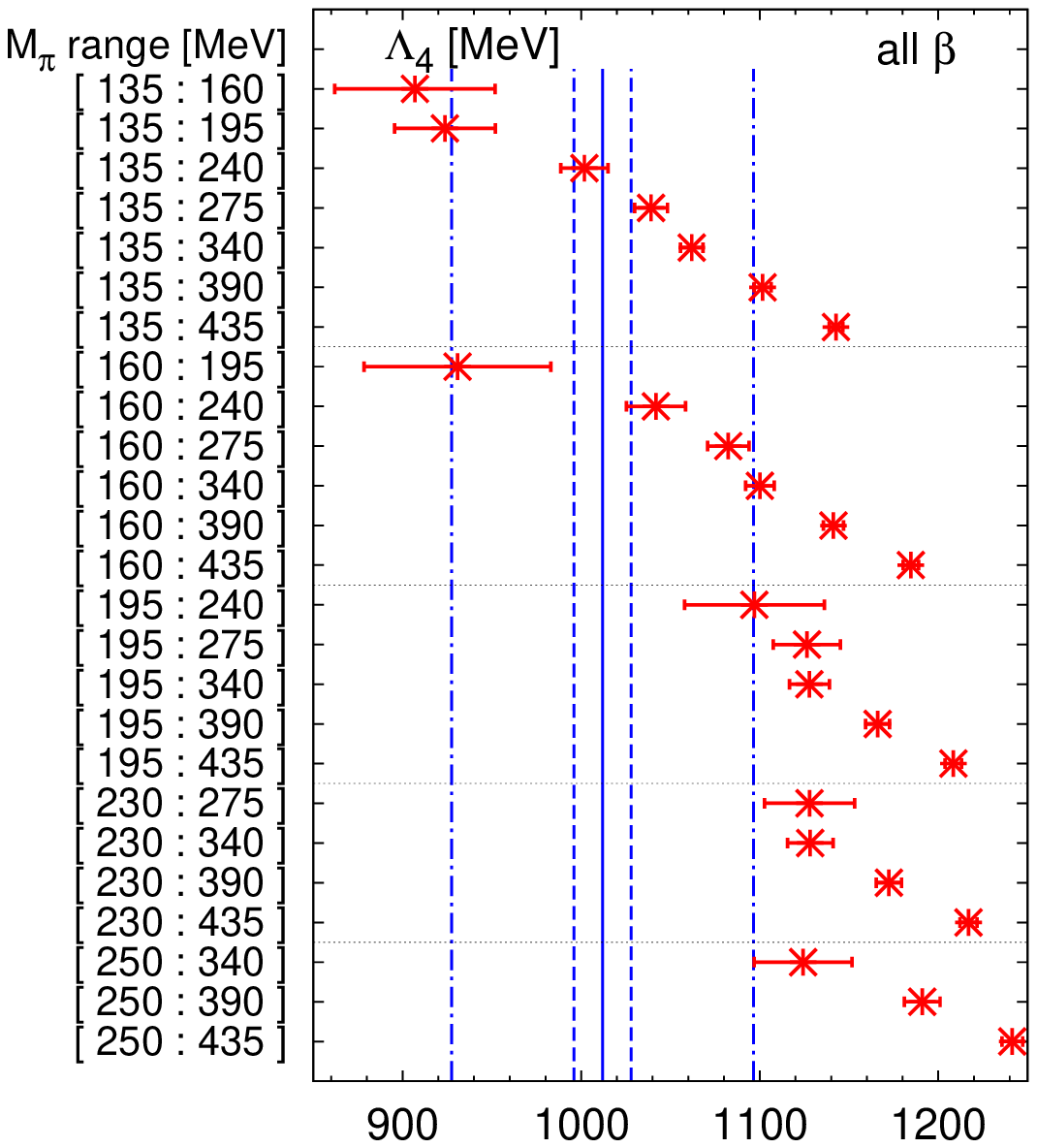}\\
\includegraphics[width=.45\textwidth]{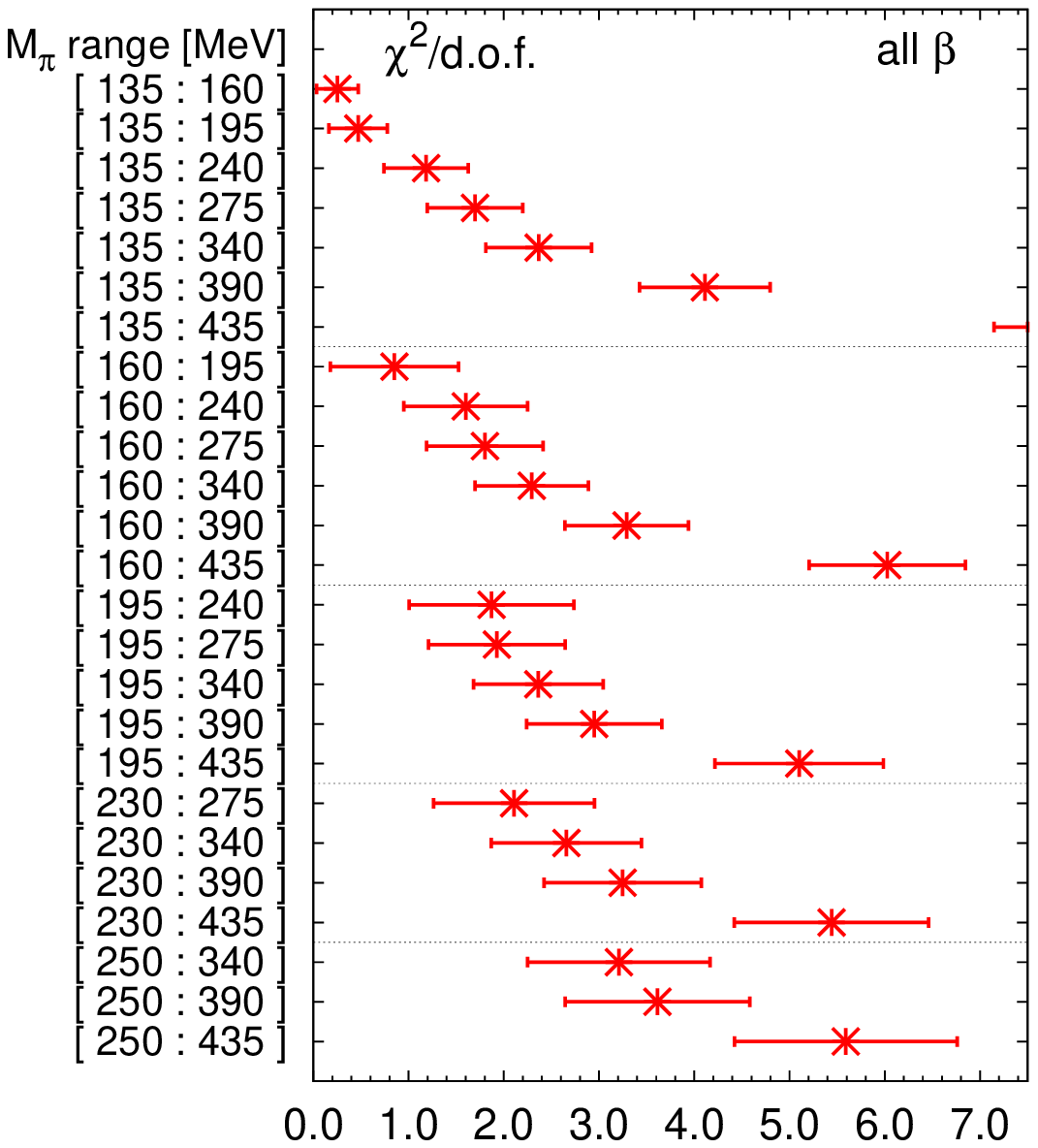}%
\begin{minipage}[b]{.45\textwidth}
\caption{Results for the fitted parameters (\textit{top} and \textit{middle panels}) and $\chi^2/\textrm{d.o.f.}$ (\textit{bottom panel}) from NLO-ChPT fits without constraints using different mass ranges but including all lattice spacings. The \textit{solid, dashed and dashed-dotted blue lines} for the fit parameters denote the central value, statistical and total (statistical plus systematic) error bands, respectively, from our preferred unconstrained fit (cf.\ left column of Table~\ref{tab:NLO:results}).}
\label{fig:NLO:massrange_allB}
\end{minipage}%
\end{center}
\end{figure}

\begin{figure}
\begin{center}
\includegraphics[width=.45\textwidth]{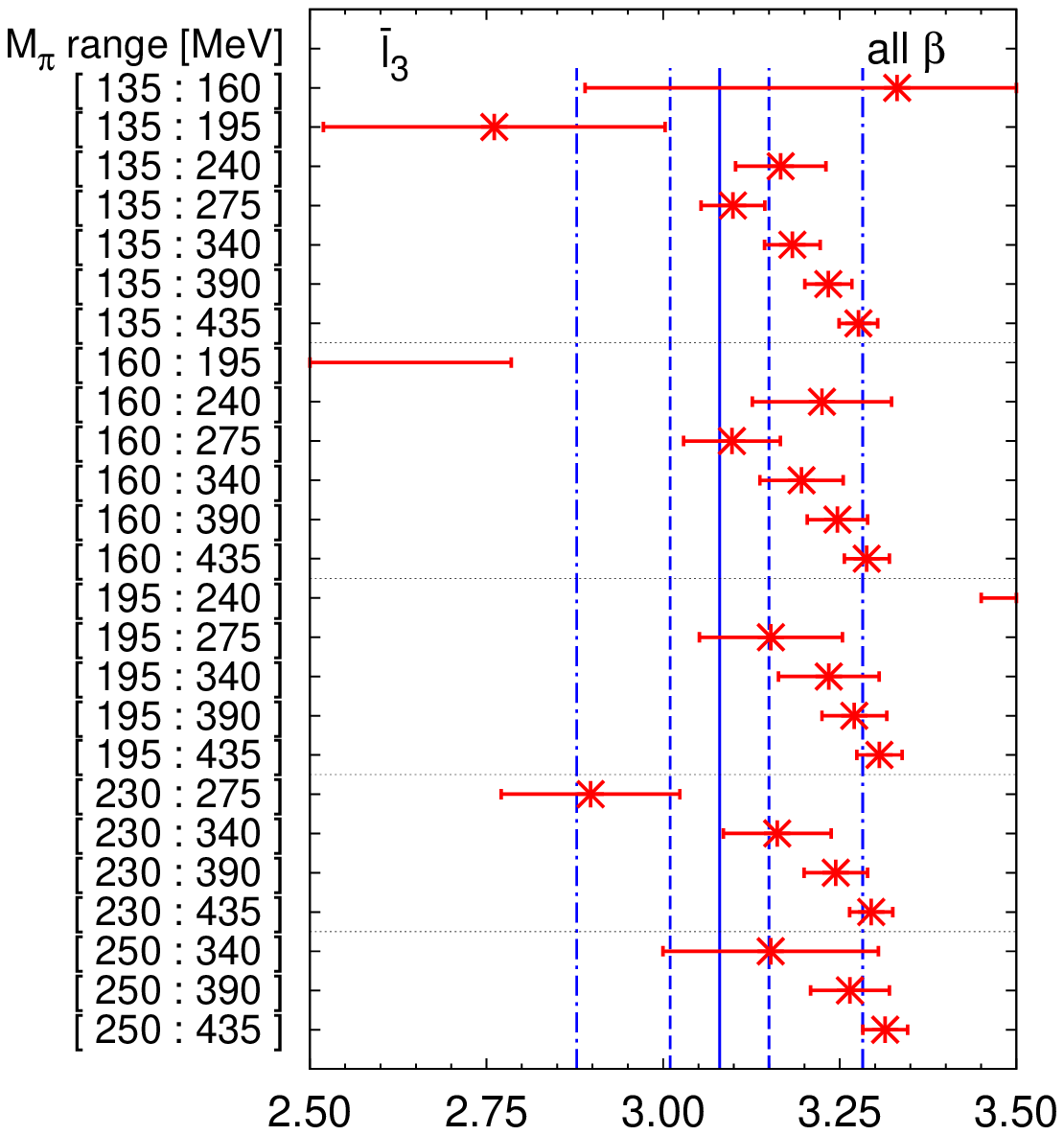}%
\includegraphics[width=.45\textwidth]{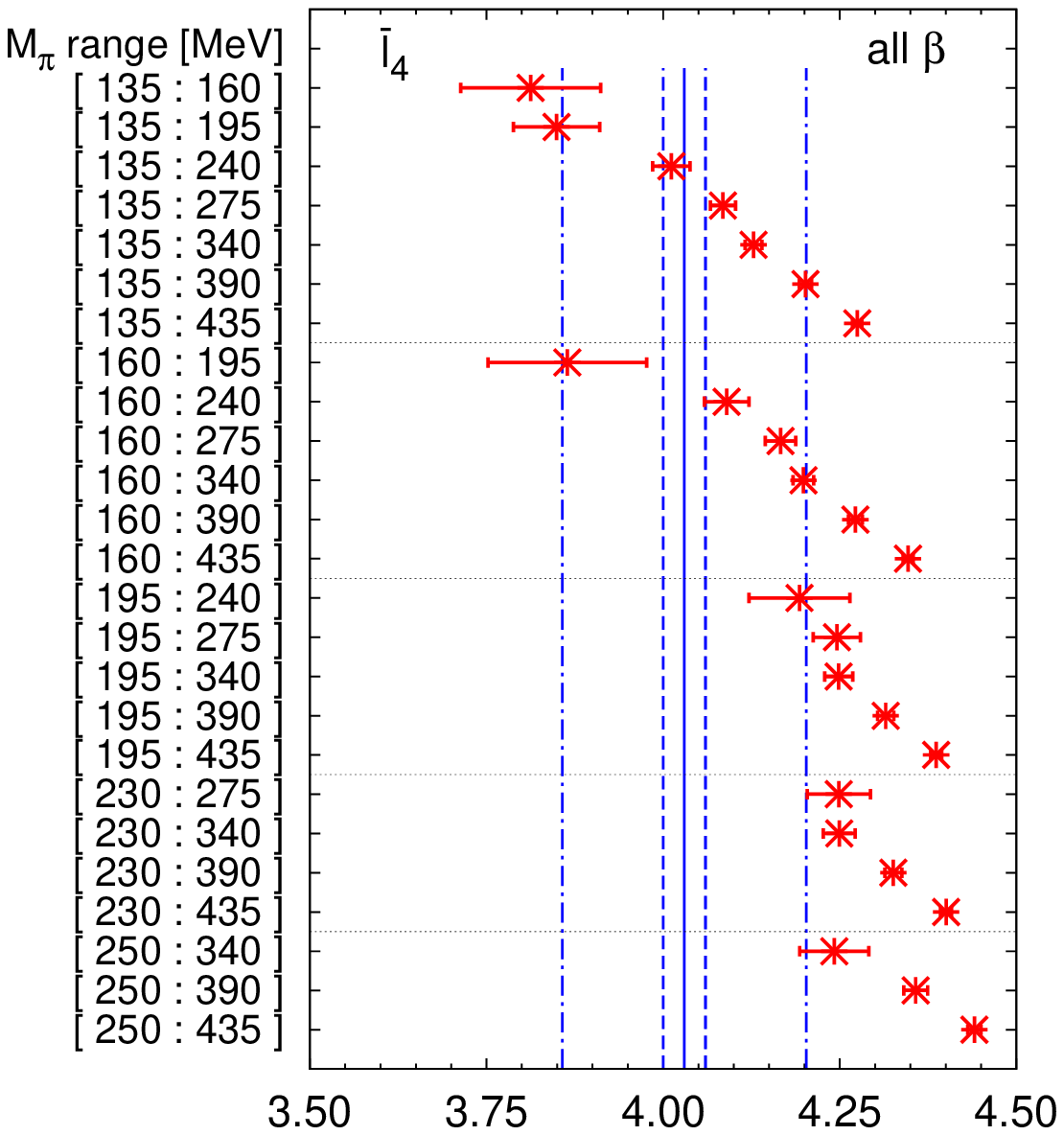}\\
\includegraphics[width=.45\textwidth]{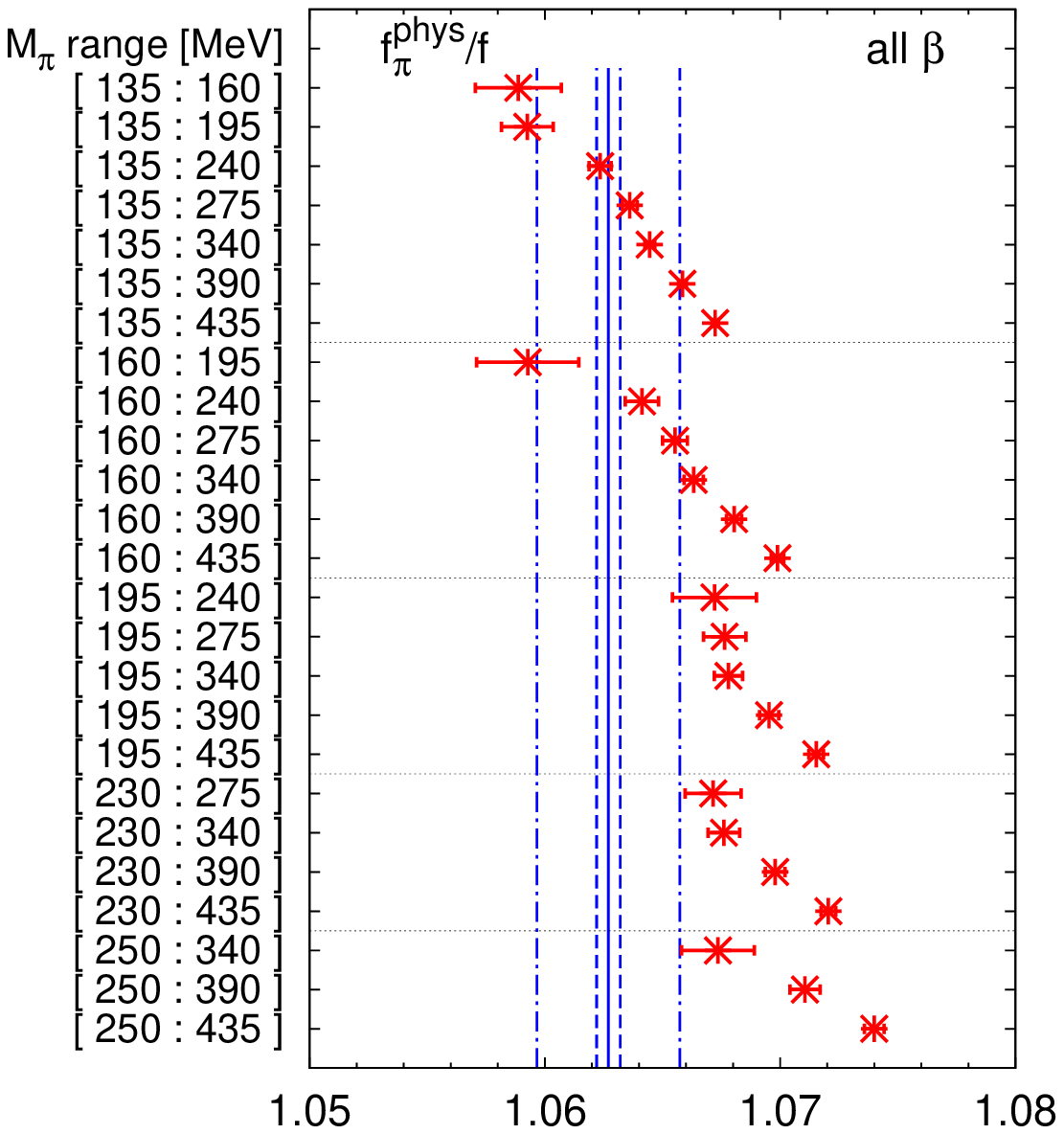}%
\begin{minipage}[b]{.45\textwidth}
\caption{Results for the parameters $\bar{\ell}_3$, $\bar{\ell}_4$ (\textit{top panels}) and the ratio $f_\pi^{\rm phys}/f$ (\textit{bottom panel}) from NLO-ChPT fits without constraints using different mass ranges but including all lattice spacings. The \textit{solid, dashed and dashed-dotted blue lines} for the fit parameters denote the central value, statistical and total (statistical plus systematic) error bands, respectively, from our preferred unconstrained fit (cf.\ left column of Table~\ref{tab:NLO:results}).}
\label{fig:NLO:massrange_allB_pheno}
\end{minipage}%
\end{center}
\end{figure}

In Fig.\,\ref{fig:NLO:allB_135-390} we show the result of a combined fit to the meson masses and decay constants to the data at all available lattice spacings and a mass range of $135\,{\rm MeV}\:\leq\:M_{\pi}\:\leq\:390\,{\rm MeV}$, i.e.\ excluding only the heaviest point at each simulated $\beta$ value. These plots show all available data points, whether or not included in the fit range. We marked those points which have been excluded by green symbols, while points included in the fit range are marked by black symbols. As one can already judge by eye, this fit gives a bad $\chi^2/\textrm{d.o.f.}$ of about 4.1 (with $\#\mr{d.o.f.}=58-4=54$), although the fitted parameters are in a reasonable range, cf.\ Fig.\,\ref{fig:NLO:massrange_allB}. In a first step, we consider reducing the mass range by excluding more and more of the heavier points as indicated by the dashed horizontal lines in our landscape plot, Fig.\,\ref{fig:landscape}. In Fig.\,\ref{fig:NLO:massrange_allB} we show how the fitted parameters and the $\chi^2/\textrm{d.o.f.}$ vary when we change the mass range in the fit, still considering the ensembles at all available lattice spacings. First we focus on those fit ranges including the near physical points which are shown above the topmost dashed horizontal line in each of the plots in Fig.\,\ref{fig:NLO:massrange_allB}. (In the fit results shown below that line, the near physical points and subsequently other points with light meson masses have been excluded; we will comment on those results below.) As one can see, the $\chi^2/\textrm{d.o.f.}$ improves by narrowing the fit range, and for the range $135\,{\rm MeV}\:\leq\:M_{\pi}\:\leq\:240\,{\rm MeV}$, it is already comparable with 1.0. The four fit parameters do not plateau yet, but their magnitude seems to settle. Note, that for the two narrowest fit ranges (upper mass limit at 160 or 195 MeV), the number of available data points might be too small as is also indicated by the larger error bars at these ranges. We already show in these plots the central values and error bands as determined from our preferred fit below, just to demonstrate that these error bands are compatible with the values obtained from fits to all ensembles. For completeness, in Fig.\,\ref{fig:NLO:massrange_allB_pheno} we show the LECs $\bar{\ell}_3$, $\bar{\ell}_4$ and the phenomenologically relevant ratio $f_\pi^{\rm phys}/f$ as determined from our fits with different mass ranges. 


\begin{figure}
\begin{center}
\includegraphics[width=.45\textwidth]{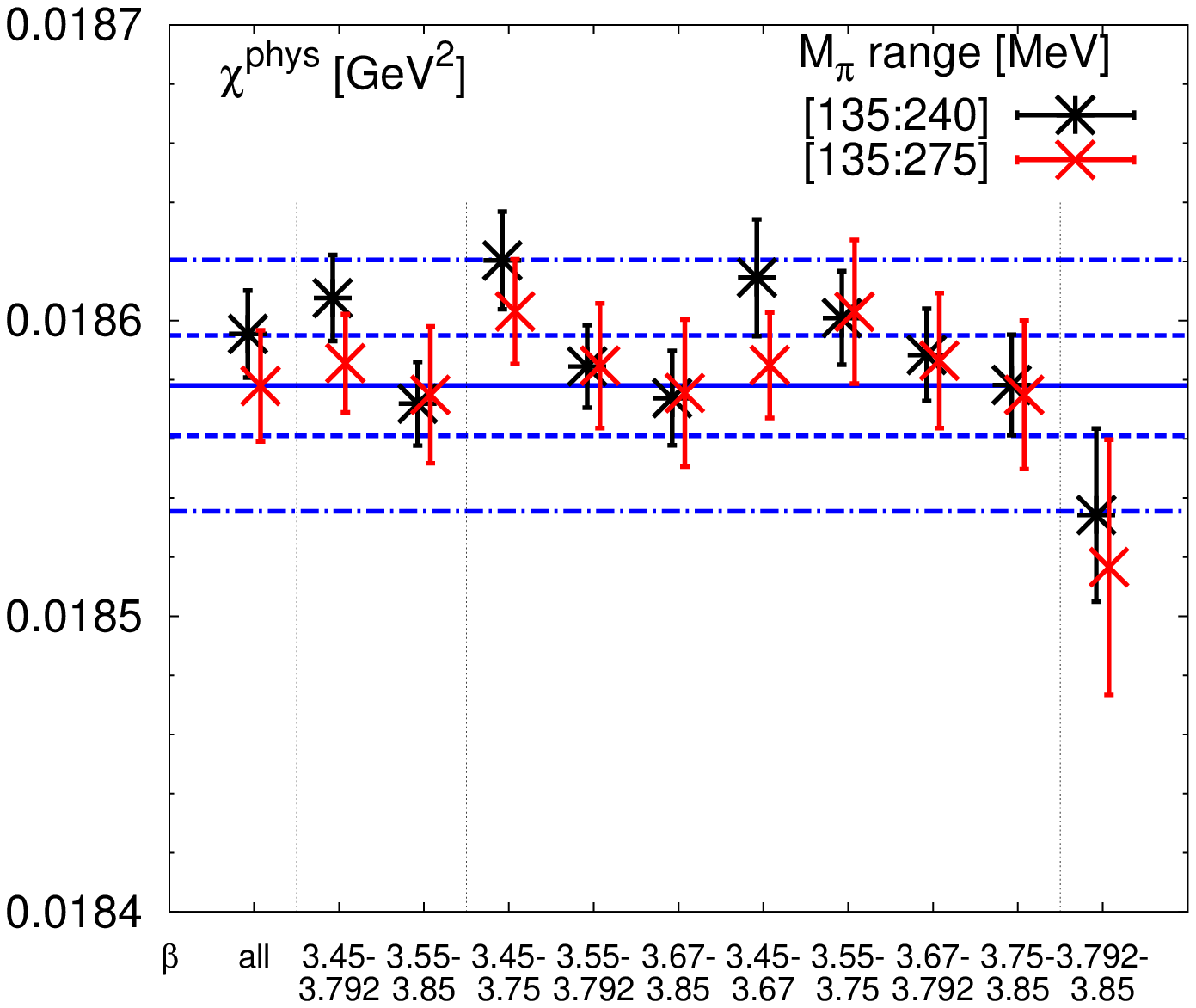}%
\includegraphics[width=.45\textwidth]{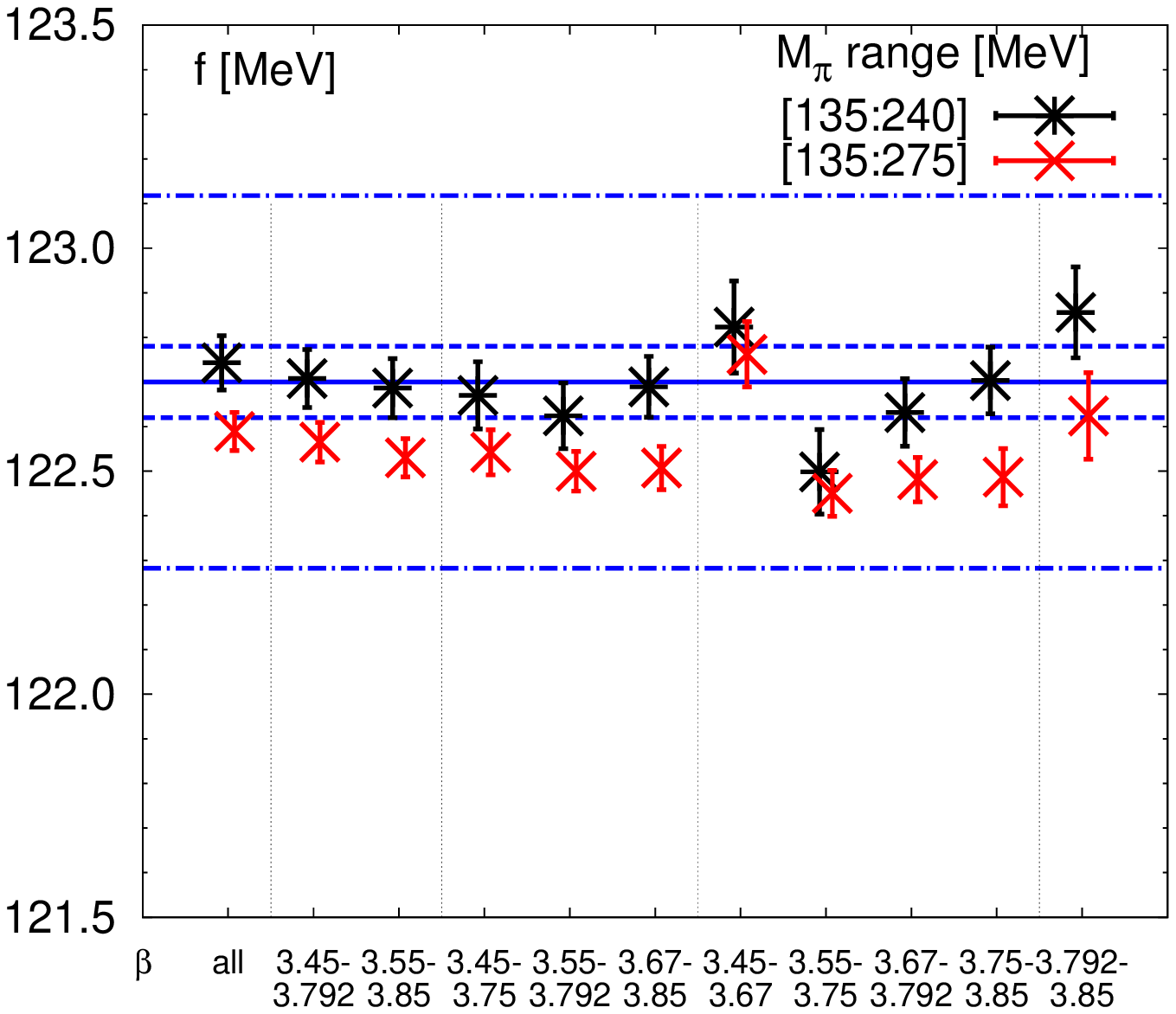}\\
\includegraphics[width=.45\textwidth]{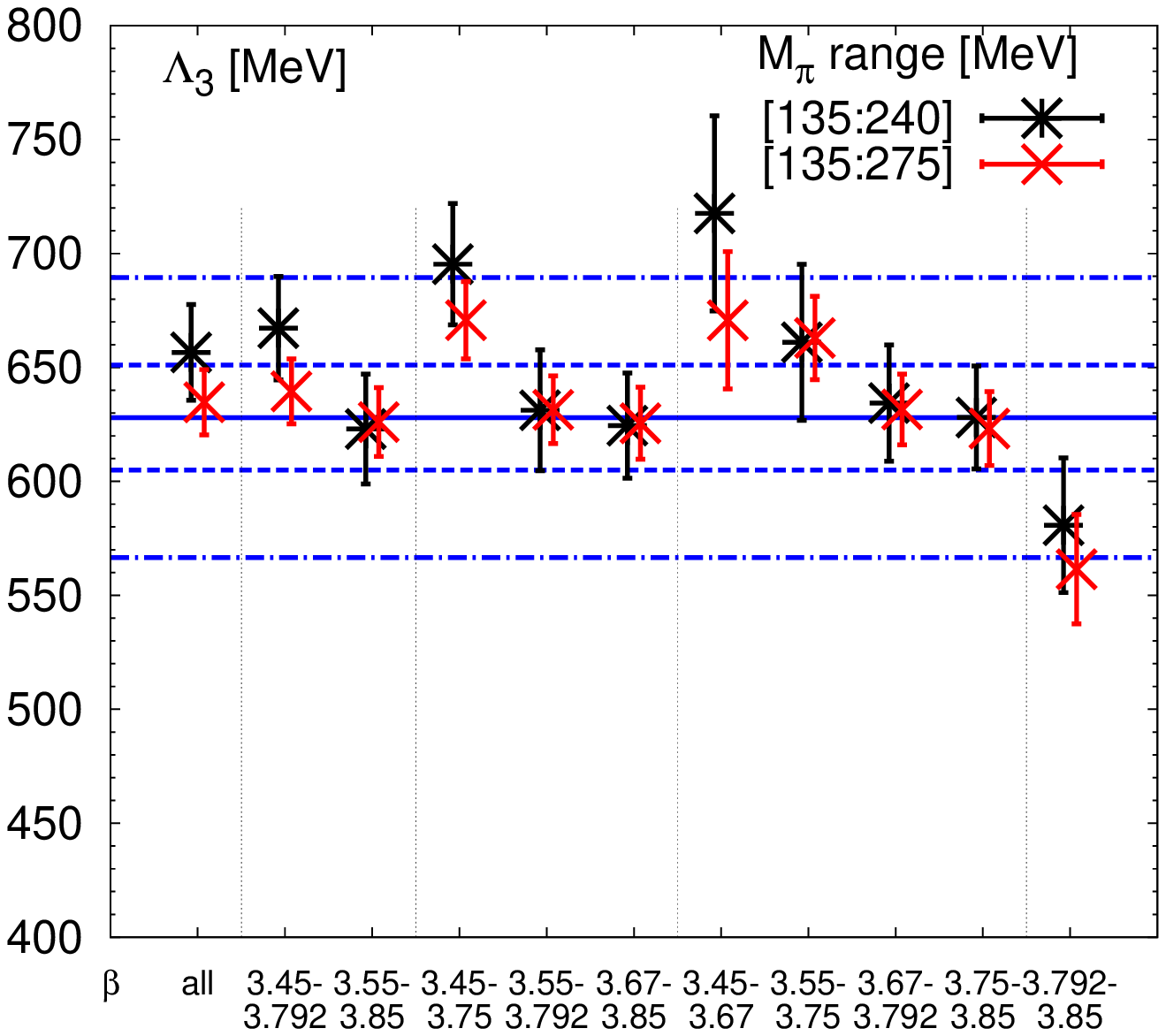}%
\includegraphics[width=.45\textwidth]{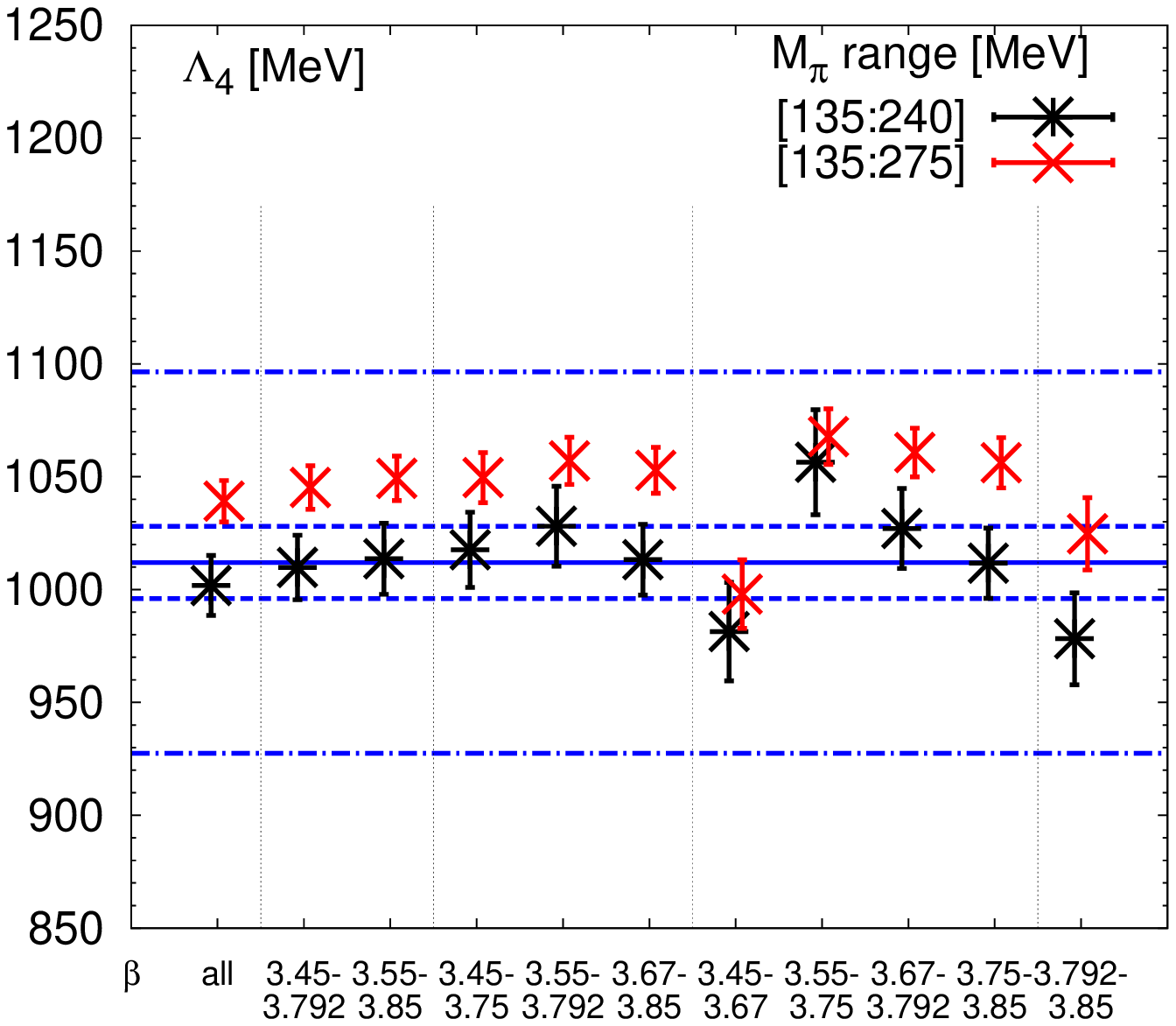}\\
\includegraphics[width=.45\textwidth]{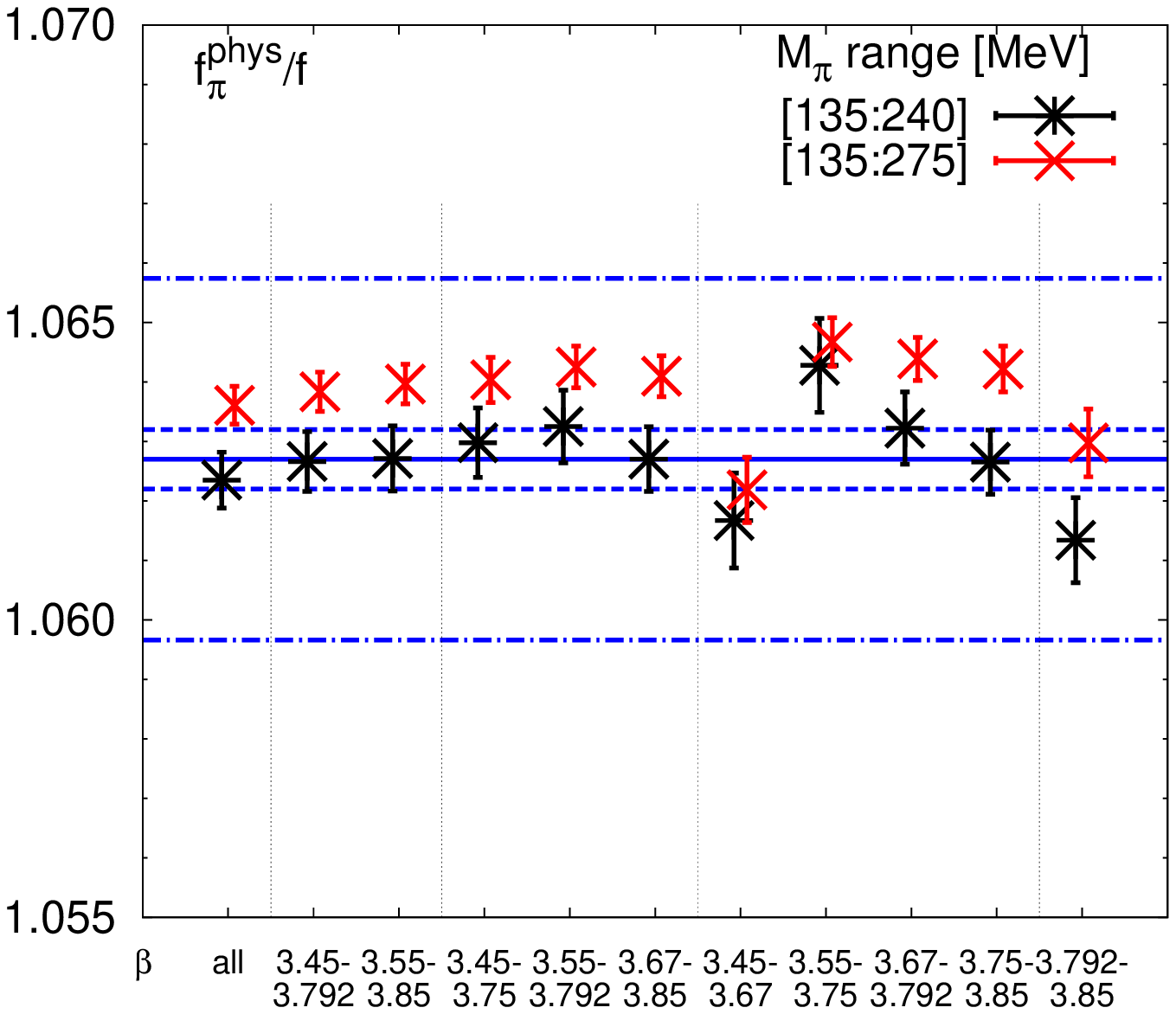}%
\includegraphics[width=.45\textwidth]{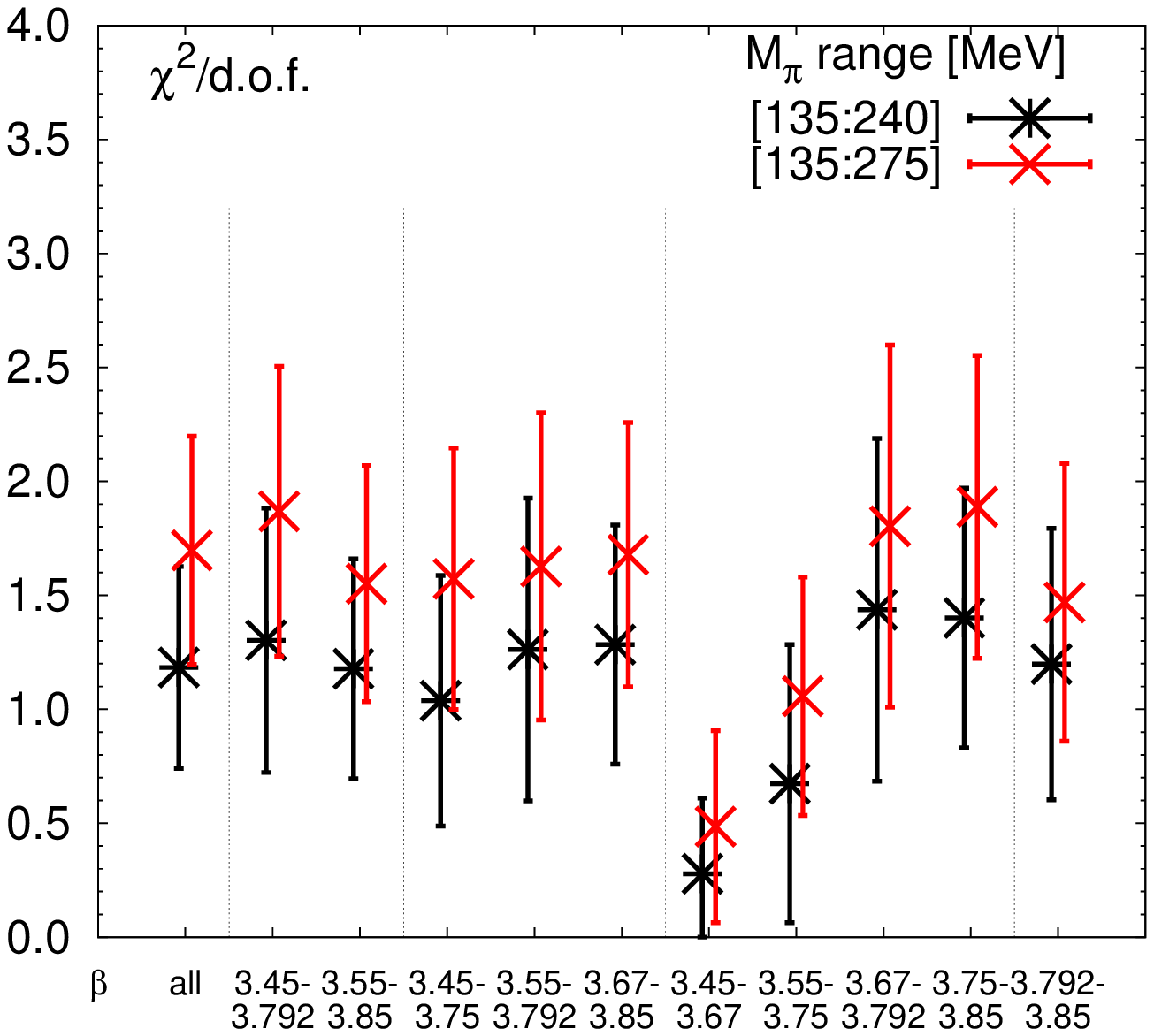}
\caption{Fit parameters and $\chi^2/\textrm{d.o.f.}$ from NLO-ChPT fits without constraints to two different mass ranges, where several lattice spacings have been excluded. \textit{Vertical dashed lines} group points where no, one, two, three or four (from left to right) lattice spacings have been excluded. The \textit{solid, dashed and dashed-dotted horizontal blue lines} denote the central value, statistical and total (statistical plus systematic) error bands, respectively, from our preferred unconstrained fit (cf.\ left column of Table~\ref{tab:NLO:results}).}
\label{fig:NLO:slidingB}
\end{center}
\end{figure}

In a second step, we will now examine whether or not all the available lattice spacings already lie in the scaling region. Remember that since the meson mass and decay constant define our scaling trajectory, no terms modelling discretization effects have been added to our chiral formulas. To test the scaling behavior, we excluded ensembles belonging to one or more gauge coupling $\beta$ from the fits. In Fig.\,\ref{fig:NLO:slidingB} we show how the fit parameters and $\chi^2/\textrm{d.o.f.}$ change with respect to which lattice spacings are included in the fits. We show this for two different mass ranges, both including the near physical points. The leftmost point on each of these plots is from a fit to all available lattice spacings. Then, separated by vertical dashed lines, groups of fits follow where one, two, three or four lattice spacings have been excluded. The horizontal blue lines show our final estimate with error bands of the quantity displayed. Especially in the group where three lattice spacings have been excluded (the second from the right), we observe the parameters to reach a plateau by excluding the coarse lattice spacings. Overall, it is also reassuring that all the points fall into the error band of the combined statistical and systematic error, which we are going to discuss in the remainder now.

\begin{figure}
\begin{center}
\includegraphics[width=.45\textwidth]{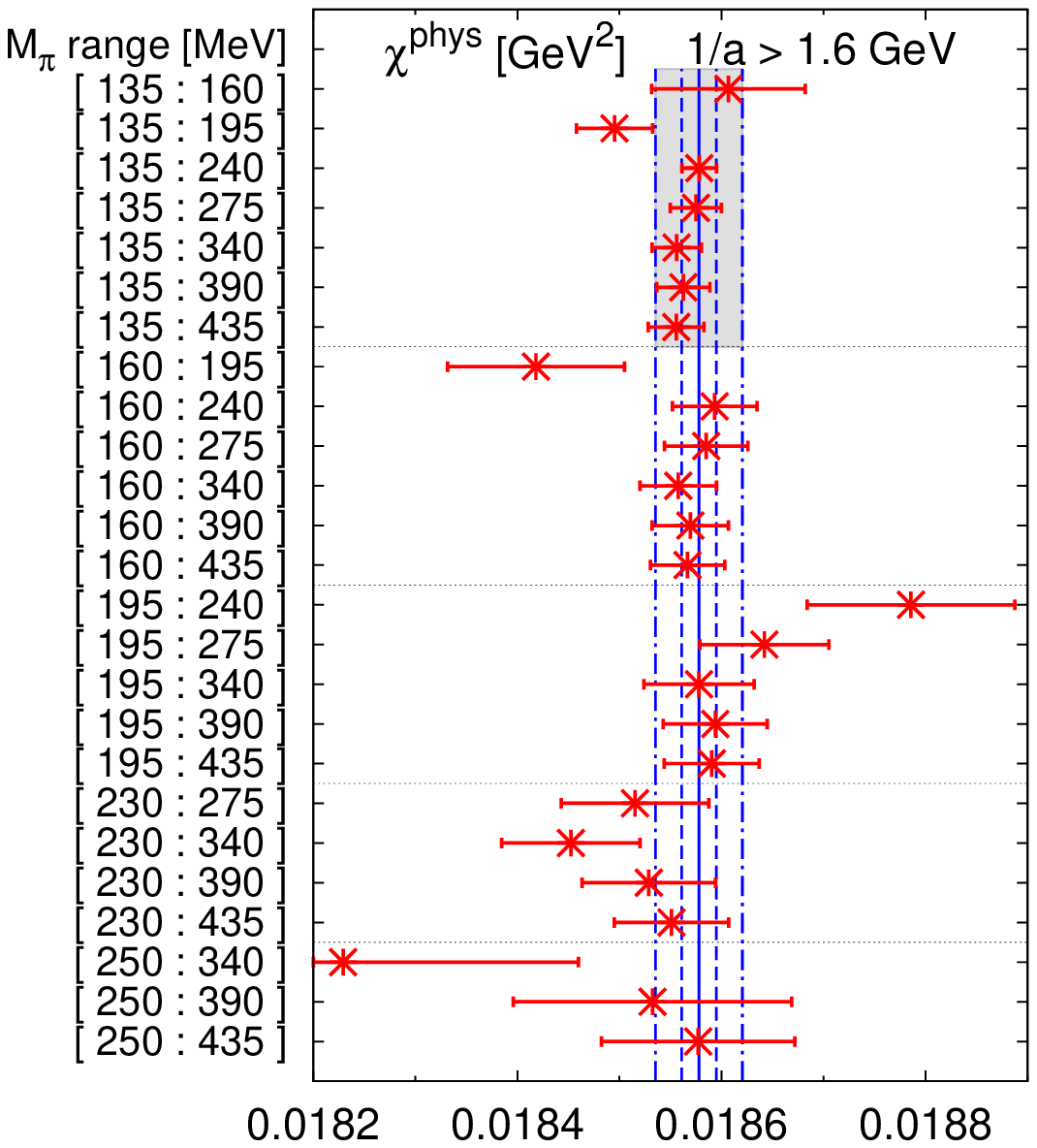}%
\includegraphics[width=.45\textwidth]{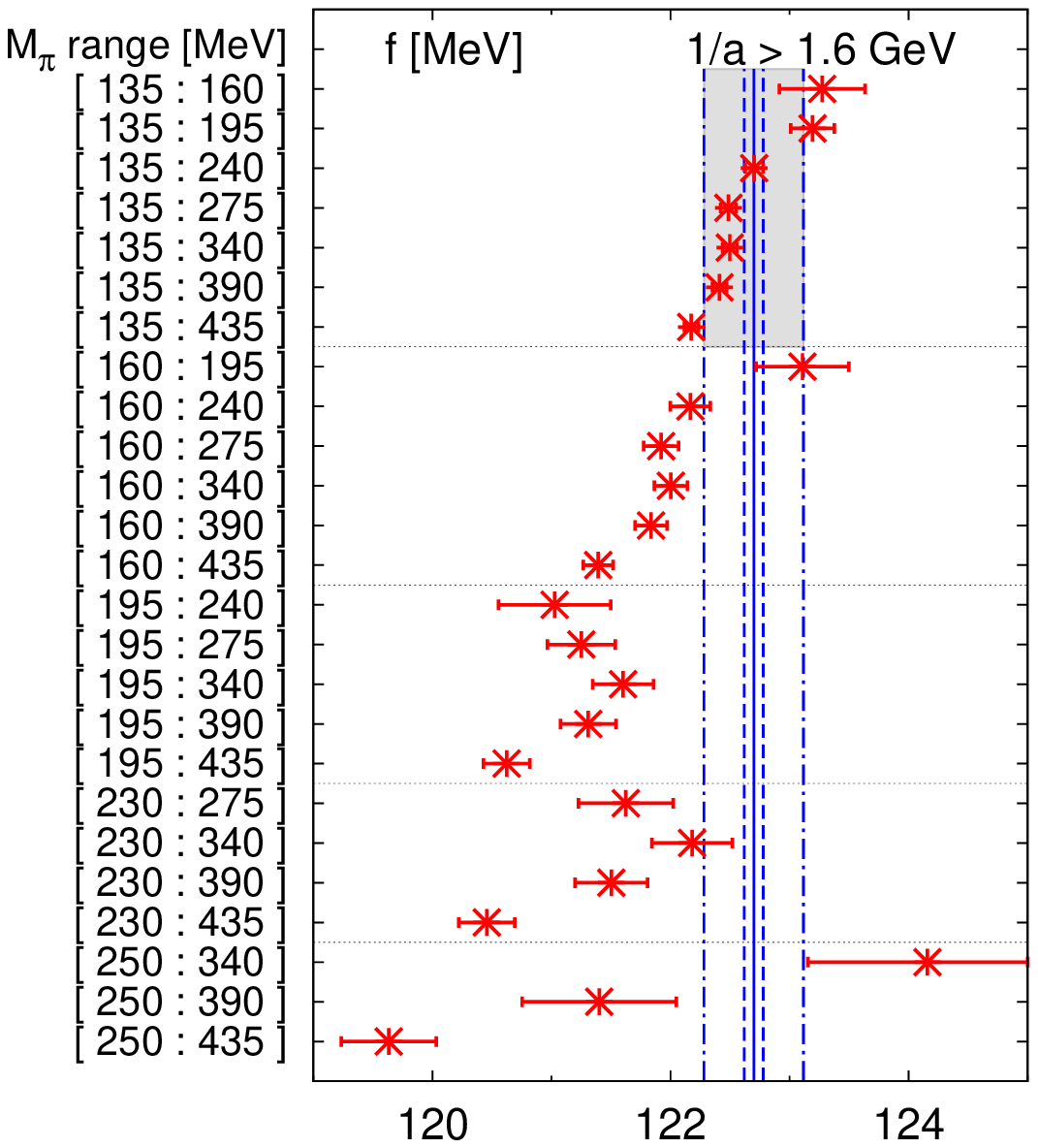}\\
\includegraphics[width=.45\textwidth]{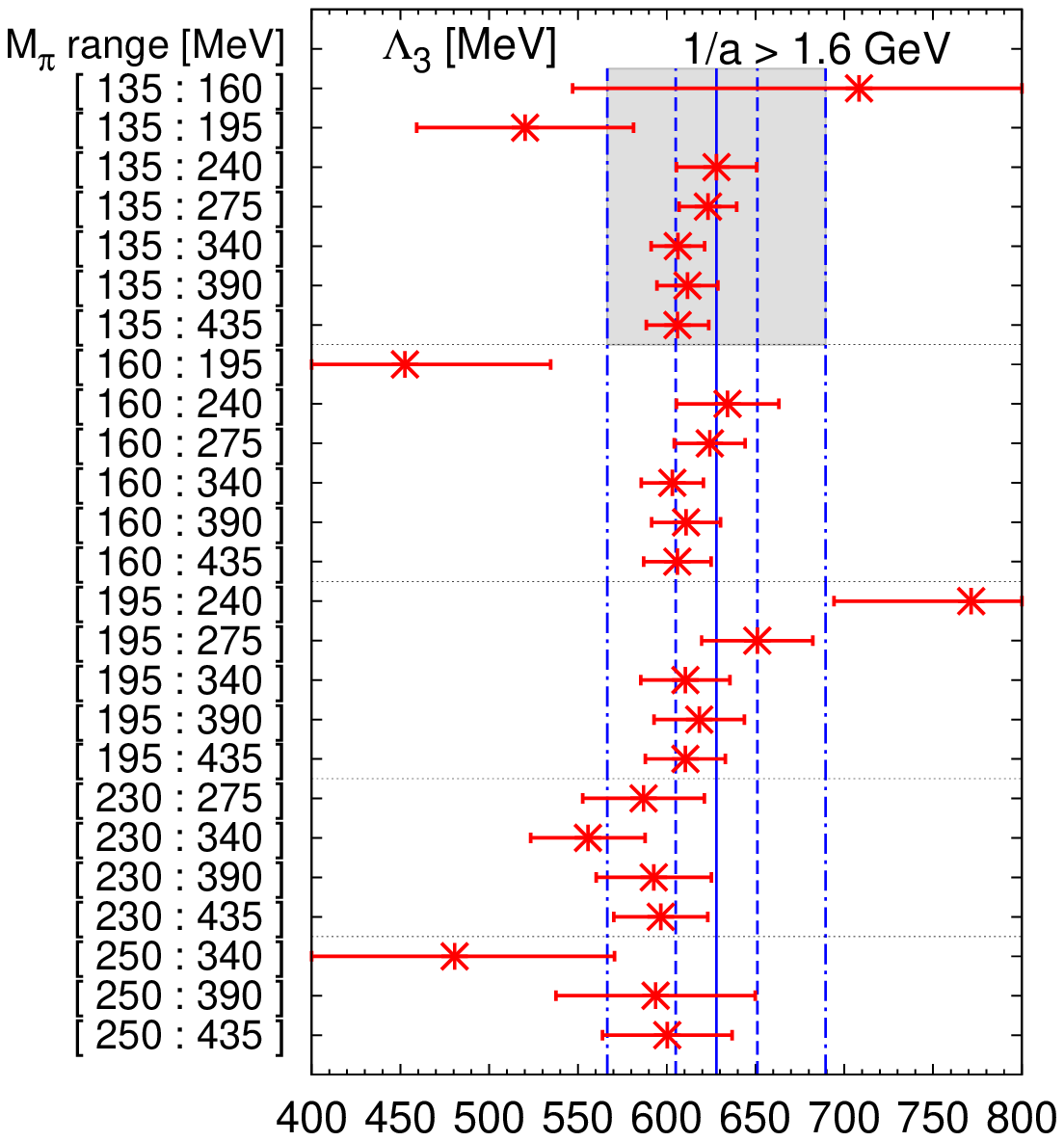}%
\includegraphics[width=.45\textwidth]{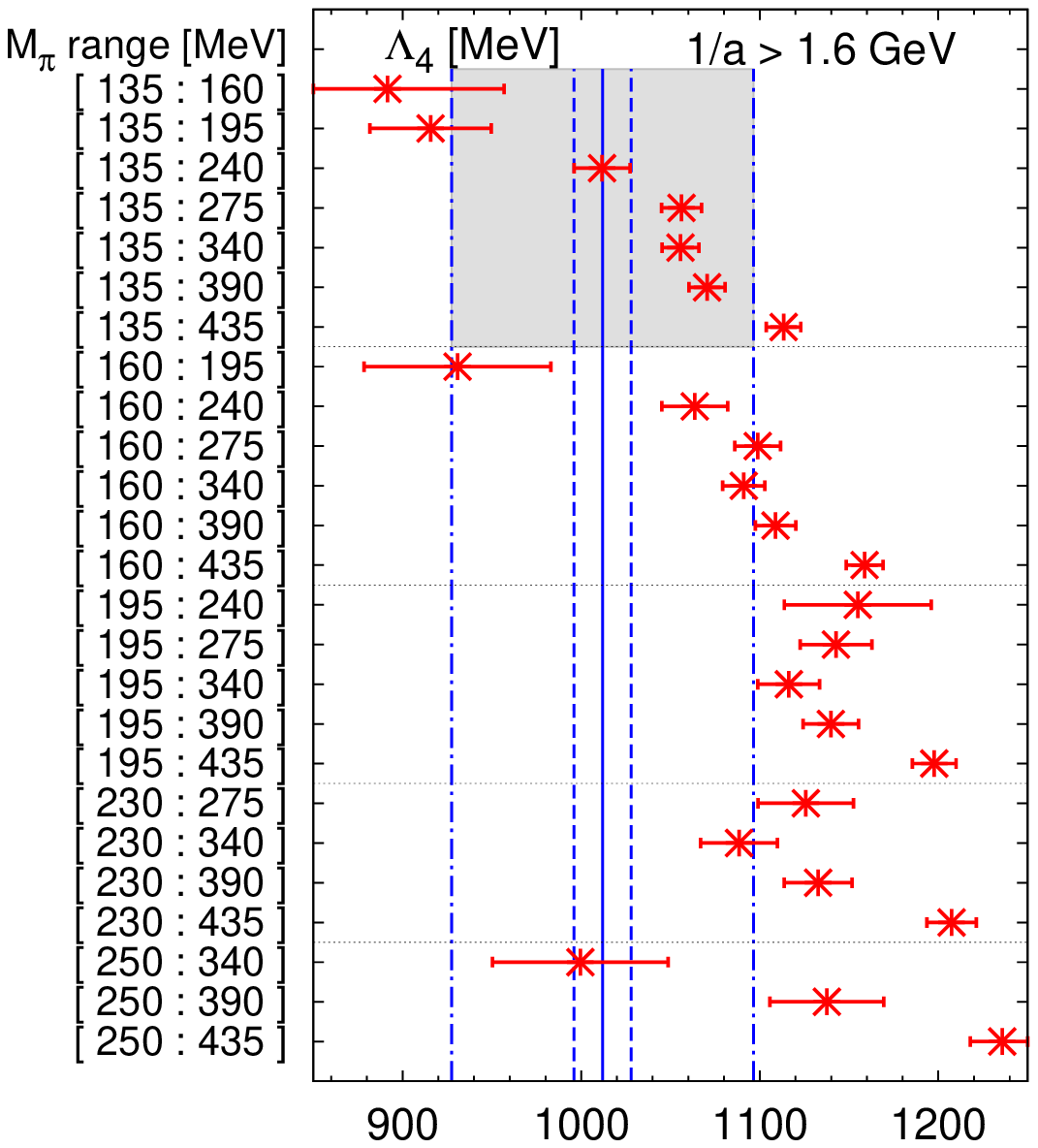}\\
\includegraphics[width=.45\textwidth]{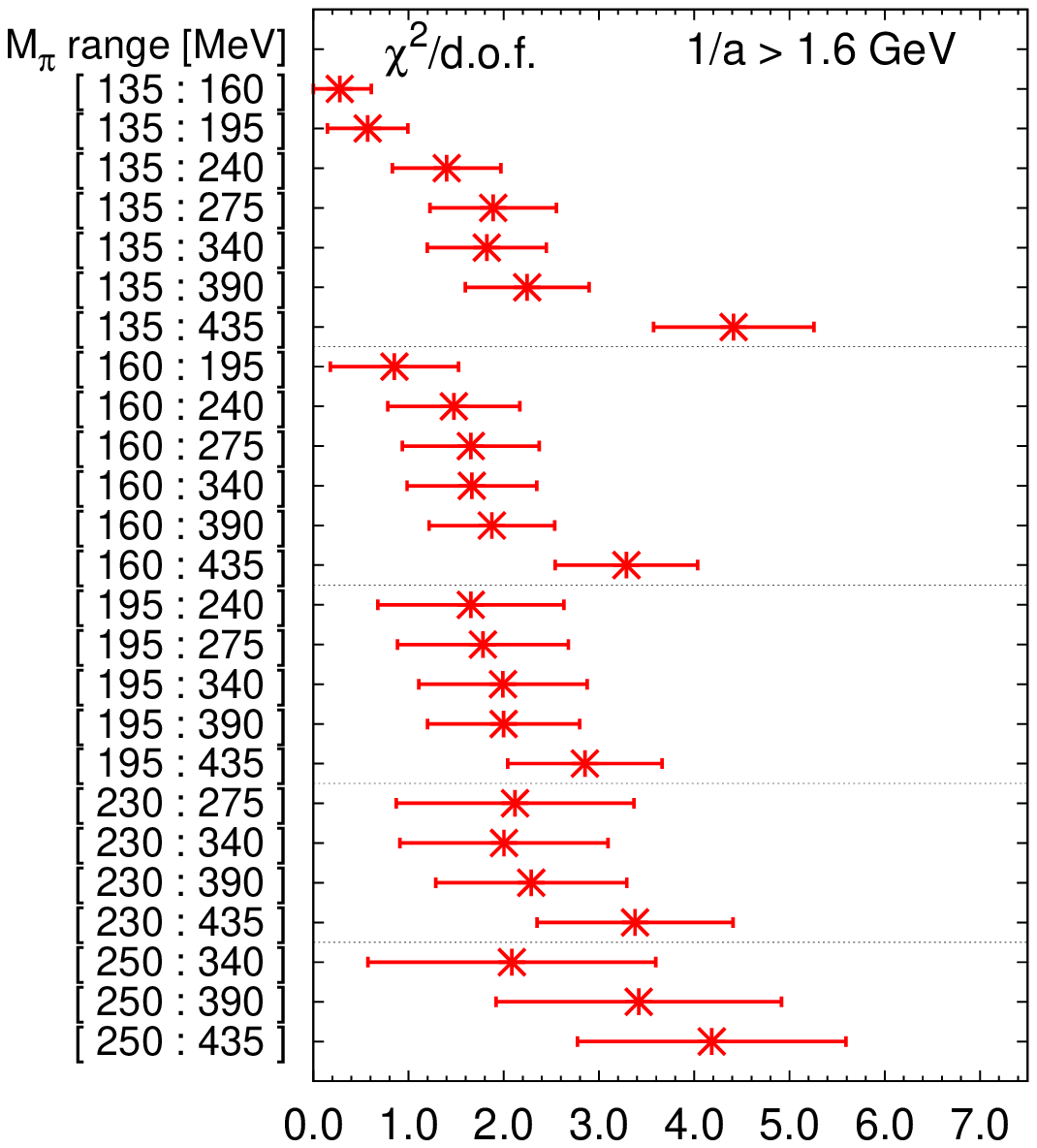}%
\begin{minipage}[b]{.45\textwidth}
\caption{Results for the fitted parameters (\textit{top} and \textit{middle panels}) and $\chi^2/\textrm{d.o.f.}$ (\textit{bottom panel}) from NLO-ChPT fits without constraints, using different mass ranges and including only lattice spacings $1/a\:>\:1.6\,{\rm GeV}$. The \textit{solid, dashed and dashed-dotted blue lines} for the fit parameters denote the central value, statistical and total (statistical plus systematic) error bands, respectively, from our preferred unconstrained fit (cf.\ left column of Table~\ref{tab:NLO:results}).}
\label{fig:NLO:massrange_fine}
\end{minipage}%
\end{center}
\end{figure}

\begin{figure}
\begin{center}
\includegraphics[width=.45\textwidth]{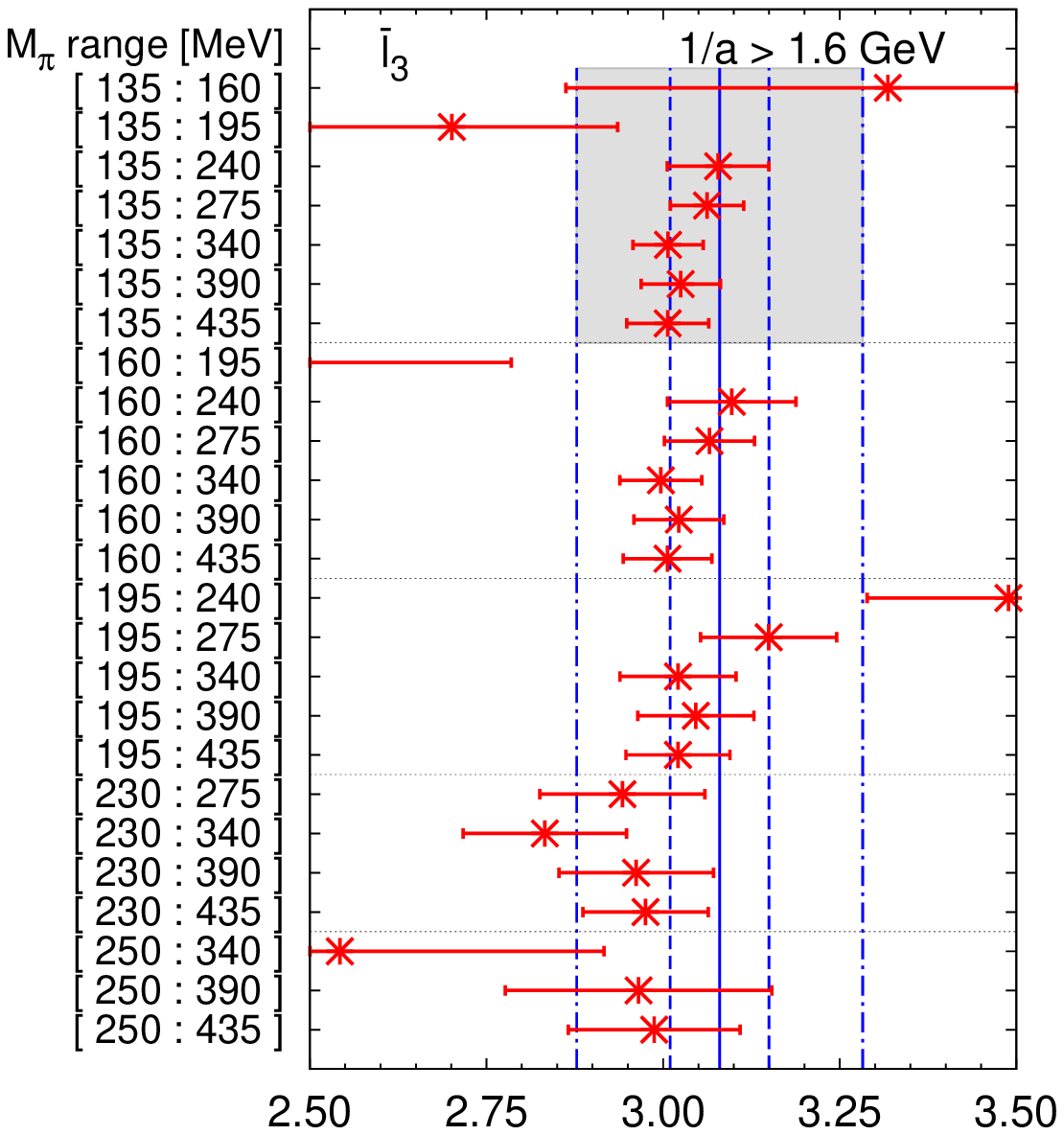}%
\includegraphics[width=.45\textwidth]{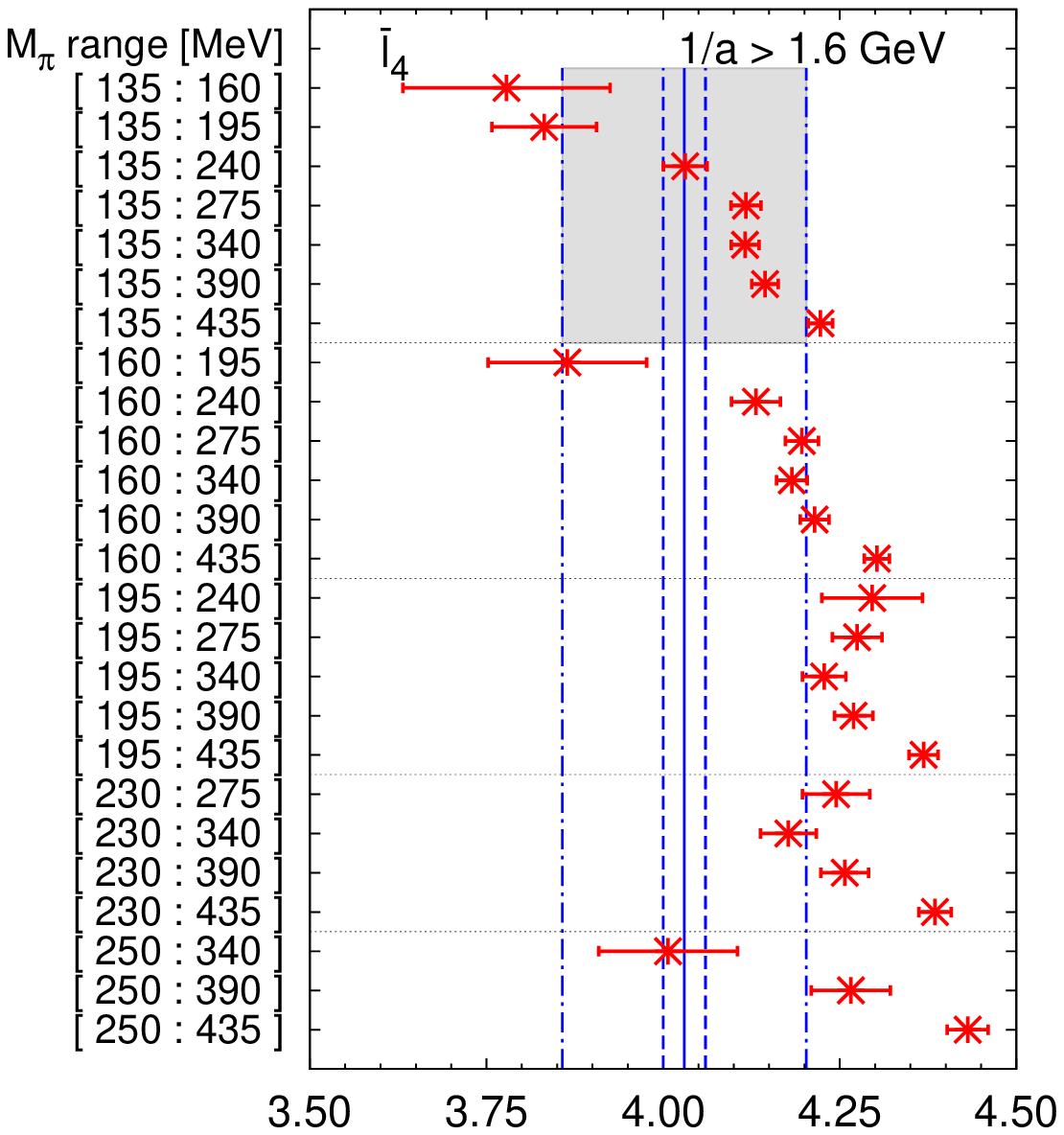}\\
\includegraphics[width=.45\textwidth]{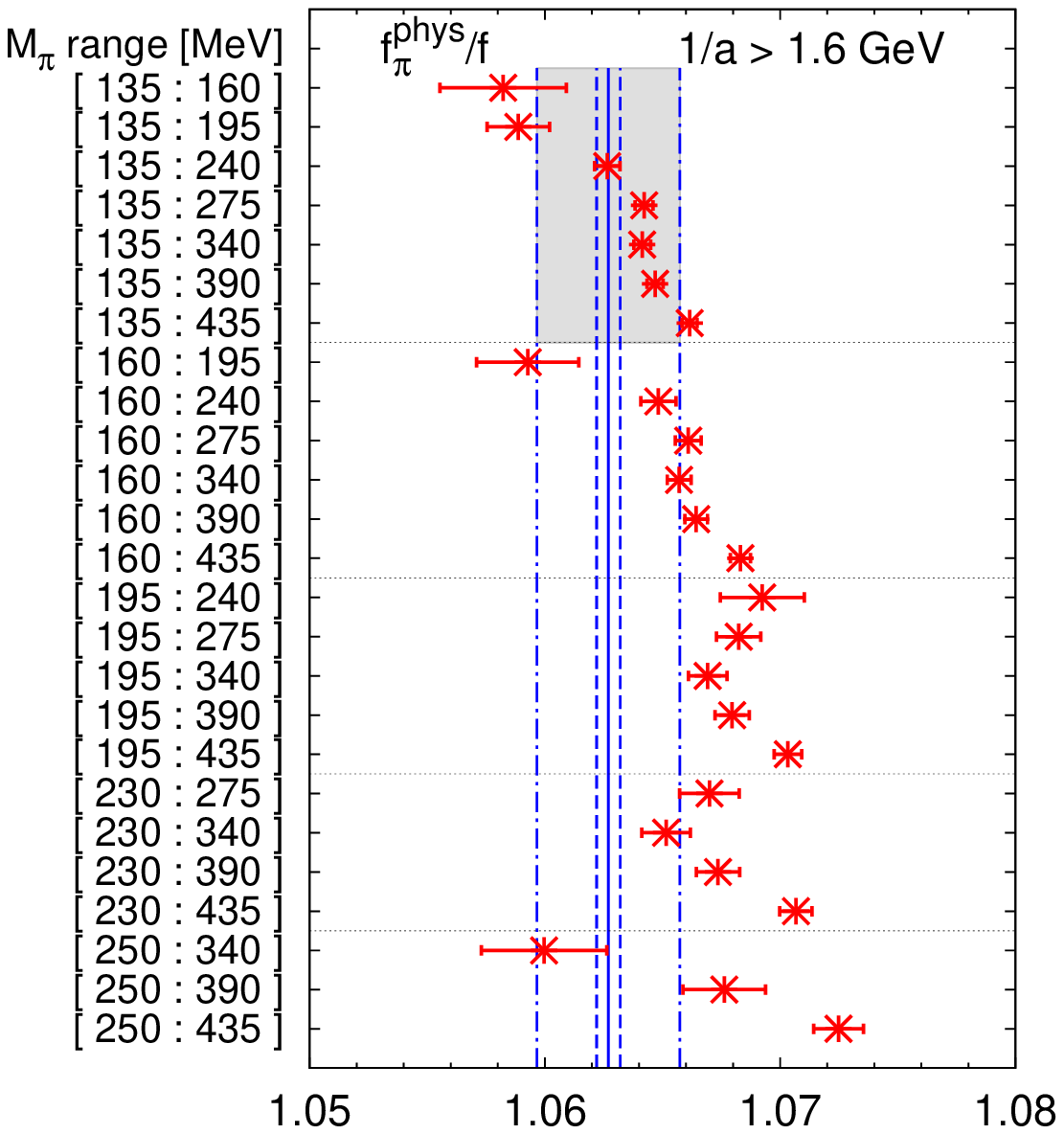}%
\begin{minipage}[b]{.45\textwidth}
\caption{Results for the parameters $\bar{\ell}_3$, $\bar{\ell}_4$ (\textit{top panels}) and the ratio $f_\pi^{\rm phys}/f$ (\textit{bottom panel}) from NLO-ChPT fits without constraints, using different mass ranges and including only lattice spacings $1/a\:>\:1.6\,{\rm GeV}$. The \textit{solid, dashed and dashed-dotted blue lines} for the fit parameters denote the central value, statistical and total (statistical plus systematic) error bands, respectively, from our preferred unconstrained fit (cf.\ left column of Table~\ref{tab:NLO:results}).}
\label{fig:NLO:massrange_fine_pheno}
\end{minipage}%
\end{center}
\end{figure}

\begin{figure}
\begin{center}
\includegraphics[width=.47\textwidth]{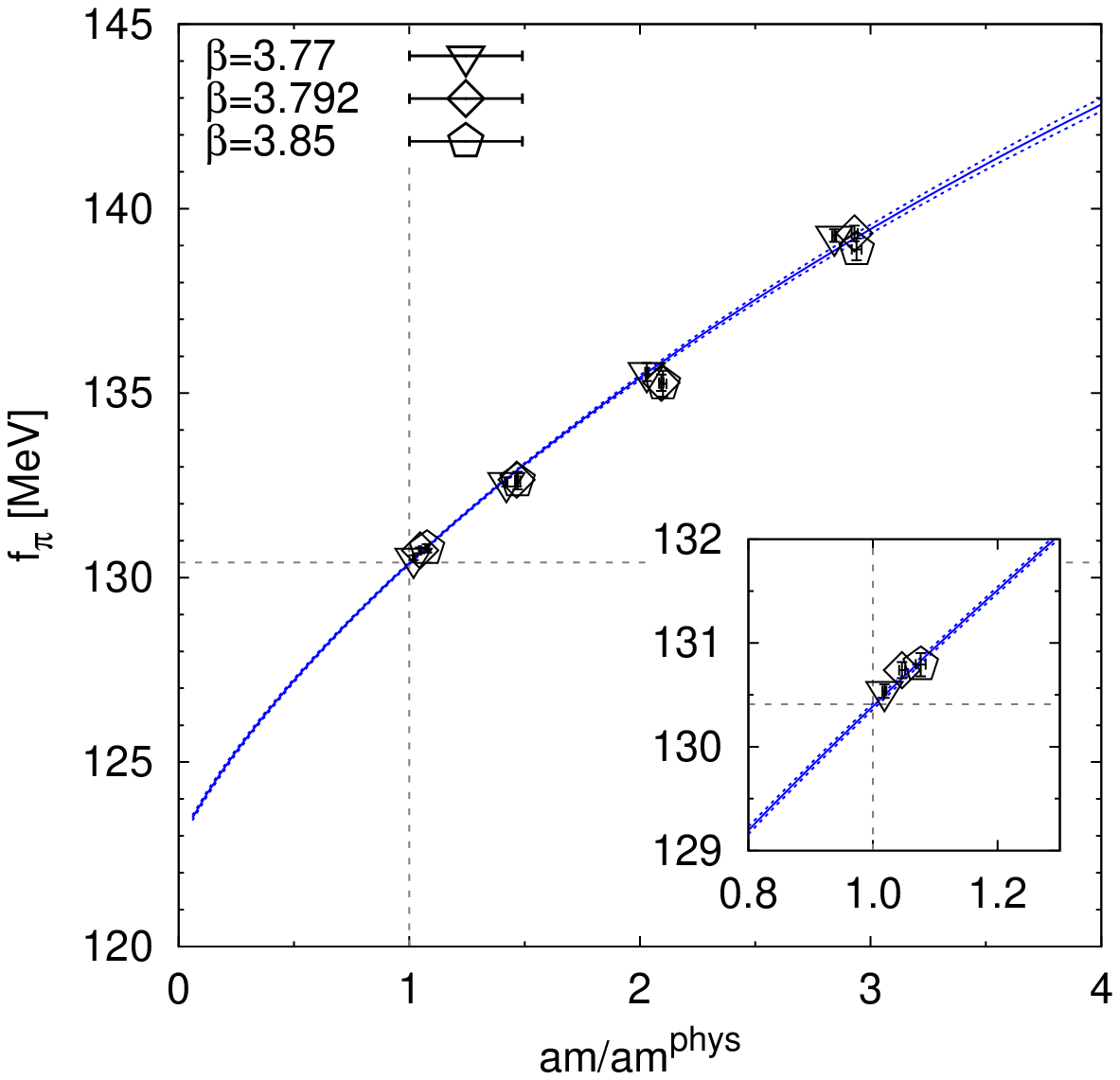}%
\includegraphics[width=.47\textwidth]{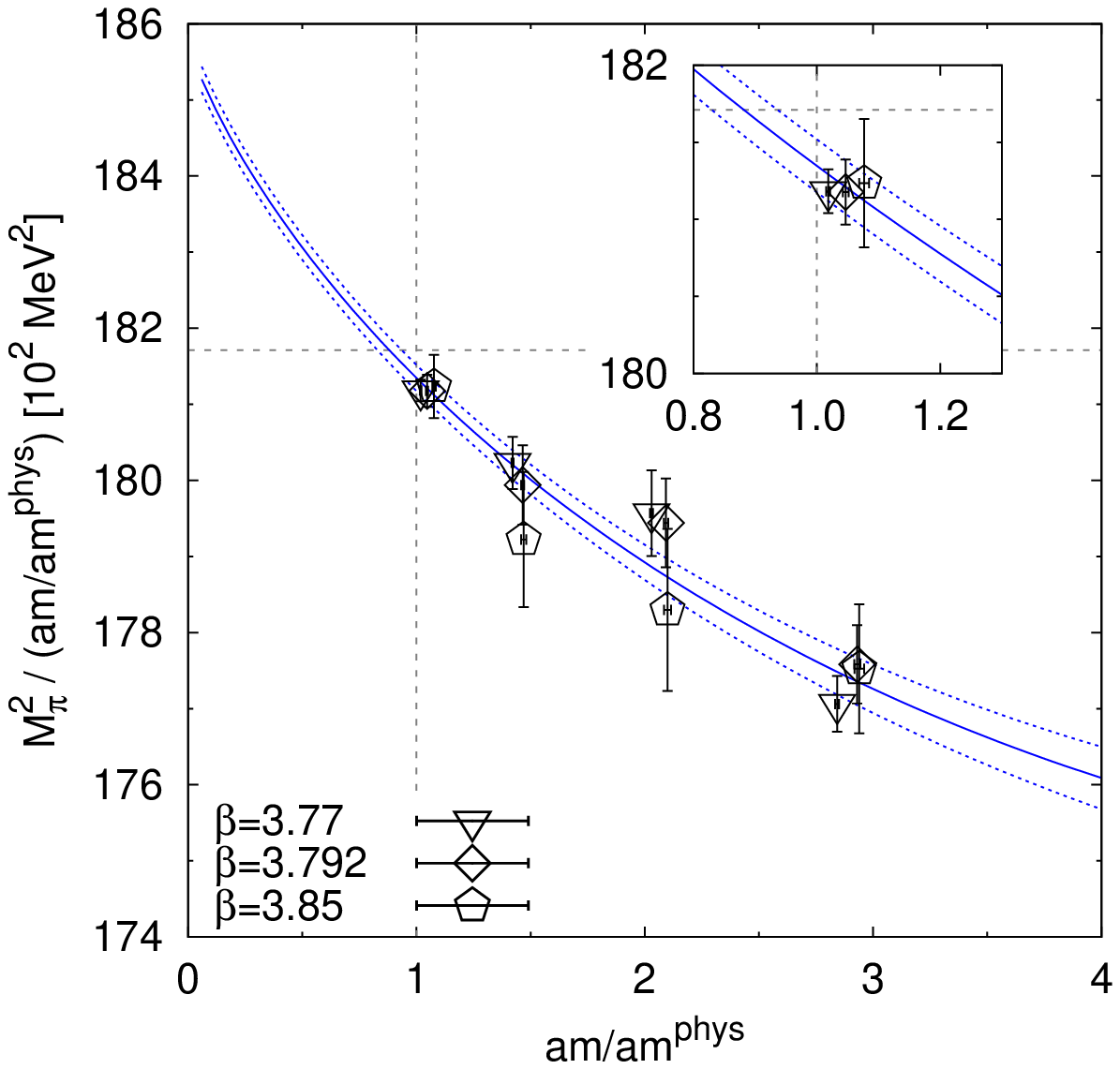}\\
\includegraphics[width=.47\textwidth]{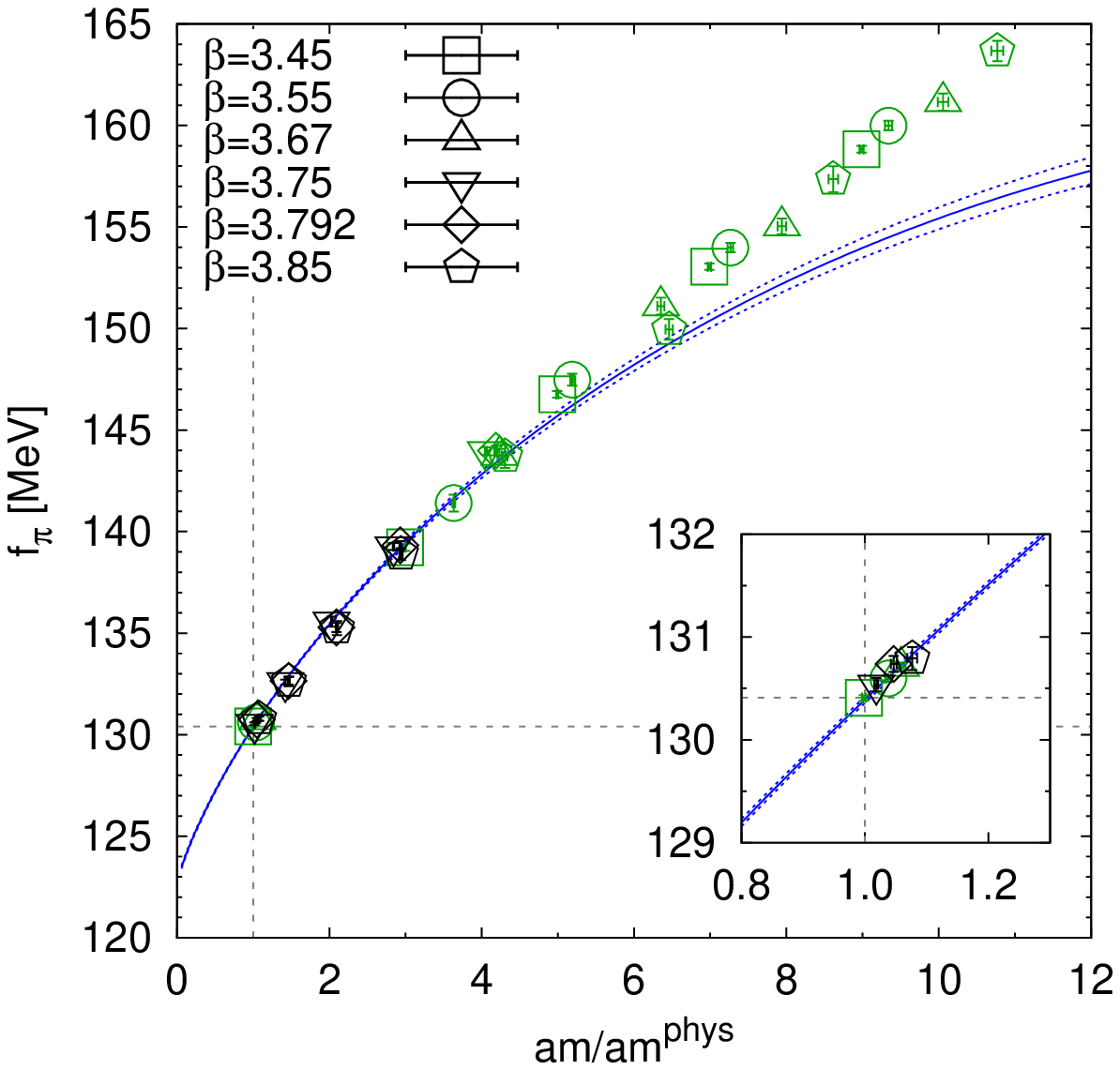}%
\includegraphics[width=.47\textwidth]{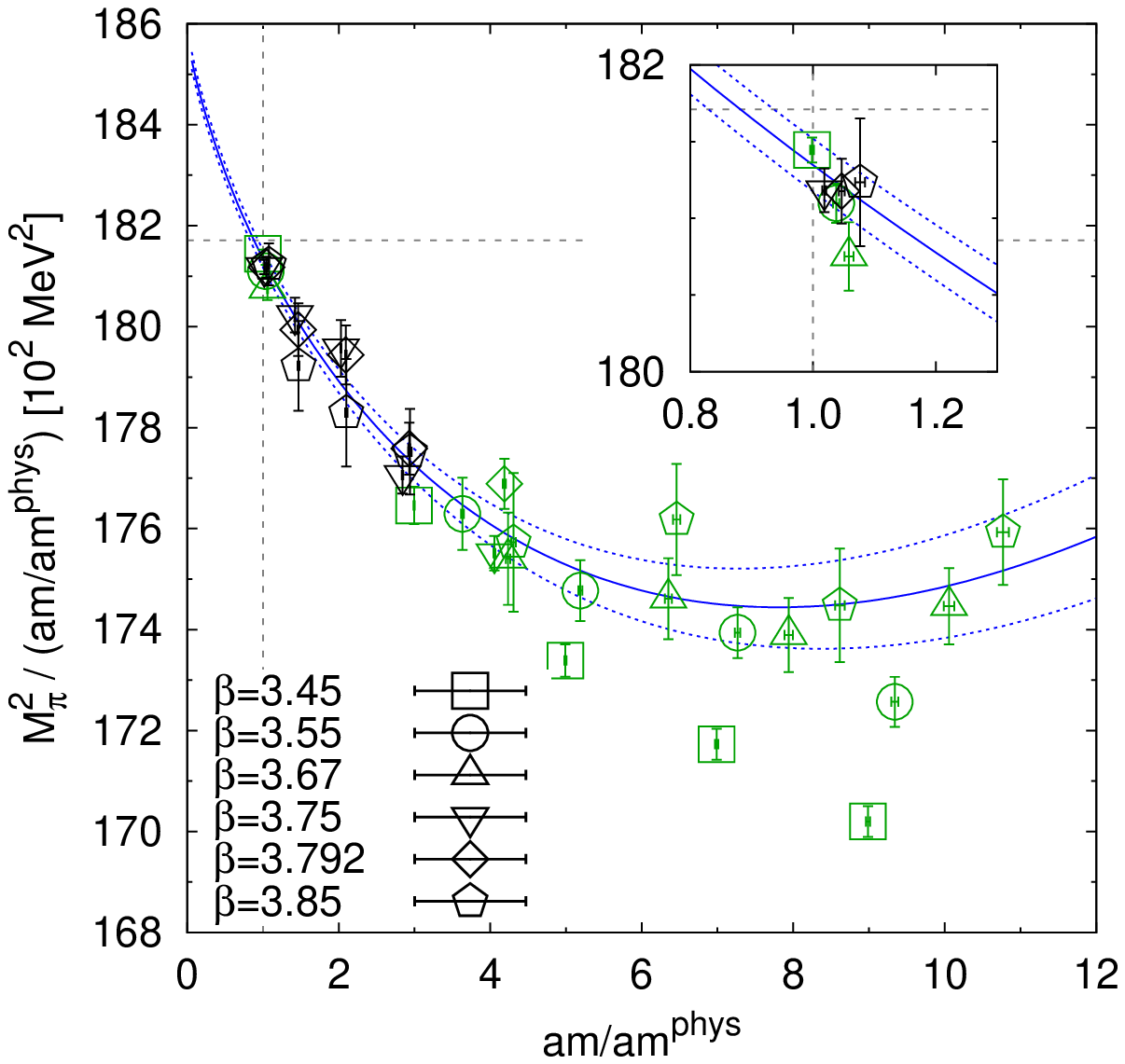}
\caption{Combined fit for lattice scales $1/a\:>\:1.6\,{\rm GeV}$ and meson masses $135\,{\rm MeV}\:\leq\:M_{\pi}\:\leq\:240\,{\rm MeV}$. \textit{Left panels:} meson decay constant, \textit{right panels:} squared meson mass divided by the quark-mass ratio. The \textit{top panels} show only the points, which have been included in the fit, while the \textit{bottom panels} show all points. There, points marked by \textit{black symbols} are included in the fit, those marked by \textit{green symbols} are excluded. The physical values are indicated by \textit{dashed gray lines}.}
\label{fig:NLO:fine_135-240}
\end{center}
\end{figure}

From the previous discussion about excluding coarser lattice spacings from the chiral analysis, we decided to restrict the range in the lattice spacings to $1/a\:>\:1.6\,{\rm GeV}$ or $a\:\leq\:0.12\,{\rm fm}$, i.e., only the ensembles at gauge couplings $\beta=3.75$, 3.792, and 3.85 will be included. Figures \ref{fig:NLO:massrange_fine} and \ref{fig:NLO:massrange_fine_pheno} show the dependence of the fitted parameters and derived quantities on the range of meson masses included in the fit range. These figures should be compared with Figs.\,\ref{fig:NLO:massrange_allB} and \ref{fig:NLO:massrange_allB_pheno}. Again, we observe the same pattern of reaching plateaus when excluding more and more heavier meson masses. Eventually, we decided to take the mass range $135\,{\rm MeV}\:\leq\:M_{\pi}\:\leq\:240\,{\rm MeV}$ as our preferred fit, from which we will quote the central values and statistical errors. This choice is also indicated in the landscape plot, Fig.\,\ref{fig:landscape}, by the full blue lines. The combined global fit has an acceptable $\chi^2/\textrm{d.o.f.}=1.4(0.6)$ (with $\#\mr{d.o.f.}=24-4=20$) and is shown in Fig.\,\ref{fig:NLO:fine_135-240}. In the top panels only the data points included in the fit are plotted, while the bottom panels show the excluded data points as well. To estimate the systematic error on a fitted parameter, we take the variance of this parameter with respect to the fits using different mass ranges, which also include the near physical points. These are the topmost points above the first horizontal dashed line in each panel of Figs.\,\ref{fig:NLO:massrange_fine} and \ref{fig:NLO:massrange_fine_pheno} (as indicated by the gray shaded areas). This procedure results in the set of LECs given in the left column of Table~\ref{tab:NLO:results}. Note, that only the first four parameters are fit parameters, while the remaining ones are subsequently derived from this set of parameters. The central values and error bands (statistical and combined statistical and systematic) have always been shown in the various compilations of fit results (Figs.\,\ref{fig:NLO:massrange_allB}--\ref{fig:NLO:massrange_fine_pheno}).
%
It is reassuring that basically all relevant results are compatible with these error bands, which {\it a posteriori} justifies our procedure of estimating the systematic error.


\begin{table}
\begin{center}
\begin{tabular}{lrr}
\hline\hline
 & \multicolumn{1}{c}{unconstrained} & \multicolumn{1}{c}{parameter-reduced} \\\hline
$\chi^{\rm phys}/(10^{-2} {\rm GeV}^2)$ & $1.8578(17)_{\rm stat}(39)_{\rm syst}$  & $1.8639(18)_{\rm stat}(44)_{\rm syst}$\\
$f/{\rm MeV}$                   & $122.70(08)_{\rm stat}(41)_{\rm syst}$  & $122.73(06)_{\rm stat}(28)_{\rm syst}$\\
$\Lambda_3/{\rm MeV}$           & $628(23)_{\rm stat}(57)_{\rm syst}$     & $678(40)_{\rm stat}(119)_{\rm syst}$\\
$\Lambda_4/{\rm MeV}$           & $1,012(16)_{\rm stat}(83)_{\rm syst}$   & $1,006(15)_{\rm stat}(71)_{\rm syst}$\\
$\bar{\ell}_3$                     & $3.08(07)_{\rm stat}(19)_{\rm syst}$    & $3.23(12)_{\rm stat}(30)_{\rm syst}$\\
$\bar{\ell}_4$                     & $ 4.03(03)_{\rm stat}(17)_{\rm syst}$   & $4.02(03)_{\rm stat}(14)_{\rm syst}$\\
$f_\pi^{\rm phys}/f$                       & $ 1.0627(05)_{\rm stat}(30)_{\rm syst}$ & $1.0626(06)_{\rm stat}(24)_{\rm syst}$\\
\hline\hline
\end{tabular}
\caption{Results for LECs from unconstrained (\textit{left column}, see Sec.~\ref{subsubsec:NLOfits_free} for details) and parameter-reduced fits (\textit{right column}, see Sec.~\ref{subsubsec:NLOfits_fix} for details). In the case of the unconstrained fits, the first four entries ($\chi^{\rm phys}$, $f$, $\Lambda_3$, $\Lambda_4$) are free fit parameters while the remaining entries are derived from these. For the parameter-reduced fit, only the first two entries ($\chi^{\rm phys}$, $f$) are free fit parameters. These two sets are used to calculate the final values as quoted in Eqs.\,(\ref{eq:result:NLO:chiphys})--(\ref{eq:result:NLO:fratio}).}
\label{tab:NLO:results}
\end{center}
\end{table}

\begin{figure}
\begin{center}
\includegraphics[width=.47\textwidth]{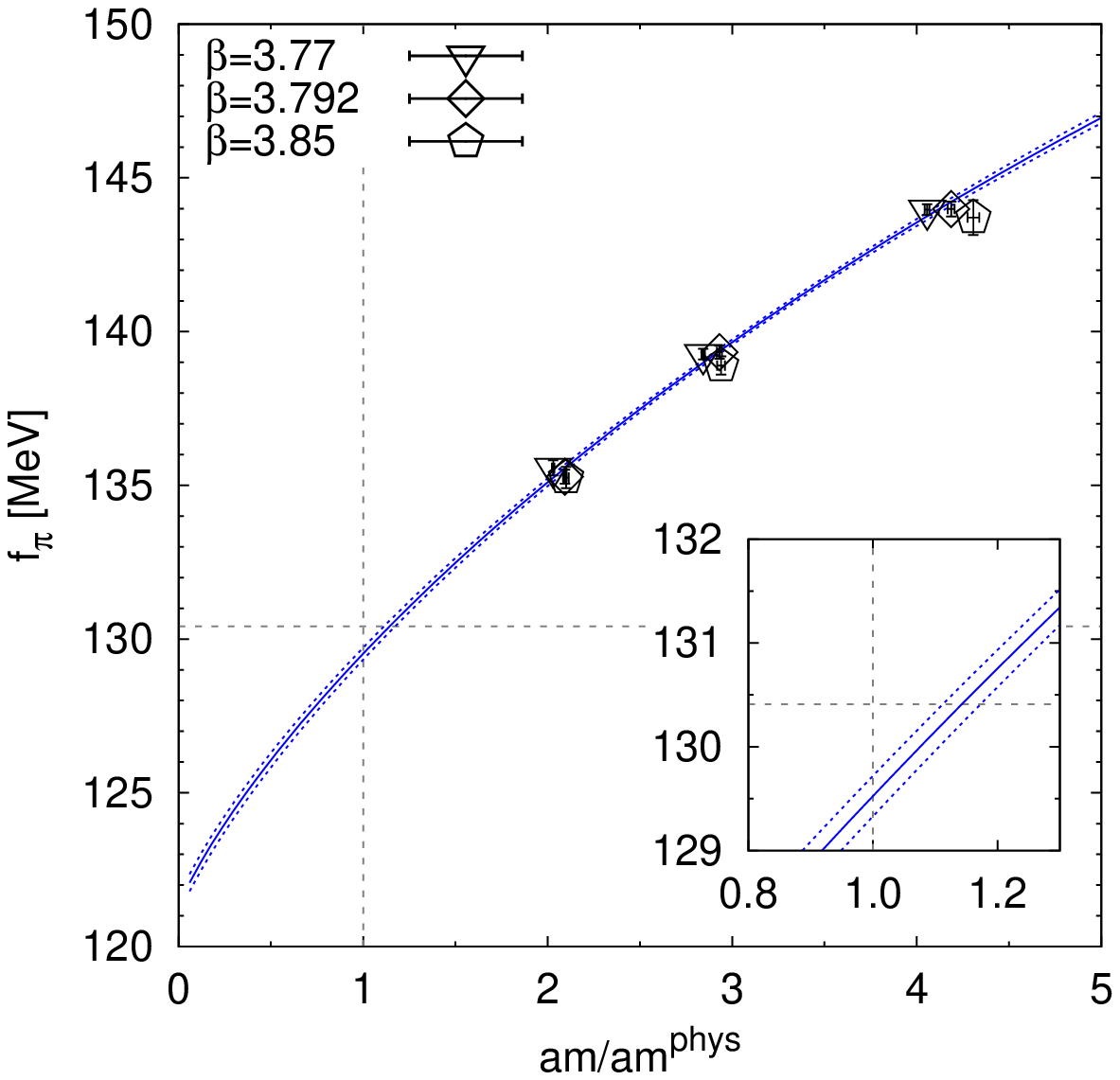}%
\includegraphics[width=.47\textwidth]{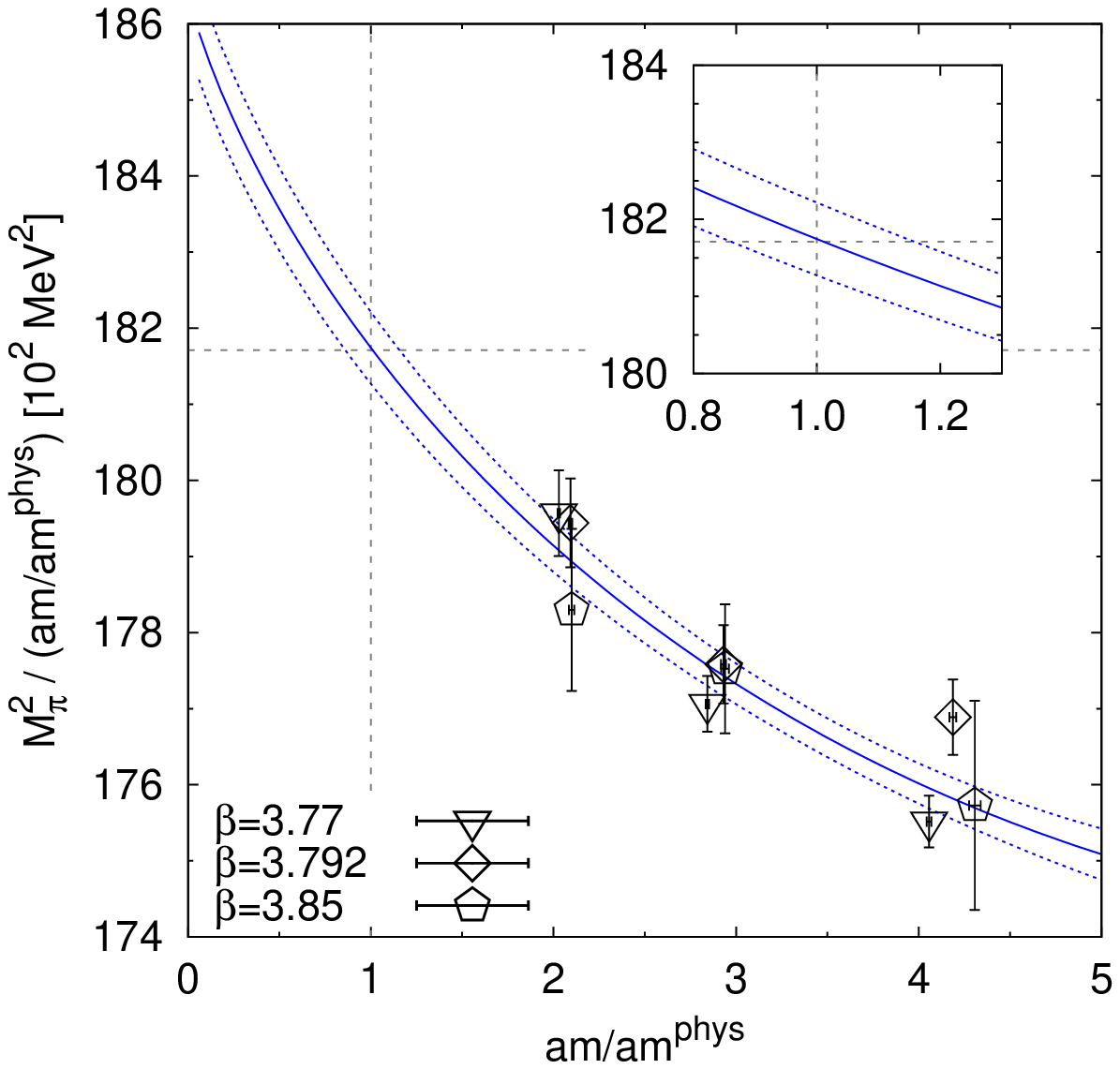}\\
\includegraphics[width=.47\textwidth]{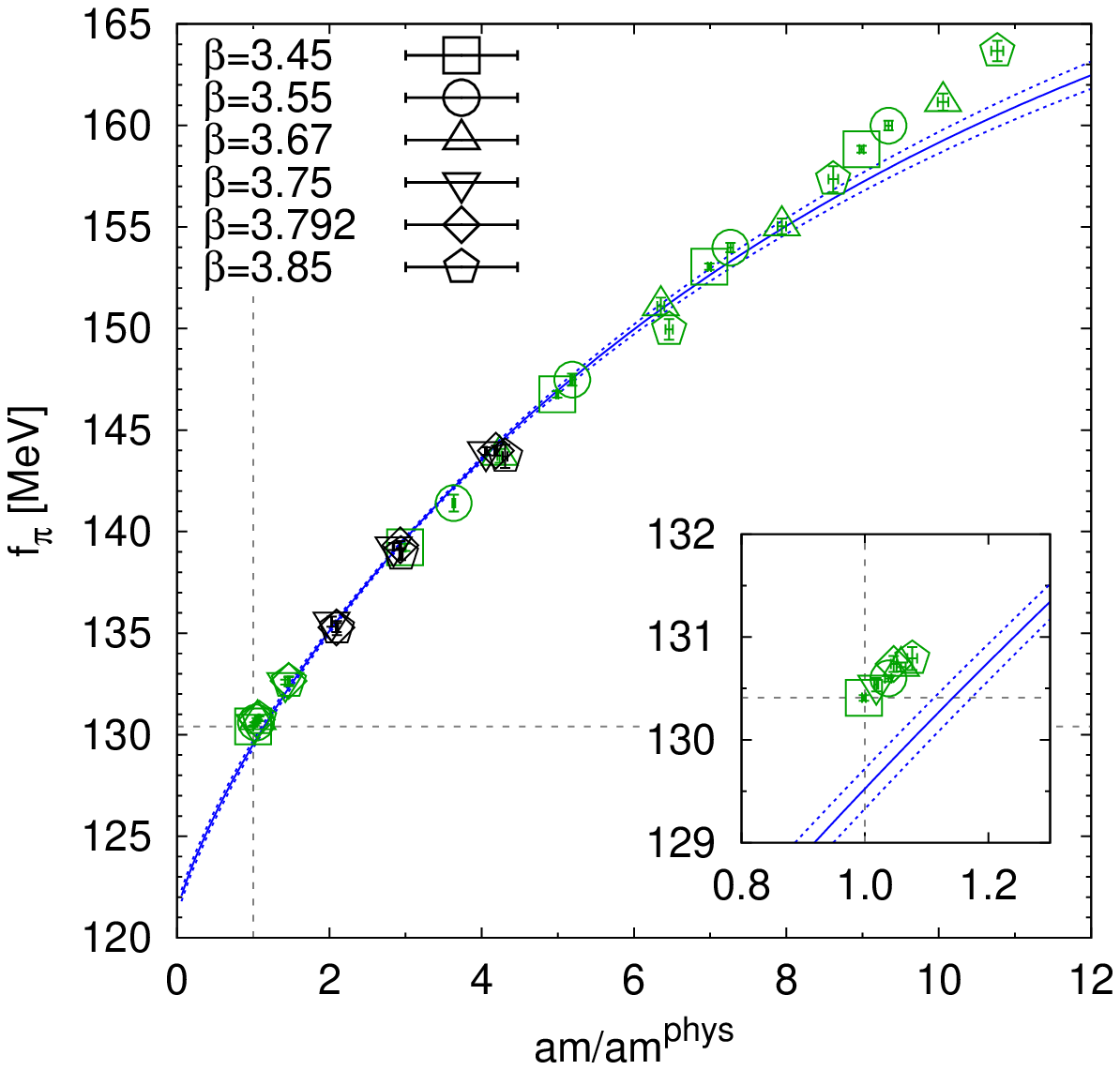}%
\includegraphics[width=.47\textwidth]{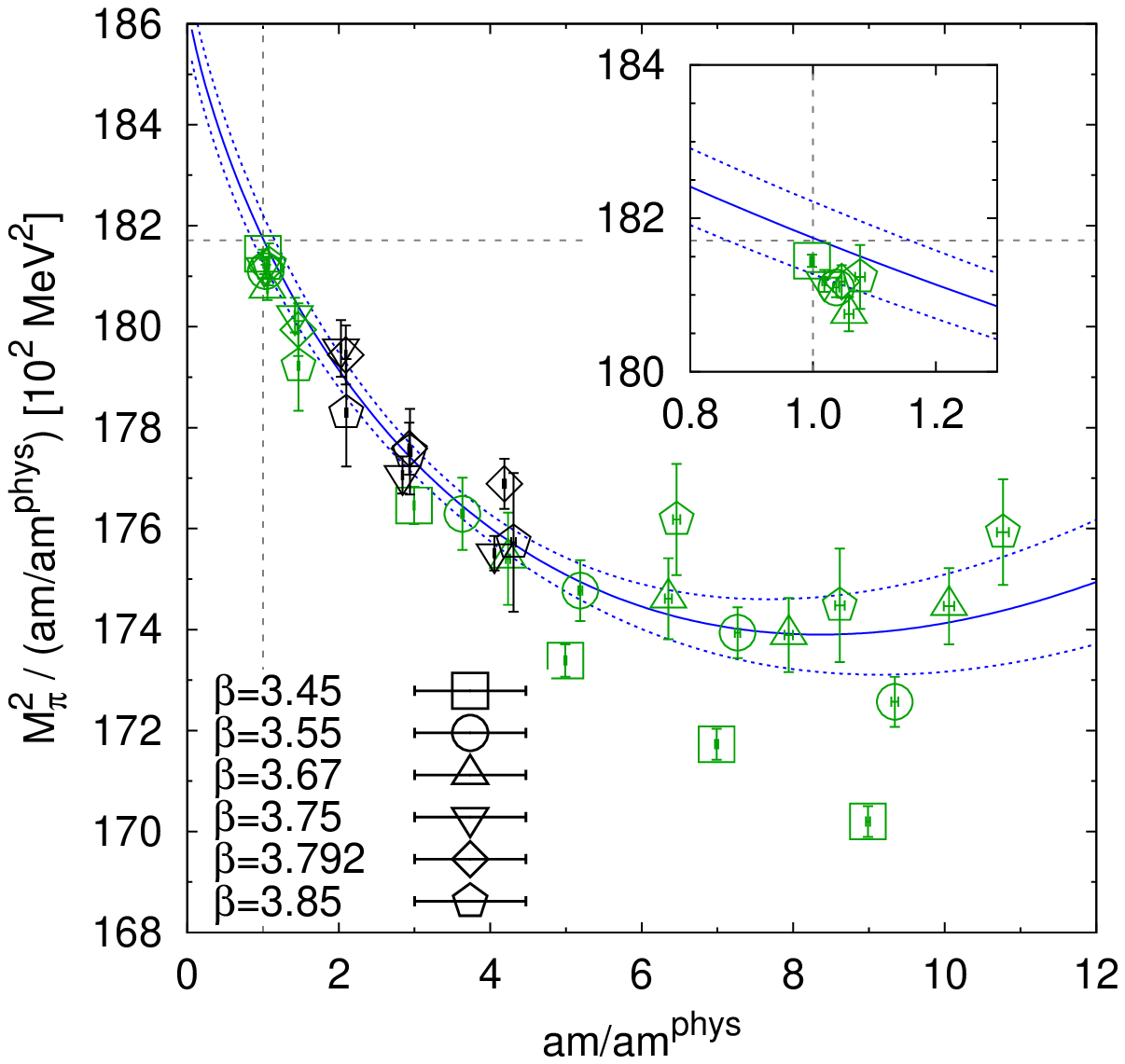}
\caption{Combined fit for lattice scales $1/a\:>\:1.6\,{\rm GeV}$ and meson masses $195\,{\rm MeV}\:\leq\:M_{\pi}\:\leq\:275\,{\rm MeV}$ excluding the nearly physical points. \textit{Left panels:} meson decay constant, \textit{right panels:} squared meson mass divided by the quark-mass ratio. The \textit{top panels} show only the points, which have been included in the fit, while the \textit{bottom panels} show all points. There, points marked  by \textit{black symbols} are included in the fit, those marked by \textit{green symbols} are excluded. The physical values are indicated by \textit{dashed gray lines}.}
\label{fig:NLO:fine_195-275}
\end{center}
\end{figure}

\begin{figure}
\begin{center}
\includegraphics[width=.45\textwidth]{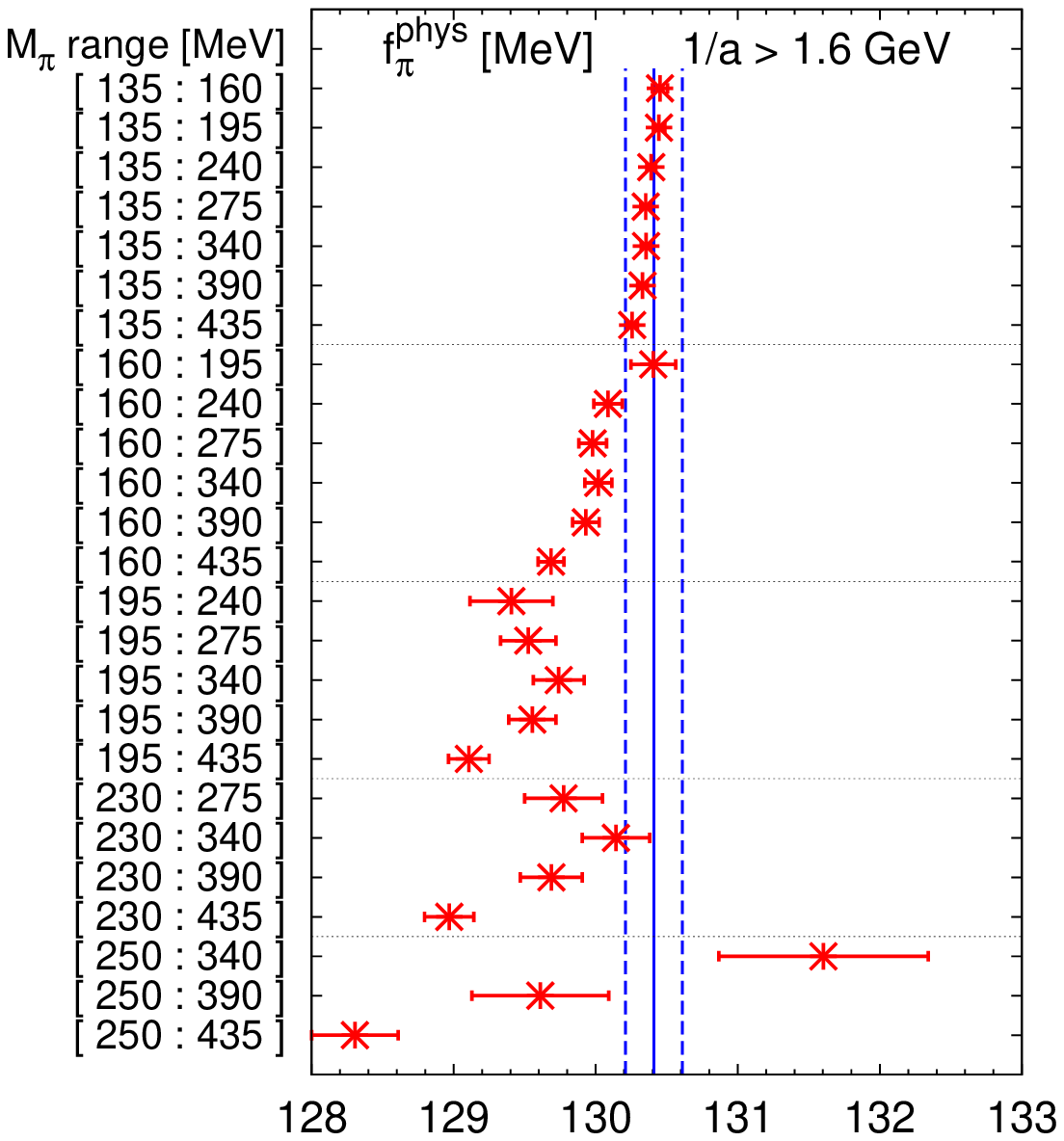}%
\includegraphics[width=.45\textwidth]{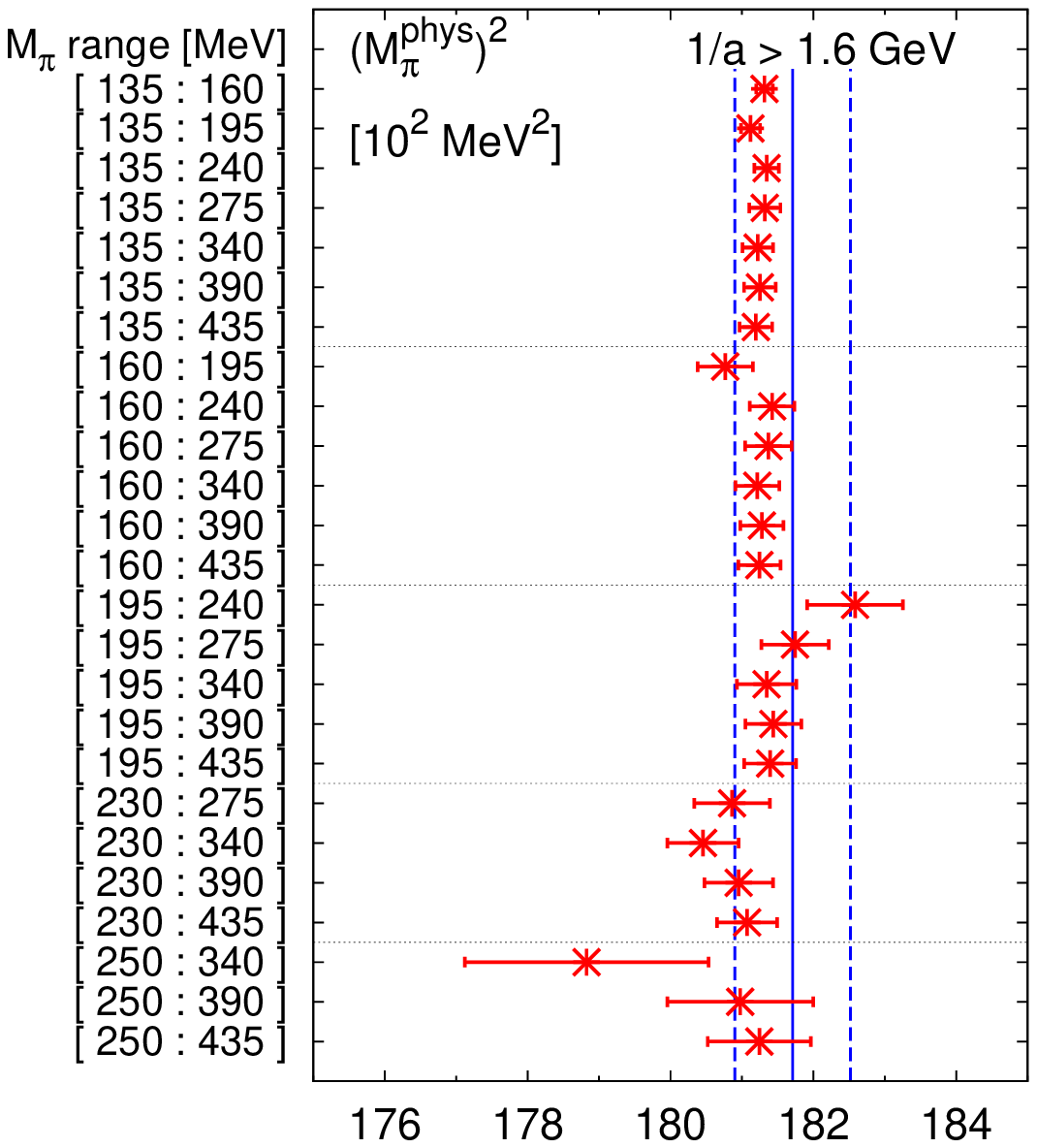}
\caption{Extrapolated $f_\pi^{\rm phys}$ (\textit{left panel}) and $(M_\pi^{\rm phys})^2$ (\textit{right panel}) from NLO-ChPT fits without constraints, using different mass ranges and including only lattice spacings $1/a\:>\:1.6\,{\rm GeV}$. The \textit{solid and dashed blue lines} denote the central value and total error bands, respectively, as quoted by Refs.\,\cite{Nakamura:2010zzi,Colangelo:2010et}.}
\label{fig:NLO:massrange_fine_phys}
\end{center}
\end{figure}

In the remainder of this section, we briefly discuss the influence of the near physical points on the fits and the fitted quantities (the effect of such low-end mass cuts on other observables, e.g., $m_{ud}$ or $m_s/m_{ud}$, have been discussed in Ref.\,\cite{Wittig:2012ha}). As an example, we provide in Fig.\,\ref{fig:NLO:fine_195-275} the result of a combined global fit in the mass range $195\,{\rm MeV}\leq M_{\pi}\leq 275\,{\rm MeV}$, using ensembles with $1/a\,>\,1.6\,{\rm GeV}$. As can be seen from the insert magnifying the region around the physical point, for the pion decay constant the fit misses the points simulated in that region by several standard deviations. This results in an extrapolated $f_\pi^{\rm phys}\:=\:129.5(0.2)\,{\rm MeV}$ (statistical error only), to be compared to the value of $130.41(0.03)(0.20)$ quoted by the PDG \cite{Nakamura:2010zzi}. For the extrapolated pion mass the situation looks somewhat better. In Fig.\,\ref{fig:NLO:massrange_fine_phys} we provide the values for $f_\pi^{\rm phys}$ and $(M_\pi^{\rm phys})^2$ extrapolated to the physical point by the NLO-ChPT fits with various mass ranges using ensembles with $1/a\:>\:1.6\,{\rm GeV}$ only. In these plots, the solid and dashed blue lines represent the central value and (total) uncertainty of these quantities as quoted by Refs.\,\cite{Nakamura:2010zzi,Colangelo:2010et}. Evidently, while the extrapolated pion mass is within errors compatible with the experimental result, the extrapolated pion decay constant starts to shift toward lower values once the nearly physical points are excluded from the fit range. This shift increases with both an increasing lower and higher bound on the fit range. This behavior is also in accordance with what could be observed in the mass range plots for the fitted parameters (Figs.\,\ref{fig:NLO:massrange_allB} and \ref{fig:NLO:massrange_fine}) and derived phenomenological quantities (Figs.\,\ref{fig:NLO:massrange_allB_pheno} and \ref{fig:NLO:massrange_fine_pheno}) shown before. These observations may be summarized as follows. For quantities like $\chi^{\rm phys}$ and $\Lambda_3$ (or $\bar{\ell}_3$), which predominantly influence the quark-mass dependence of $M_{\pi}^2$, the values from fits excluding ensembles with low meson masses are still in agreement with our estimate. This is no longer the case for the quantities $f$ and $\Lambda_4$ (or $\bar{\ell}_4$) which predominantly affect the quark-mass dependence of $f_{\pi}$. It is noteworthy that the ratio $f_\pi^{\rm phys}/f$ (where $f_\pi^{\rm phys}$ is the value extrapolated from the fit) tends to shift toward higher values once more and more lighter masses are excluded (see, e.g., the bottom panel of Fig.\,\ref{fig:NLO:massrange_fine_pheno}), although both $f$ and $f_\pi^{\rm phys}$ are shifting toward lower values (see Figs.\,\ref{fig:NLO:massrange_fine} and \ref{fig:NLO:massrange_fine_phys}, respectively). In our opinion, these observations illustrate the danger inherent in applying NLO-ChPT formulas to lattice data with $\Mpi^\mr{min}$ too large (and our data suggest that, at least for some channels, this might be the case with $\Mpi^\mr{min}\sim200\MeV$ already).

\subsubsection{Parameter-reduced NLO ChPT}
\label{subsubsec:NLOfits_fix}

\begin{figure}
\begin{center}
\includegraphics[width=.45\textwidth]{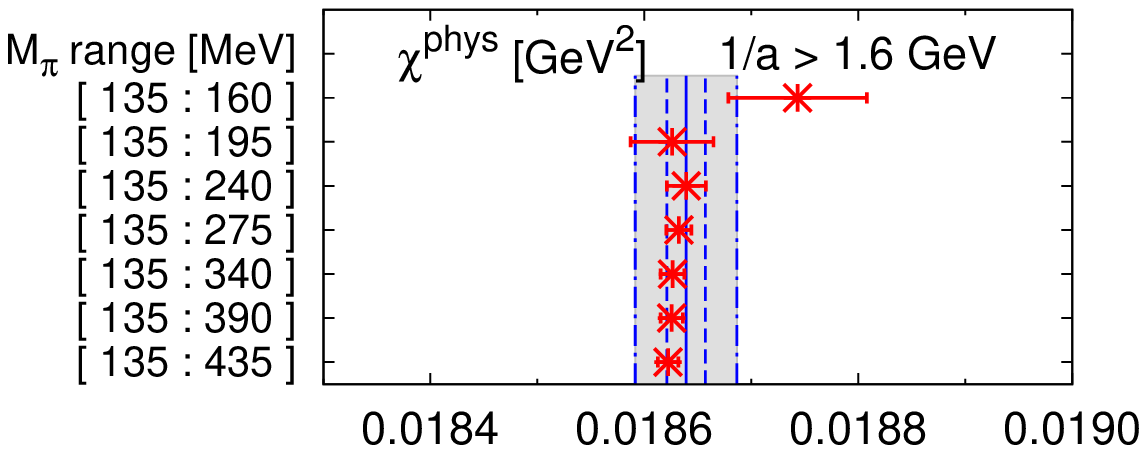}%
\includegraphics[width=.45\textwidth]{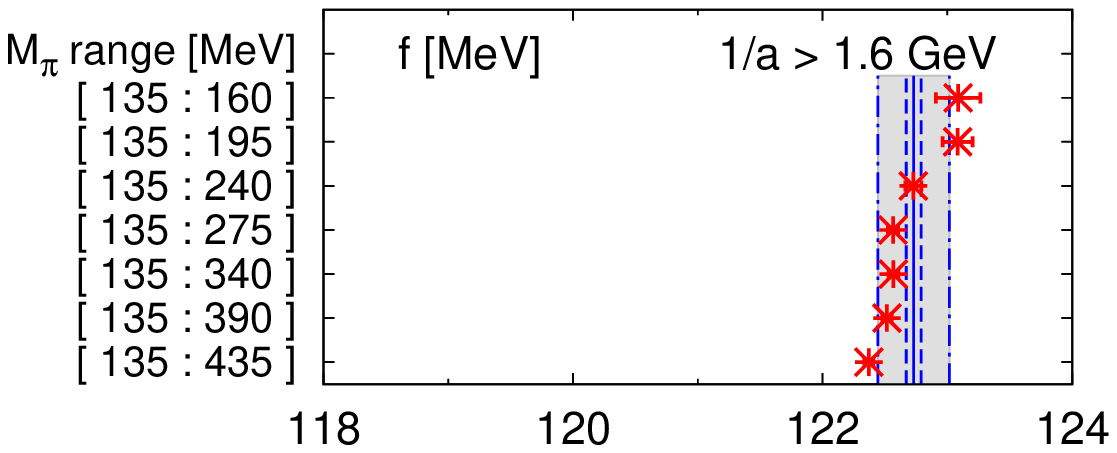}\\
\includegraphics[width=.45\textwidth]{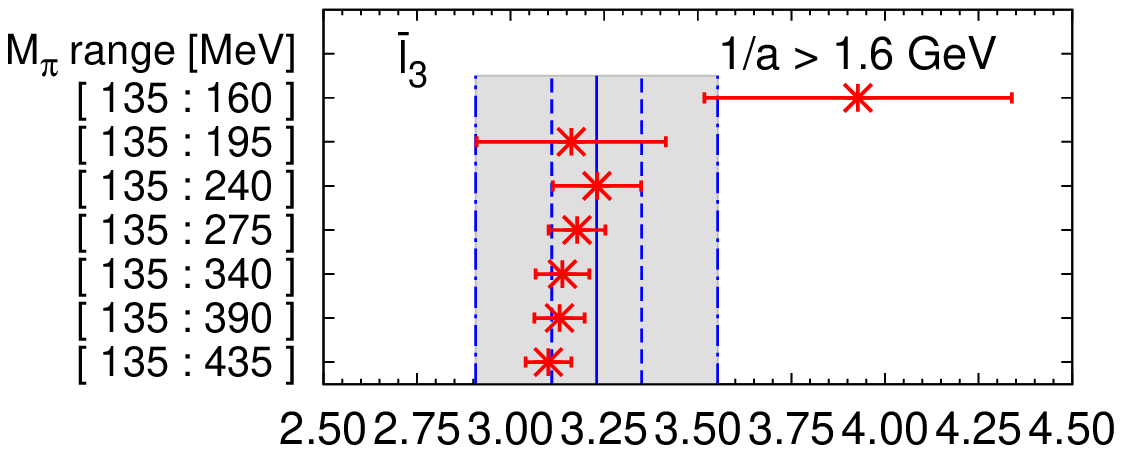}%
\includegraphics[width=.45\textwidth]{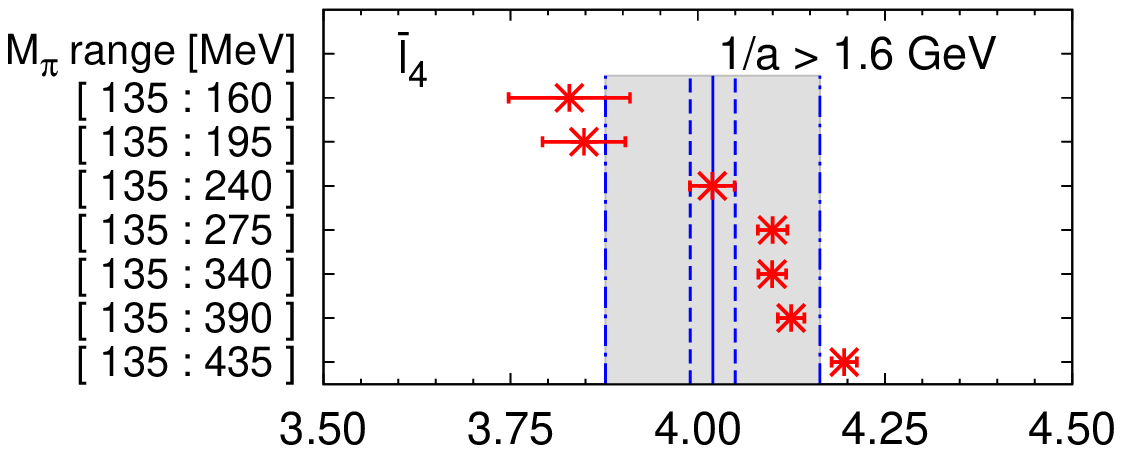}\\
\includegraphics[width=.45\textwidth]{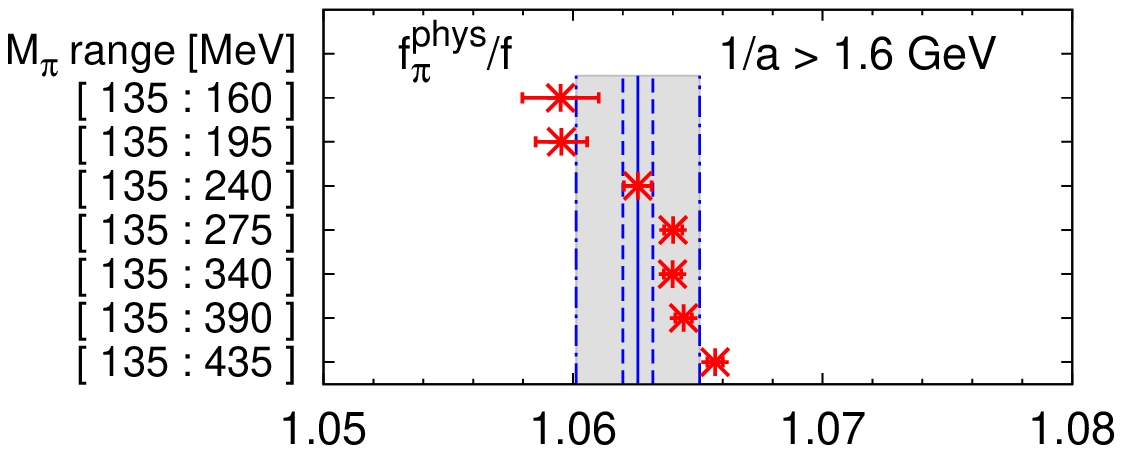}%
\includegraphics[width=.45\textwidth]{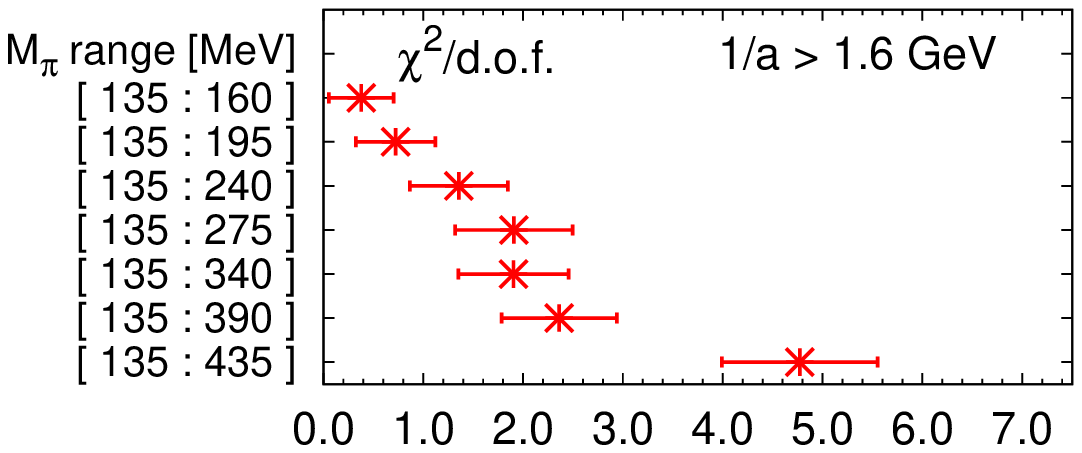}
\caption{Results for the fitted parameters (\textit{top panels}) and $\chi^2/\textrm{d.o.f.}$ (\textit{bottom right panel}) from parameter-reduced NLO-ChPT fits, using different mass ranges and including only lattice spacings $1/a\:>\:1.6\,{\rm GeV}$. In this setup $\bar{\ell}_3$, $\bar{\ell}_4$ and $f_\pi^{\rm phys}/f$ are derived quantities (see text). The \textit{solid, dashed and dashed-dotted blue lines} for the fit parameters denote the central value, statistical and total (statistical plus systematic) error bands, respectively, from our preferred parameter-reduced fit, cf.\ right column of Table~\ref{tab:NLO:results}.}
\label{fig:NLOfix:massrange_fine}
\end{center}
\end{figure}

\begin{figure}
\begin{center}
\includegraphics[width=.45\textwidth]{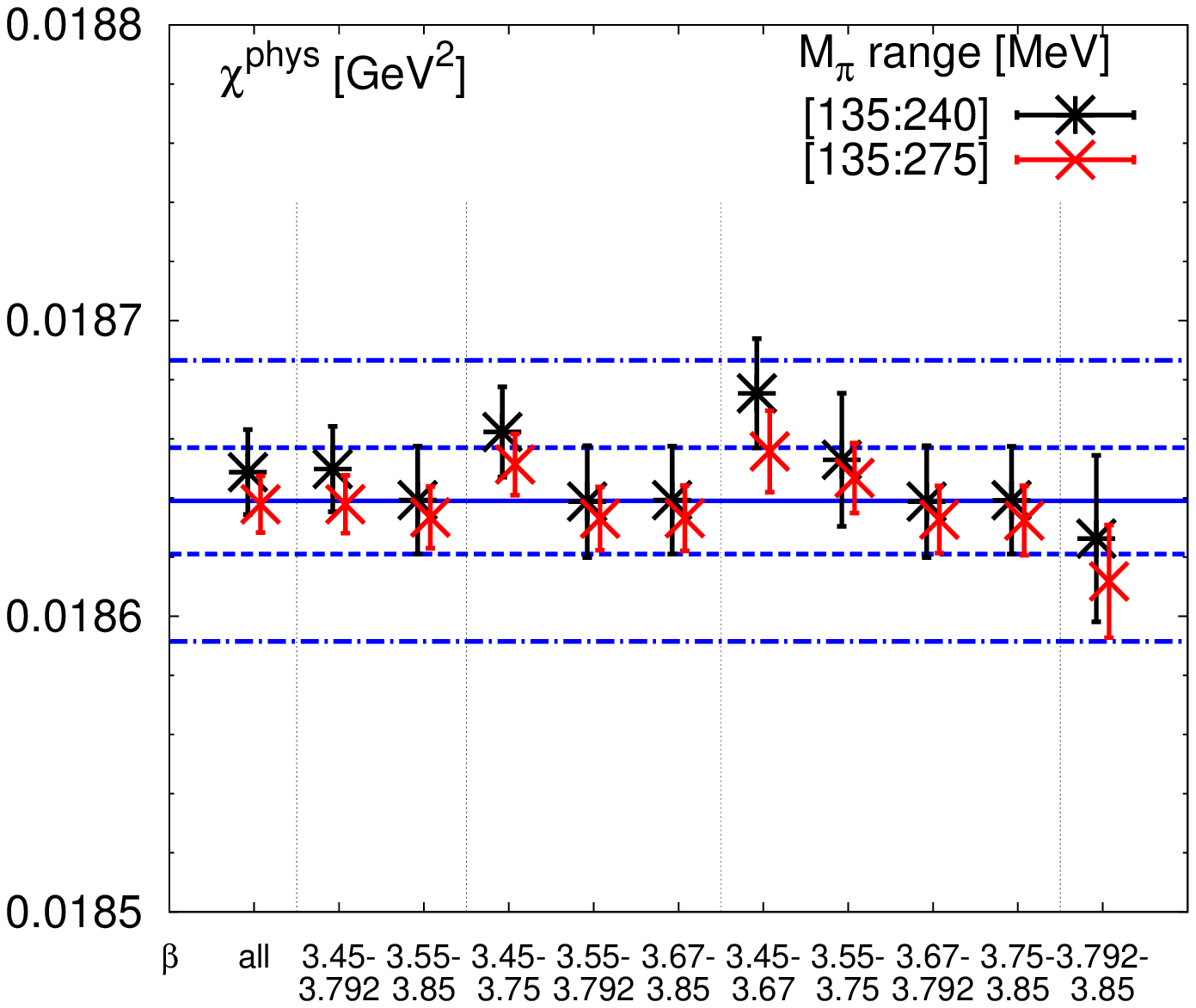}%
\includegraphics[width=.45\textwidth]{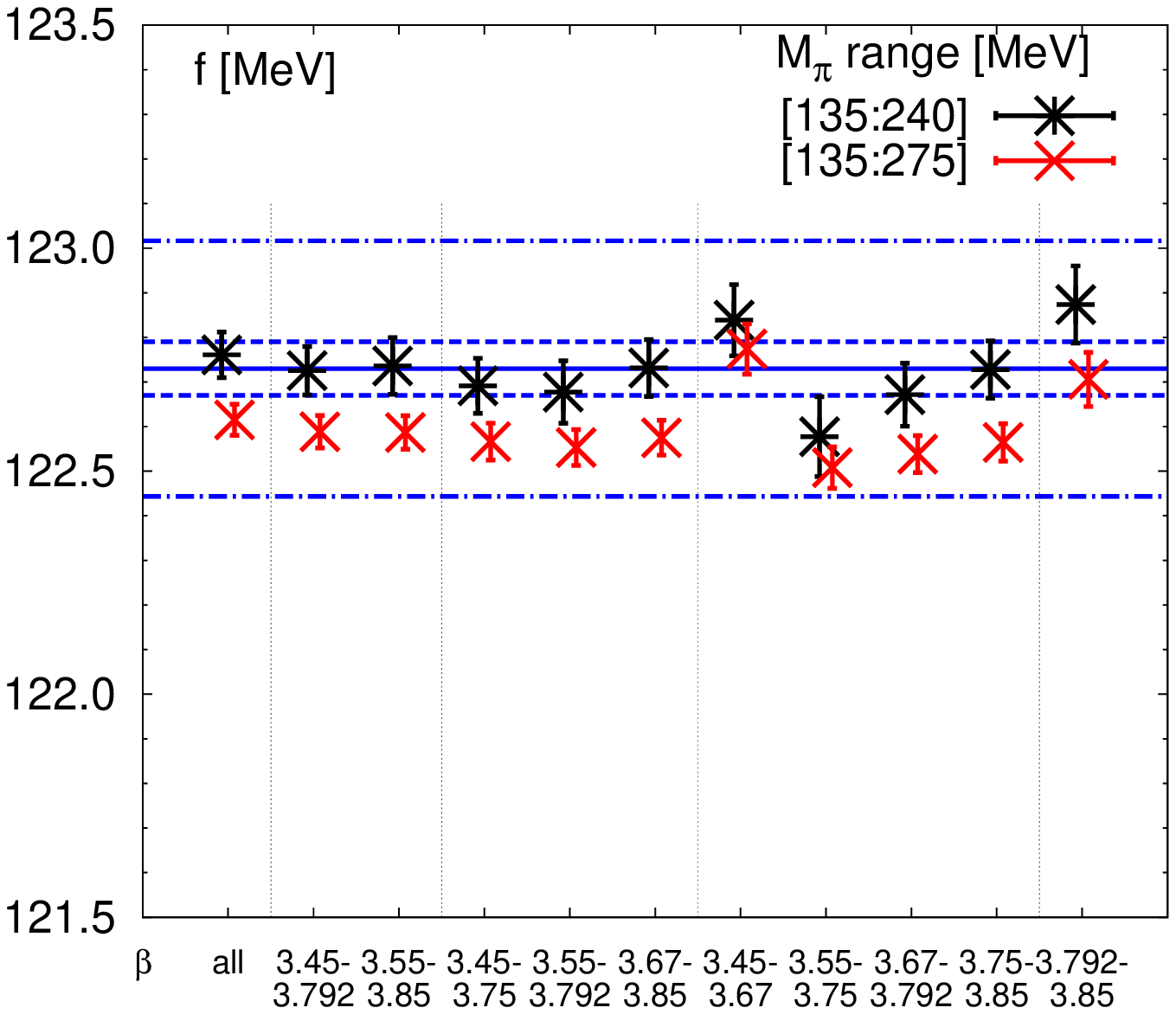}\\
\includegraphics[width=.45\textwidth]{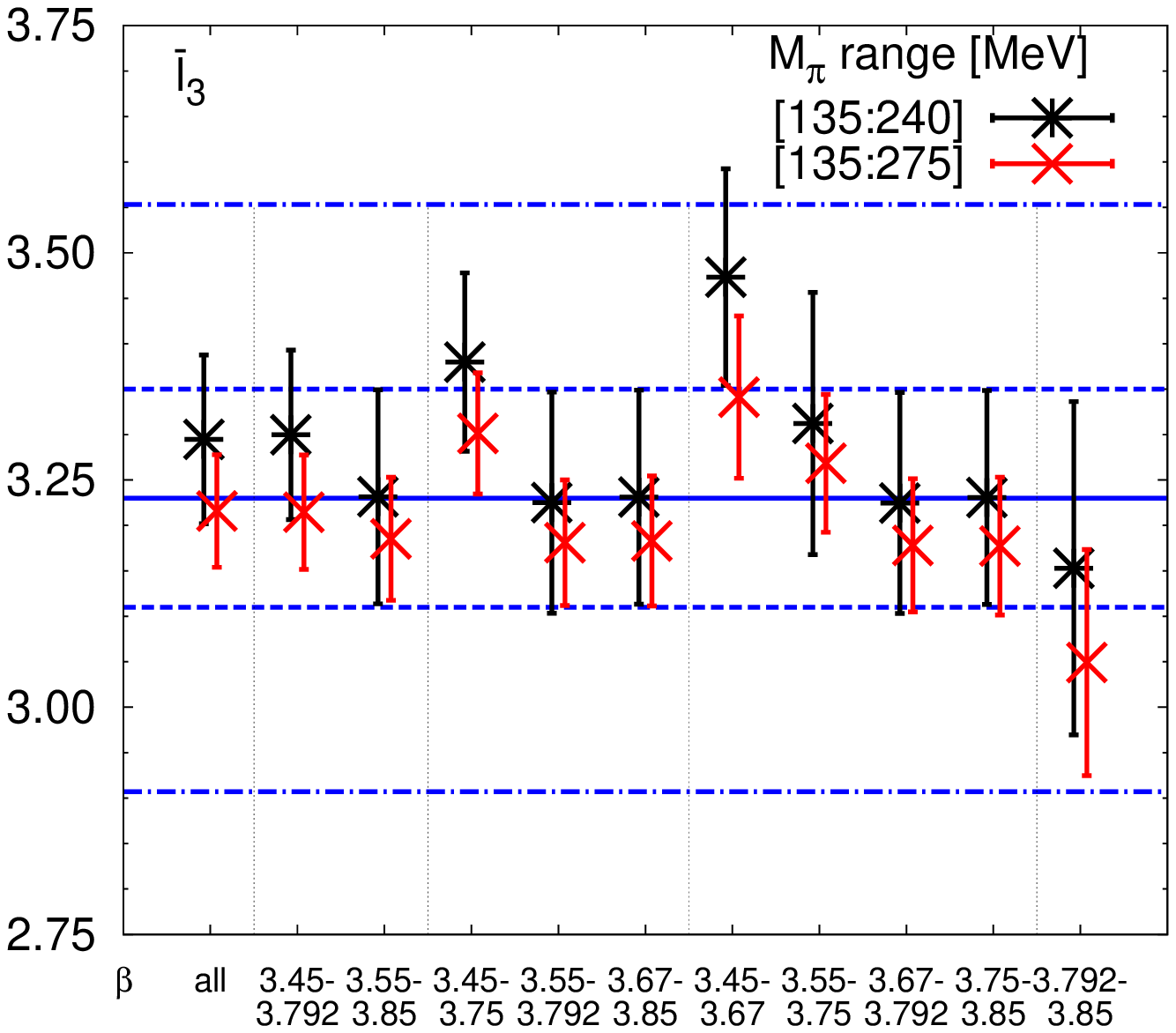}%
\includegraphics[width=.45\textwidth]{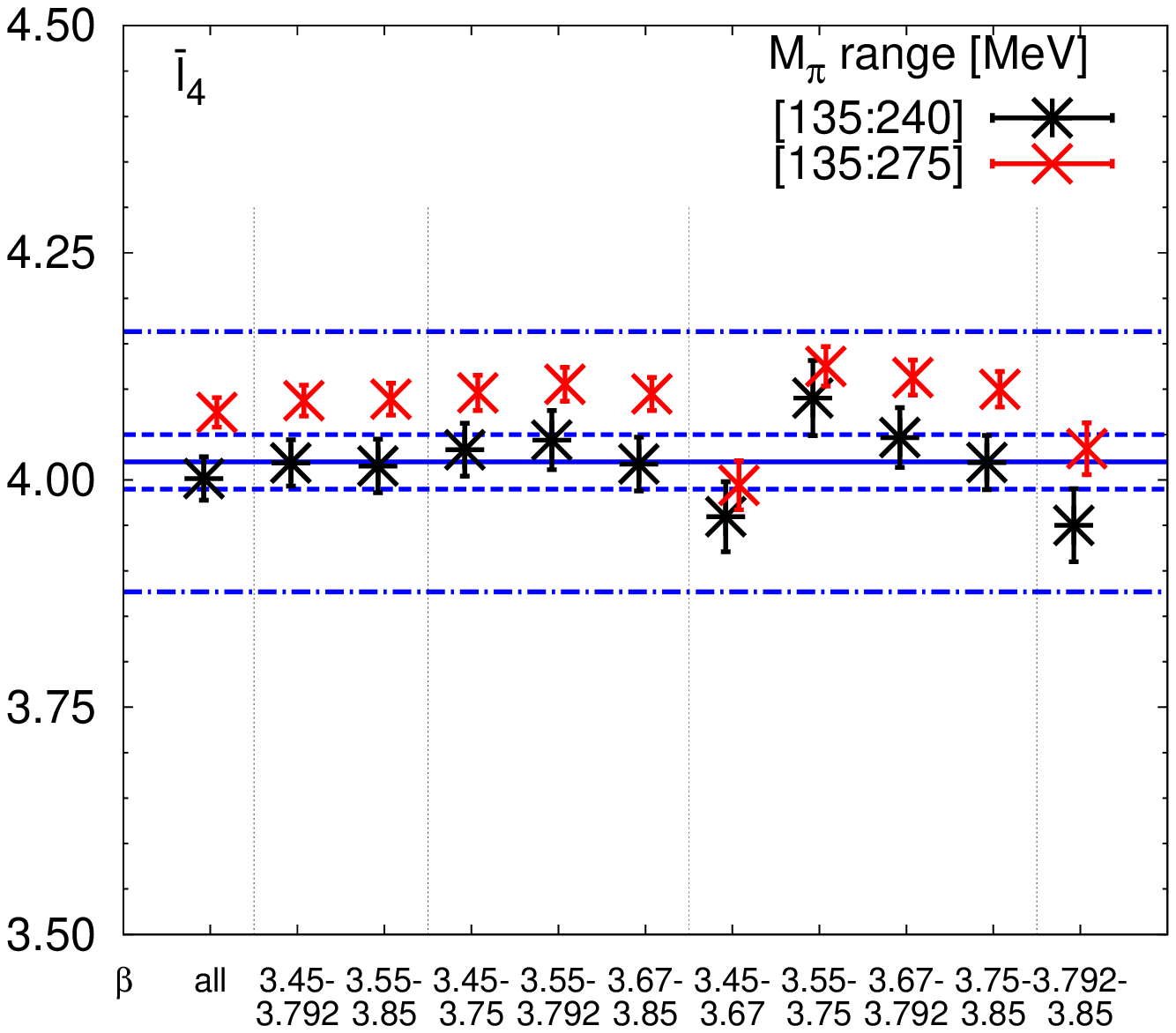}\\
\includegraphics[width=.45\textwidth]{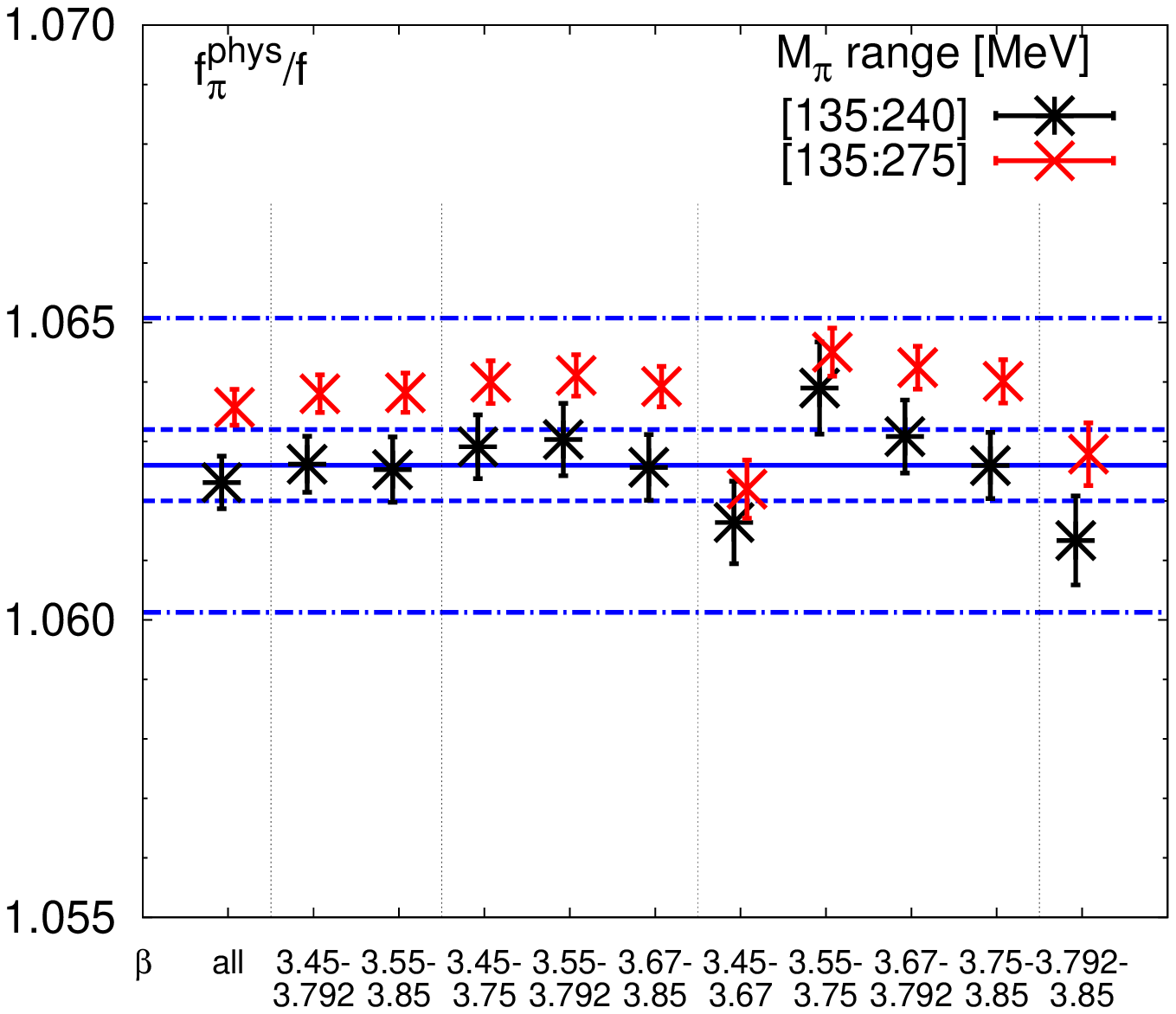}%
\includegraphics[width=.45\textwidth]{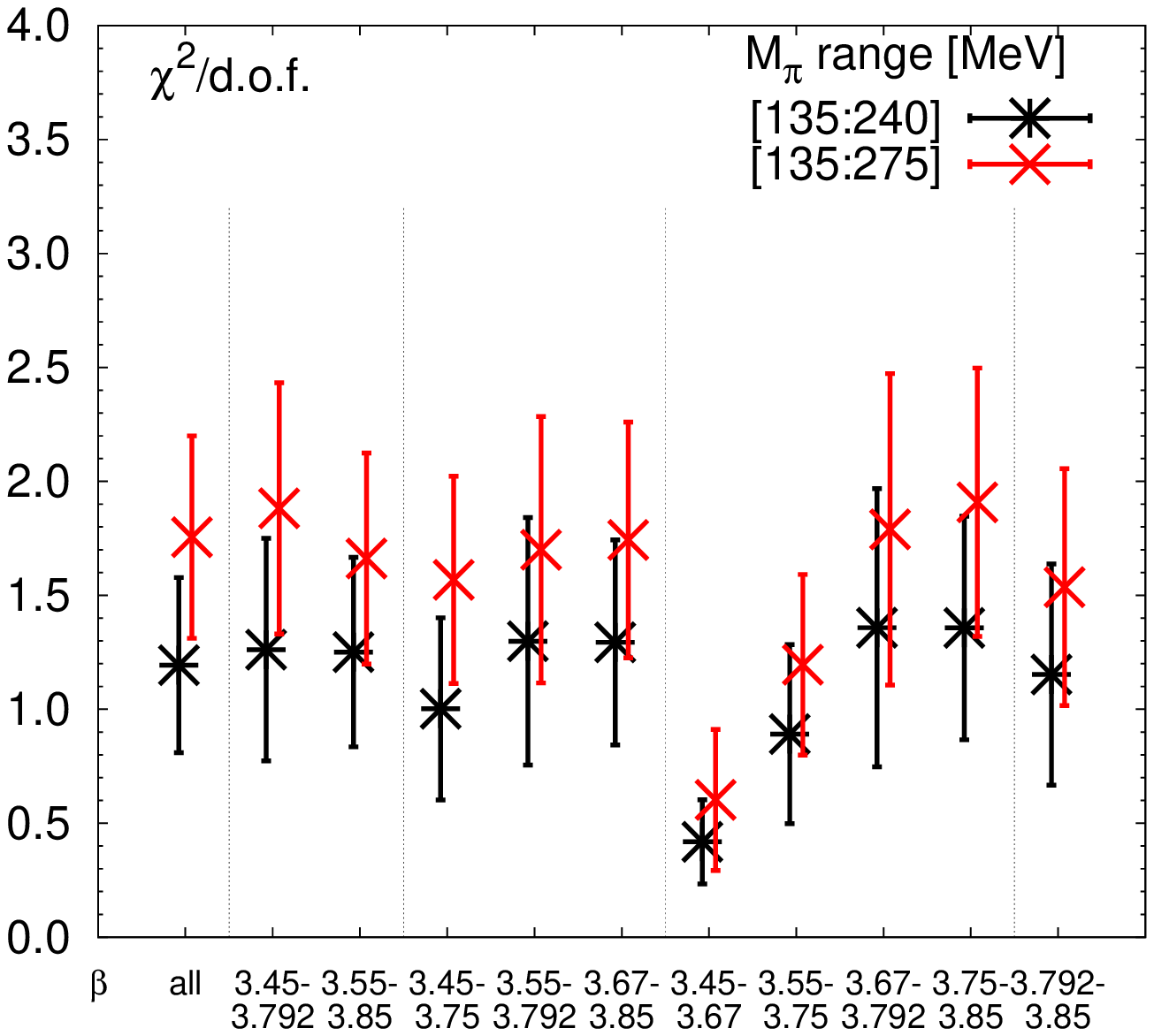}
\caption{Fit parameters and $\chi^2/\textrm{d.o.f.}$ from parameter-reduced NLO-ChPT fits to two different mass ranges, where several lattice spacings have been excluded. \textit{Vertical dashed lines} group points where no, one, two, three or four (from left to right) lattice spacings have been excluded. The \textit{solid, dashed and dashed-dotted horizontal blue lines} denote the central value, statistical and total (statistical plus systematic) error bands, respectively, from our preferred parameter-reduced fit, cf.\ right column of Table~\ref{tab:NLO:results}.}
\label{fig:NLOfix:slidingB}
\end{center}
\end{figure}

As mentioned in the beginning of this section, we also considered parameter-reduced NLO-ChPT fits which are constrained to reproduce the physical values of the pion mass and decay constant at the physical quark mass, see Eqs.~(\ref{eq:NLO:Msqconstr}) and (\ref{eq:NLO:fconstr}). This will reduce the number of fit parameters in our combined global NLO fits from four to two, leaving us with $\chi^{\rm phys}=2Bm^{\rm phys}$ and $f$ as fit parameters in the constrained fits. Note that in this approach the low-energy constants $\bar\ell_3$ and $\bar\ell_4$ can still be determined, but they are no longer independent, since they are linked to the leading-order constants as stated in Eqs.~(\ref{eq:NLO:Lam3constr}) and (\ref{eq:NLO:Lam4constr}). The main purpose of this exercise is to show that our unconstrained fits did not need the additional degrees of freedom to work, given that we used the pion mass and decay constants to set the scales for our lattice data. We performed fits with different mass ranges and by limiting the lattice spacings included in our fits as before. In Fig.~\ref{fig:NLOfix:massrange_fine} we show the results for the fitted parameters and derived quantities as well as the $\chi^2/\textrm{d.o.f.}$ for different mass ranges including only ensembles with $1/a\:>\:1.6\,{\rm GeV}$. The central values and error bands shown in these plots have been obtained as before: the central value and statistical error is the one from the fit with the mass range $135\,{\rm MeV}\:\leq\:M_{\pi}\:\leq\:240\,{\rm MeV}$, while the systematic error has been obtained from the variation with respect to the results from other fitting ranges in the meson mass (as indicated by the gray shaded areas). (We do not show the results from fits excluding the nearly physical points in this case, since due to the constraints these results now more or less agree with the results shown.) Figure \ref{fig:NLOfix:slidingB} contains the plot showing the influence of the included lattice spacings on the fit results. We would like to state, that these parameter-reduced fits work equally well as the unconstrained ones, as measured by the resulting values for $\chi^2/\textrm{d.o.f.}$. Our estimates for the LECs and derived quantities from this fitting procedure are given in the right column of Table~\ref{tab:NLO:results} (the first two parameters were fitted, the remaining ones were derived from these two). These values are in good agreement with the results from the unconstrained NLO fits, given in the left column of that same table, but show a twice as large statistical and systematic error for $\Lambda_3$ or equivalently $\bar{\ell}_3$, whereas the remaining uncertainties are roughly the same or slightly reduced.

\subsection{Results for LECs} 
\label{subsec:LEC}

As our final set of low-energy constants determined from the NLO SU(2)-ChPT fits we quote the following values (which supersede the preliminary results in Ref.\,\cite{Scholz:2011rk}).
%
\begin{eqnarray}
\chi^{\rm phys}\:=\:2Bm^{\rm phys} &=& 1.8609(18)_{\rm stat}(74)_{\rm syst}\,\cdot\,10^{-2}\,{\rm GeV}^2 \label{eq:result:NLO:chiphys}\,,\\
f &=& 122.72(07)_{\rm stat}(35)_{\rm syst}\,{\rm MeV}\label{eq:result:NLO:f}\,,\\
\Lambda_3 &=& 653(32)_{\rm stat}(101)_{\rm syst}\,{\rm MeV}\label{eq:result:NLO:Lam3}\,,\\
\Lambda_4 &=& 1,009(16)_{\rm stat}(77)_{\rm syst}\,{\rm MeV} \label{eq:result:NLO:Lam4}\,,\\
\bar{\ell}_3 &=& 3.16(10)_{\rm stat}(29)_{\rm syst} \label{eq:result:NLO:lbar3}\,,\\
\bar{\ell}_4 &=& 4.03(03)_{\rm stat}(16)_{\rm syst} \label{eq:result:NLO:lbar4}\,,\\
f_\pi^{\rm phys}/f &=& 1.0627(06)_{\rm stat}(27)_{\rm syst} \label{eq:result:NLO:fratio}\,.
\end{eqnarray}
%
Here we averaged the central values and statistical uncertainties from the unconstrained (Sec.~\ref{subsubsec:NLOfits_free}) and parameter-reduced (Sec.~\ref{subsubsec:NLOfits_fix}) fits (both summarized in Table~\ref{tab:NLO:results}). For the square of the systematic uncertainty, we sum the squares of the average systematic uncertainty and of the spread of the central values.

\section{Fits to NNLO-SU(2) ChPT}
\label{sec:NNLOChPT}

As a further step, we examined the effects of fitting our data to SU(2) ChPT including terms up to NNLO. The motivation for this exercise is twofold. On the one hand, we would like to see whether our results obtained for the LECs by fitting to NLO ChPT are stable when considering the next-higher order in ChPT. On the other hand, there is some general interest in whether the amount of data is sufficient to reliably determine the additional fit parameters that appear at NNLO and whether NNLO ChPT is superior to NLO-ChPT in terms of describing the data (as measured, e.g., by the $\chi^2/{\rm d.o.f.}$).

\subsection{Fit formulas}
\label{subsec:NNLOformulae}

The formulas for the squared meson mass and the meson decay constant up to NNLO read (cf.\,Refs.\,\cite{Colangelo:2001df,Colangelo:2010et})
%
\begin{eqnarray}
\label{eq:NNLO:Msq}
M_{\pi}^2 &=& \chi\left[1\:+\:\frac{\chi}{16\pi^2f^2}\log\frac{\chi}{\Lambda_3^2}\:+\:{\rm NNLO}_{M^2}\right]\,,\\
\label{eq:NNLO:f}
f_{\pi} &=& f\,\left[1\:-\:\frac{\chi}{8\pi^2f^2}\log\frac{\chi}{\Lambda_4^2}\:+\:{\rm NNLO}_{f}\right]\,,\\
\label{eq:NNLO:NNLO_Msq}
{\rm NNLO}_{M^2} &=& \left(\frac{\chi}{16\pi^2f^2}\right)^2\left[\frac{1}{306}\left(60\log\frac{\chi}{\Lambda_{12}^2}\:-\:9\log\frac{\chi}{\Lambda_3^2}\:-\:49\right)^2\:+\:4k_{M^2}\right]\,,\\
\label{eq:NNLO:NNLO_f}
{\rm NNLO}_{f} &=& \left(\frac{\chi}{16\pi^2f^2}\right)^2\left[-\frac{1}{180}\left(30\log\frac{\chi}{\Lambda_{12}^2}\:+\:6\log\frac{\chi}{\Lambda_3^2}\:-\:6\log\frac{\chi}{\Lambda_4^2}\:-\:23\right)^2\:+\:4k_f\right]\,.\nonumber\\
\end{eqnarray}
%
Up to this order three new fit parameters enter: one additional low-energy scale $\Lambda_{12}$ and the two NNLO LECs $k_{M^2}$, $k_f$. The low-energy scale $\Lambda_{12}$ is related to the low-energy scales usually denoted by $\Lambda_1$ and $\Lambda_2$ in the literature via
\begin{equation}
\label{eq:NNLO:Lambda12}
\log\Lambda_{12}^2 \;=\; \frac{7}{15}\log\Lambda_1^2\:+\:\frac{8}{15}\log\Lambda_2^2\,.
\end{equation} 
The low-energy scales $\Lambda_1$, $\Lambda_2$ already appear separately in the NLO formulas for other quantities (e.g., scattering lengths in $\pi\pi$ scattering \cite{Colangelo:2001df}), but since in our case only the combination $\Lambda_{12}$ appears, we will not be able to distinguish between them. As before, the low-energy scales can also be expressed via the LECs $\bar{\ell}_i$ as in Eq.~(\ref{eq:lbar34:def}), i.e.\ 
\begin{equation}
\label{eq:NNLO:lbar:def}
\bar{\ell}_i \;=\; \log\frac{\Lambda_i^2}{(M_\pi^{\rm phys})^2}\,,\quad i=1,2,3,4,12\,.
\end{equation}

\begin{figure}
\begin{center}
\includegraphics[width=.45\textwidth]{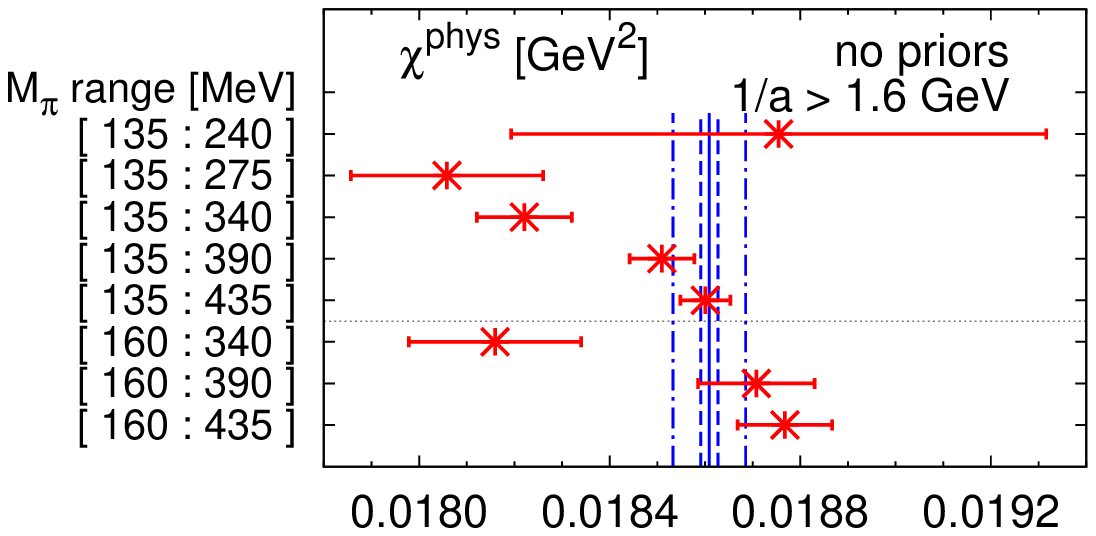}%
\includegraphics[width=.45\textwidth]{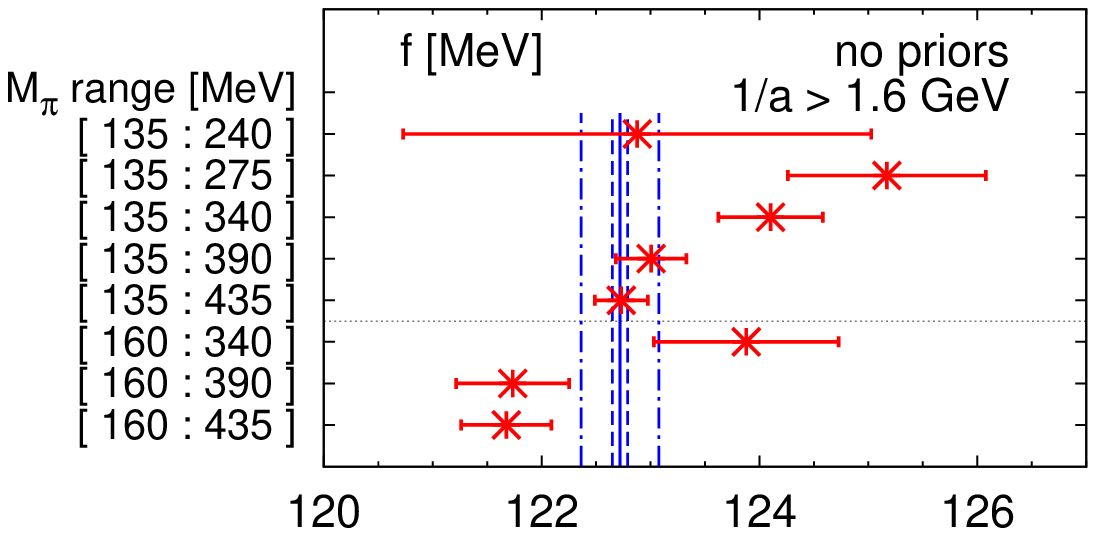}\\
\includegraphics[width=.45\textwidth]{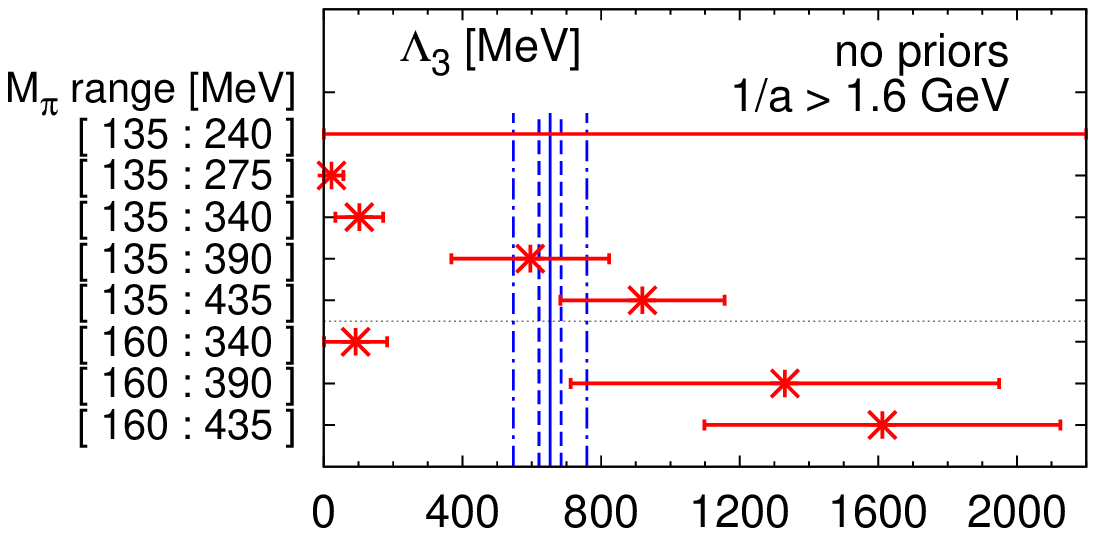}%
\includegraphics[width=.45\textwidth]{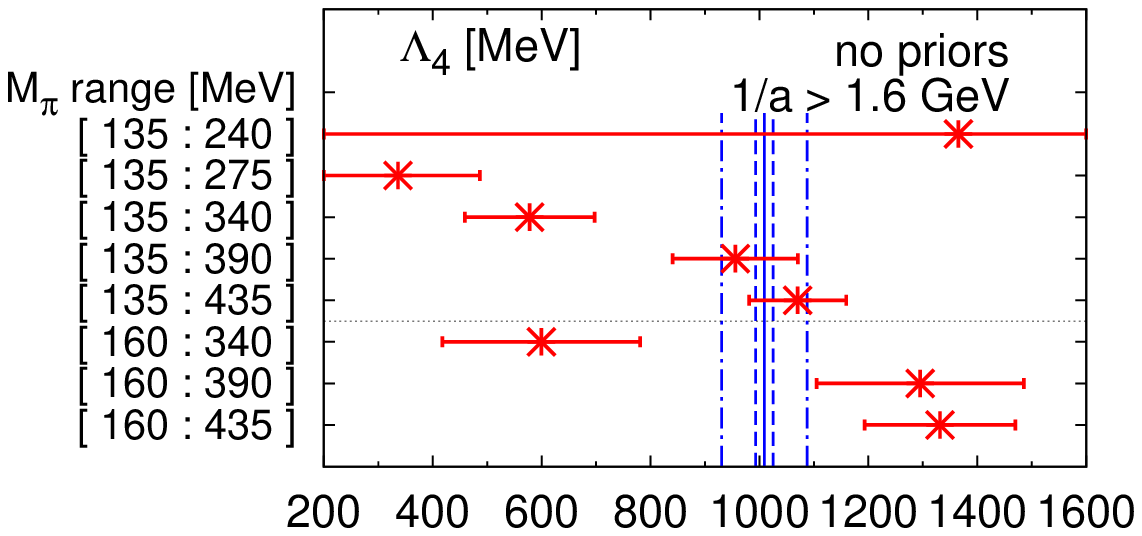}\\
\includegraphics[width=.45\textwidth]{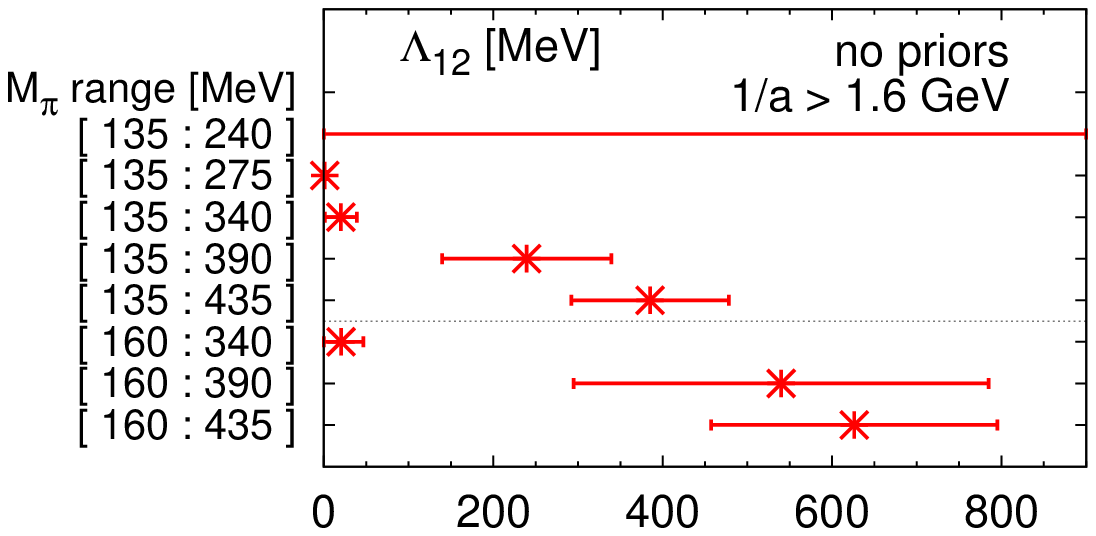}%
\includegraphics[width=.45\textwidth]{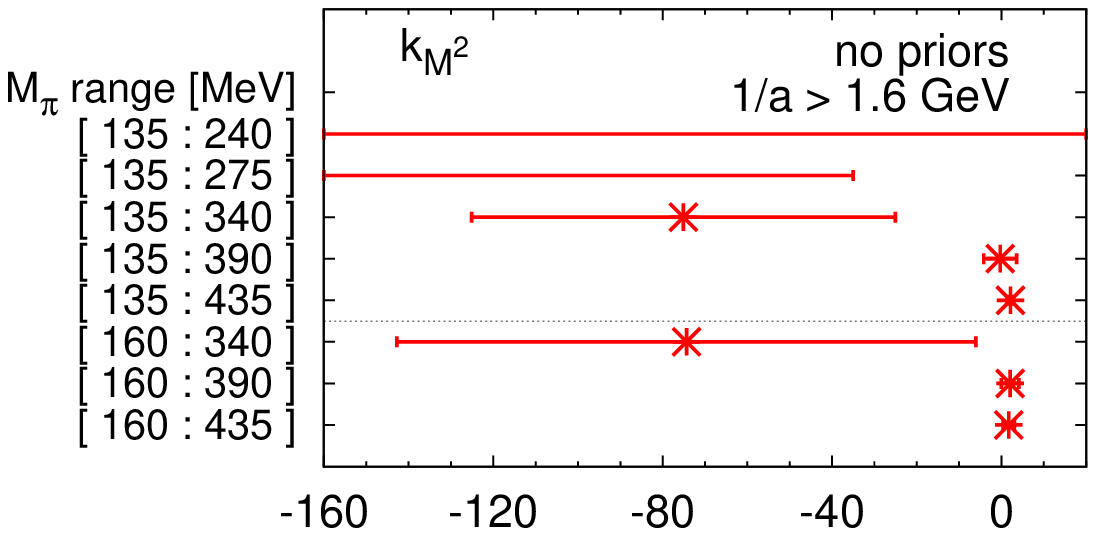}\\
\includegraphics[width=.45\textwidth]{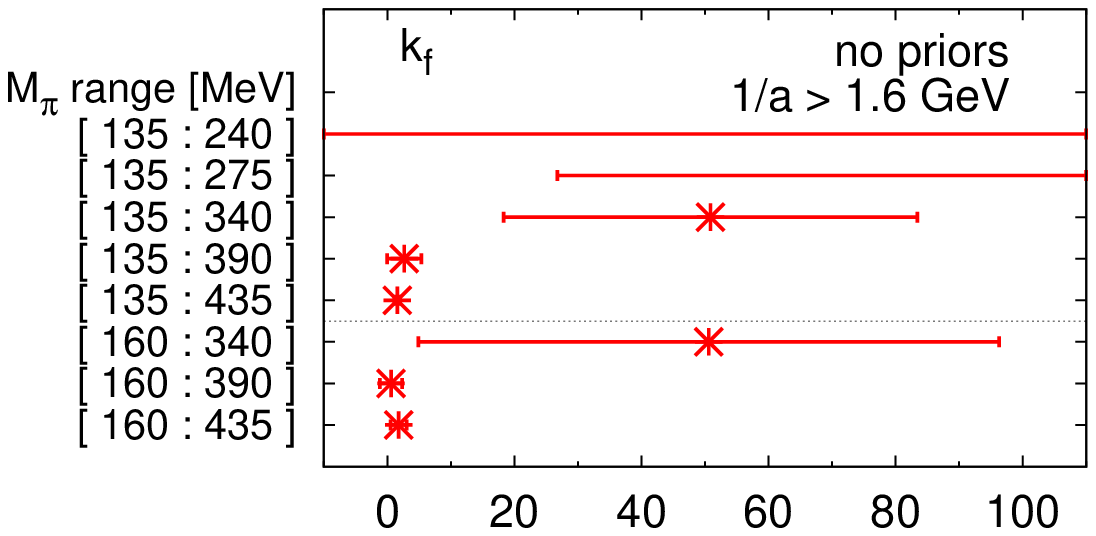}%
\includegraphics[width=.45\textwidth]{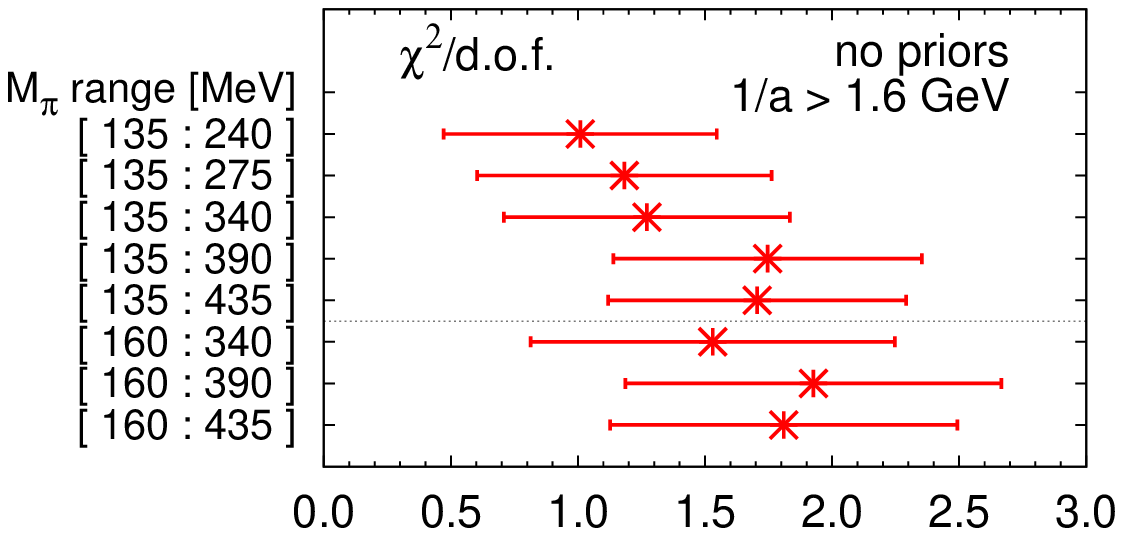}
\caption{Results for the fitted parameters and $\chi^2/\textrm{d.o.f.}$ from NNLO ChPT fits without constraints using different mass ranges and including lattice spacings $1/a\:>\:1.6\,{\rm GeV}$. The \textit{solid, dashed and dashed-dotted blue lines} for the fit parameters denote the central value, statistical and total (statistical plus systematic) error bands, respectively, from our NLO fits, cf.\ Eqs.\,(\ref{eq:result:NLO:chiphys})--(\ref{eq:result:NLO:fratio}).}
\label{fig:NNLO:massrange_fine}
\end{center}
\end{figure}

\begin{figure}
\begin{center}
\includegraphics[width=.47\textwidth]{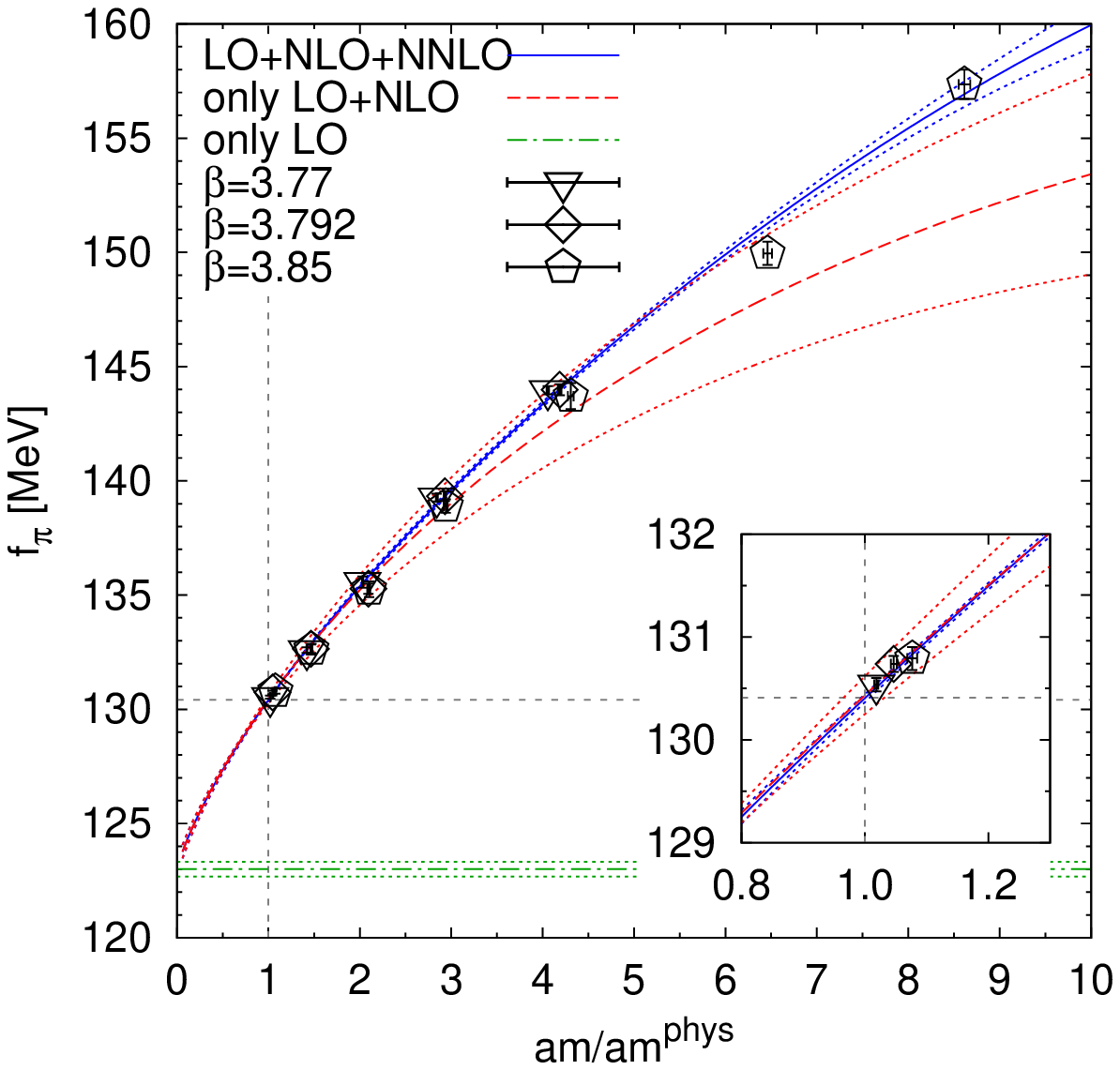}%
\includegraphics[width=.47\textwidth]{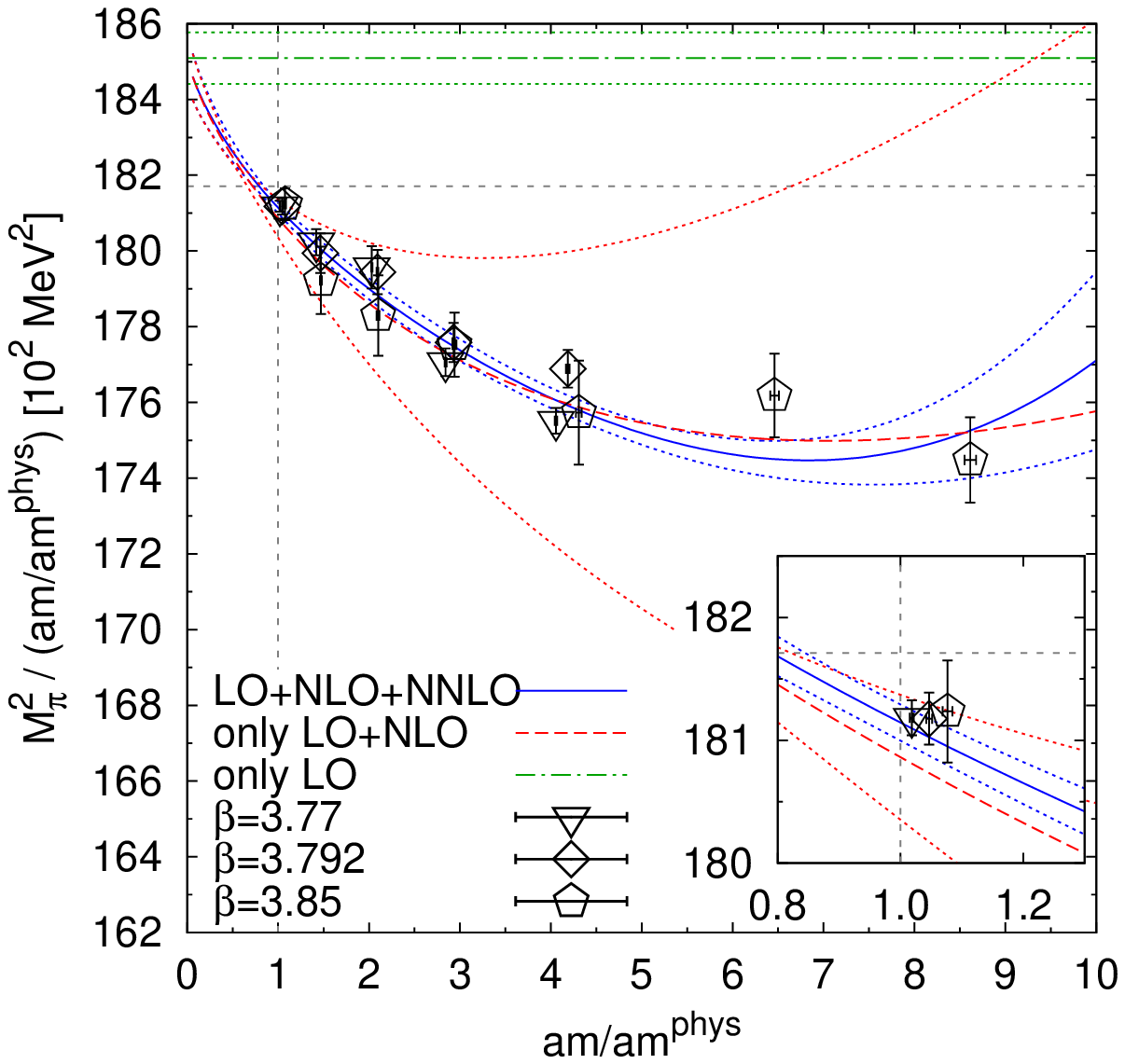}
\caption{Combined NNLO fit for lattice scales $1/a\:>\:1.6\,{\rm GeV}$ and meson masses $135\,{\rm MeV}\:\leq\:M_{\pi}\:\leq\:390\,{\rm MeV}$. \textit{Left panel:} meson decay constant, \textit{right panel:} squared meson mass divided by the quark-mass ratio. The \textit{solid blue lines} show the complete (up to NNLO) fit, whereas the \textit{dashed red lines} show the LO+NLO contribution of the full NNLO fit, and the \textit{dash-dotted green lines} show the LO contribution. Only data points included in the fit range are depicted in the plots, the physical values are marked by \textit{dashed gray lines}.}
\label{fig:NNLO:fine_135-390}
\end{center}
\end{figure}

\subsection{Combined global fits}
\label{subsec:NNLOfits}

We applied the same fit strategy for the NNLO fits as for the NLO fits (see the previous section): the scale and light quark mass at the physical point as determined in Sec.\,\ref{sec:scale} are used, and combined global fits to our data for the meson masses and decay constants are performed. First, we will discuss fits without any assumptions on the fit parameters (Sec.\,\ref{subsubsec:NNLOfits_noprior}). Later we will also constrain the additional fit parameters entering at NNLO by phenomenologically motivated estimates (Sec.\,\ref{subsubsec:NNLOfits_priors}).

\subsubsection{NNLO fits without priors}
\label{subsubsec:NNLOfits_noprior}

We will start our discussion with fit ranges including the nearly physical points and only consider varying the upper mass limit of the fit range. Since now three more parameters have to be determined, it can be expected that we will have to include more data points, i.e., including higher meson masses, compared to the NLO case. Indeed, with too few data points either the fitter could not find a solution at all or some of the fitted parameters had big numerical uncertainties. This can be seen in the compilation in Fig.\,\ref{fig:NNLO:massrange_fine} of fit results using ensembles with $1/a\,>\,1.6\,{\rm GeV}$, e.g., for the fit ranges $135\,{\rm MeV}\,\leq\,M_\pi\,\leq\,240\,{\rm MeV}$, $135\,{\rm MeV}\,\leq\,M_\pi\,\leq\,275\,{\rm MeV}$ and maybe also  $135\,{\rm MeV}\,\leq\,M_\pi\,\leq\,340\,{\rm MeV}$.

The plots of a sample fit with the range $135\,{\rm MeV}\:\leq\:M_{\pi}\:\leq\:390\,{\rm MeV}$, $1/a\:>\:1.6\,{\rm GeV}$ are given in Fig.\,\ref{fig:NNLO:fine_135-390} with a breakup into LO, NLO, and NNLO contributions (only data points included in the fit range are shown in these plots). Although technically the fit seems to work well, resulting in an acceptable $\chi^2/{\rm d.o.f.}=1.8(0.6)$ (with $\#\mr{d.o.f.}=34-7=27$) one should take the result with a grain of salt. Besides the fitted curve up to NNLO (solid blue line with error band indicated by blue dotted lines), each plot also shows the LO+NLO contribution separately (dashed red line with error band indicated by red dotted lines), i.e., the NNLO contribution being the difference between the solid blue and the dashed red line. In this case, the NLO and NNLO contributions taken separately seem to have big uncertainties. In other cases, we found the NNLO contribution to have an unnaturally large effect even at small quark masses.

At this point we have to conclude that NNLO fits to our current data are not convincing for the reasons outlined above. This situation might improve once more data points in the region between the physical point (or below) and, say, 200 or 250 MeV will be added to the analysis, allowing for fits using a smaller mass range (like we were able to do for the NLO fits). Since the generation of such data points is not planned for the near future, we will in the remainder of this section examine whether constraining the additional NNLO fit parameters can serve as a remedy to this situation (and which side effects this remedy has).

\begin{figure}
\begin{center}
\includegraphics[width=.45\textwidth]{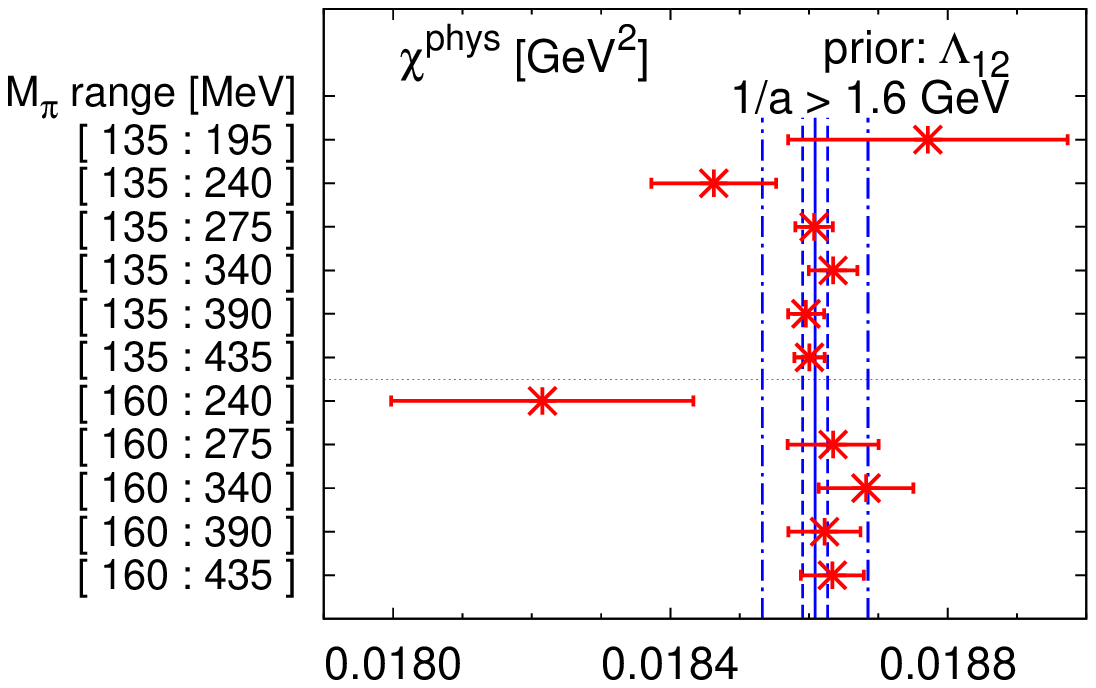}%
\includegraphics[width=.45\textwidth]{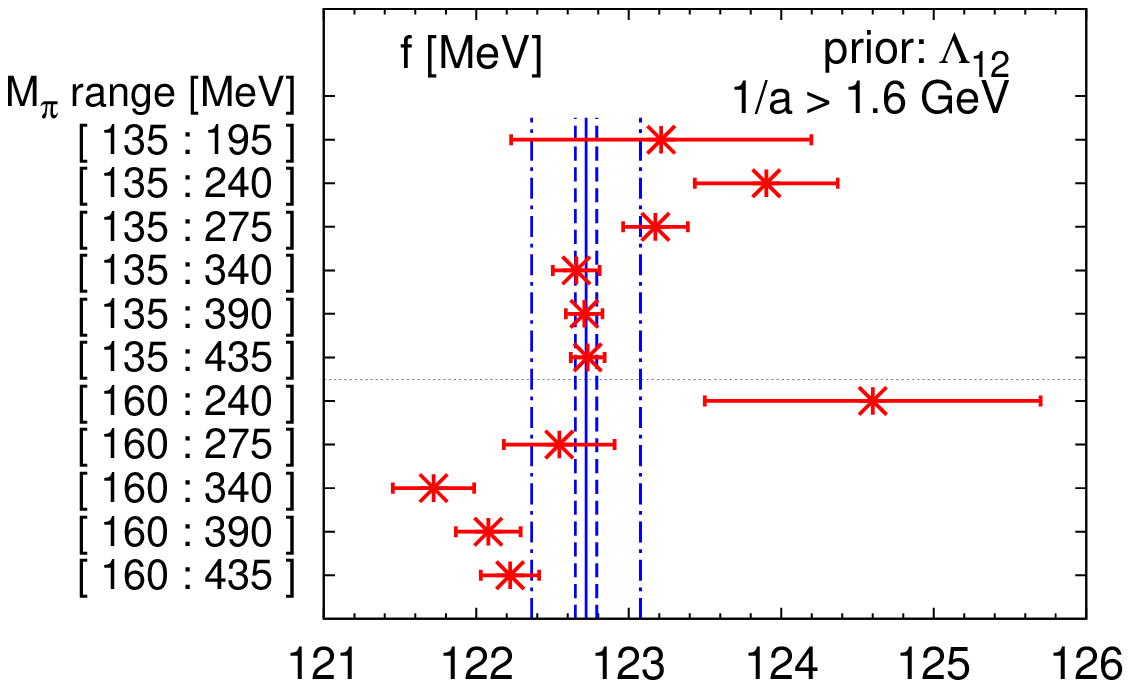}\\
\includegraphics[width=.45\textwidth]{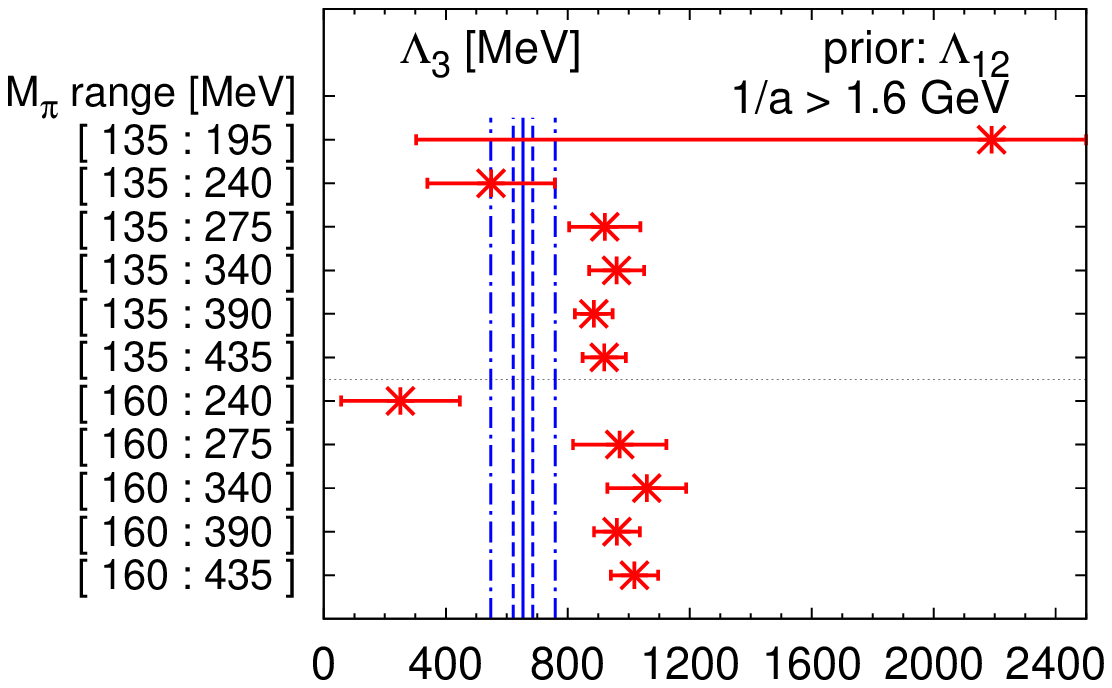}%
\includegraphics[width=.45\textwidth]{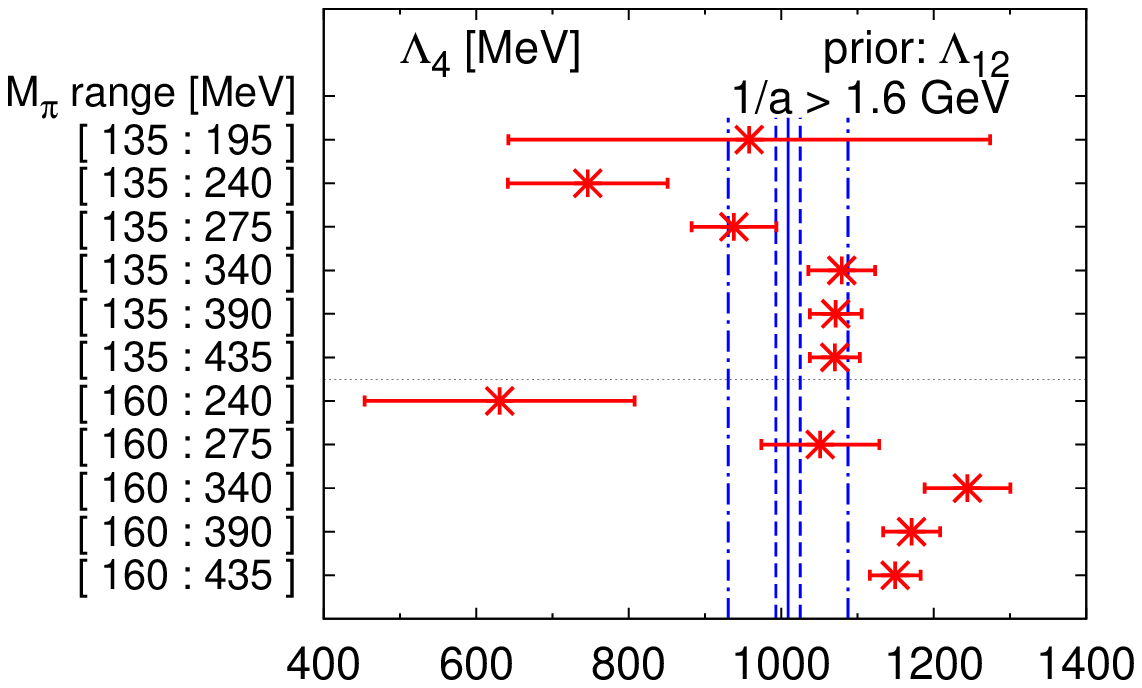}\\
\includegraphics[width=.45\textwidth]{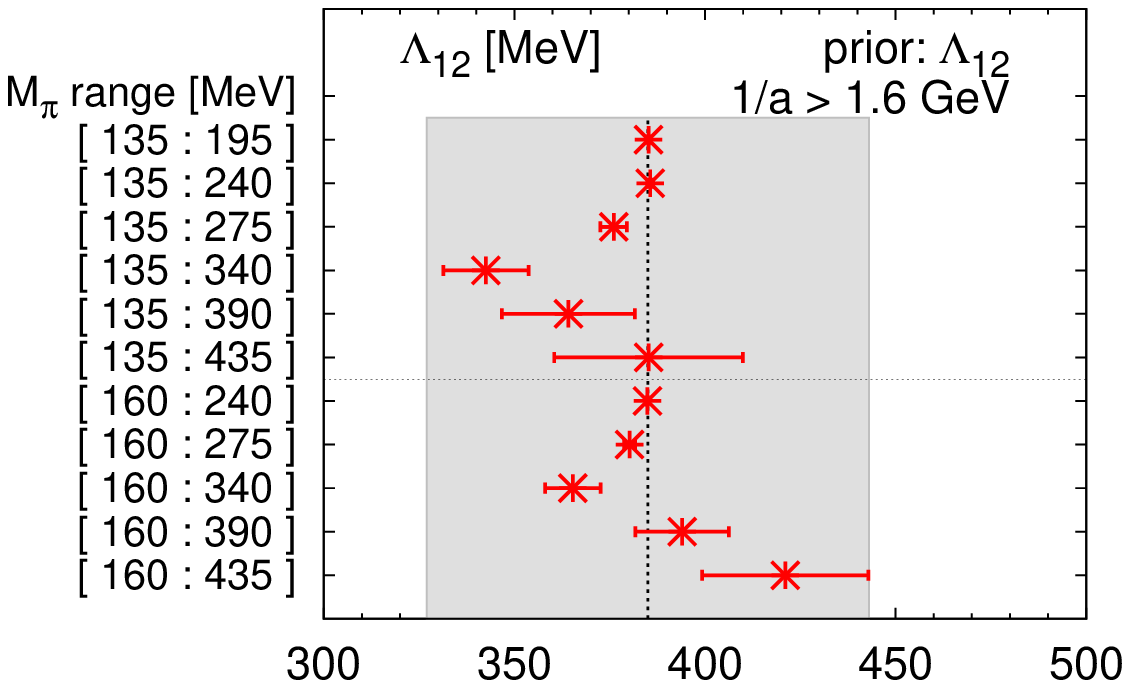}%
\includegraphics[width=.45\textwidth]{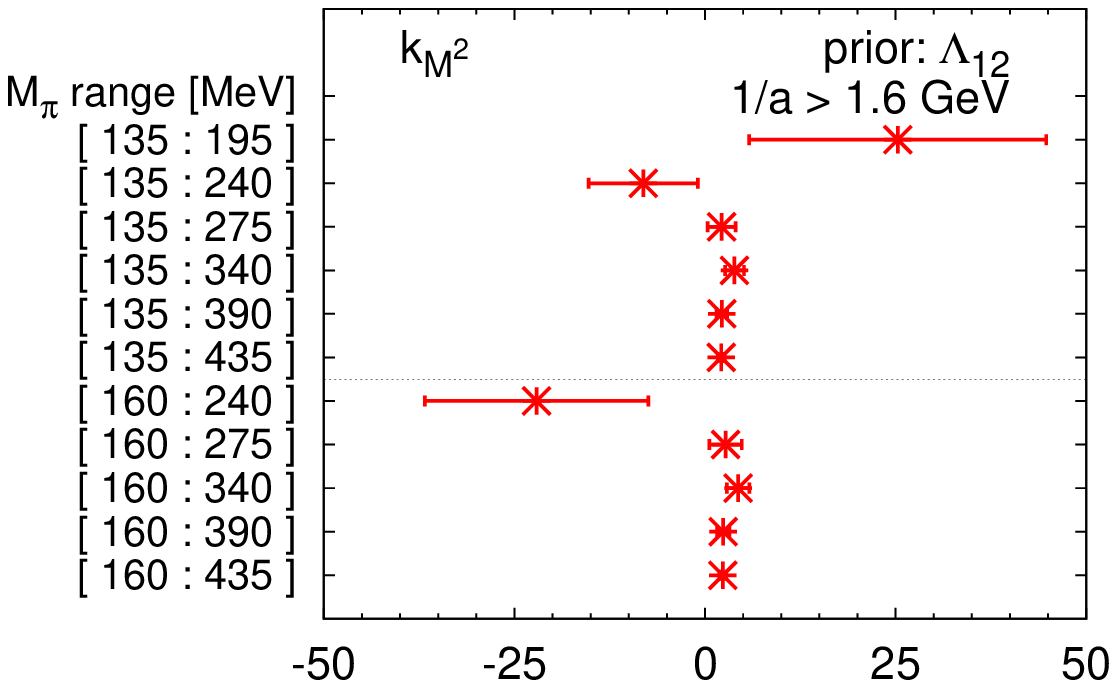}\\
\includegraphics[width=.45\textwidth]{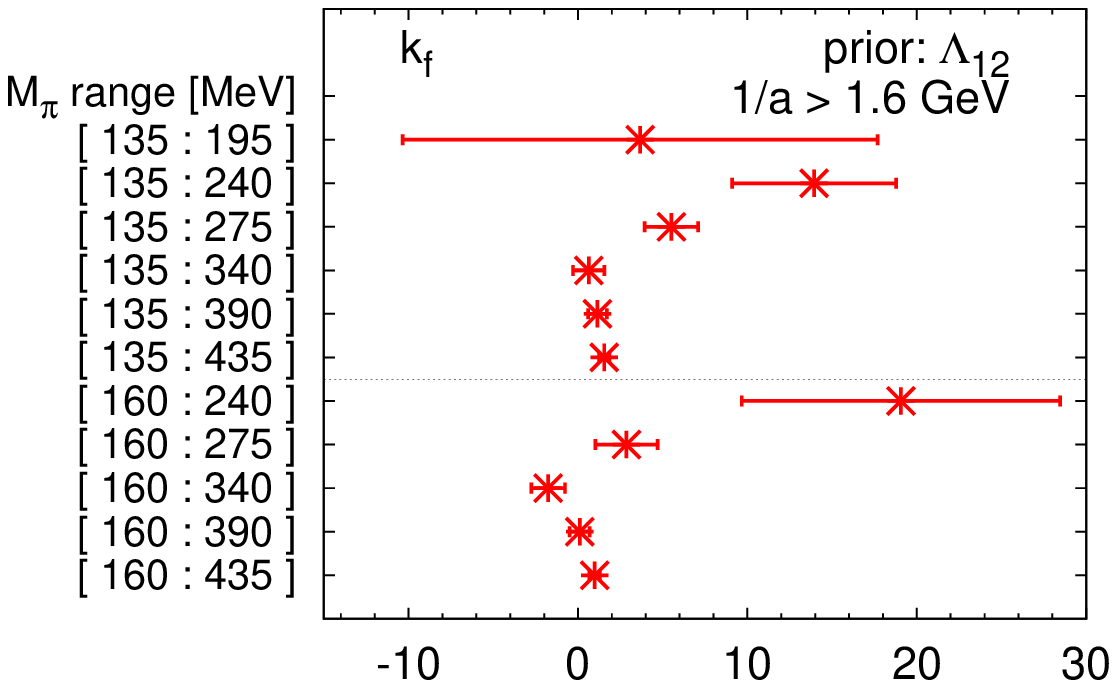}%
\includegraphics[width=.45\textwidth]{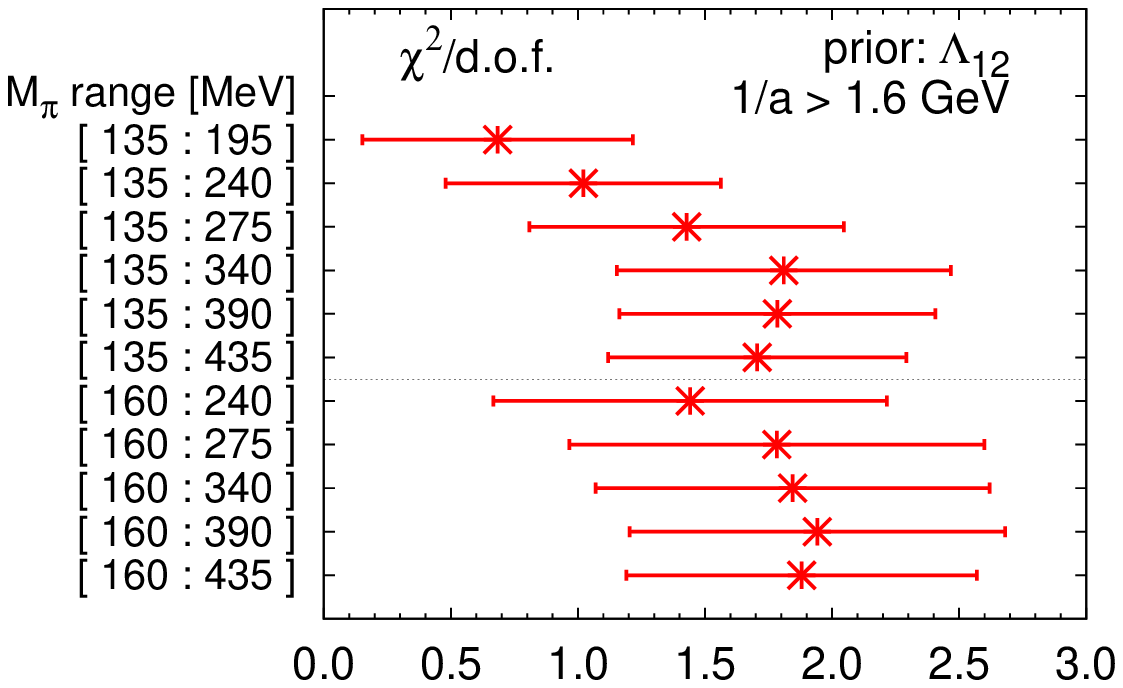}
\caption{Results for the fitted parameters and $\chi^2/\textrm{d.o.f.}$ from NNLO-ChPT fits with a prior for $\Lambda_{12}$ using different mass ranges and including lattice spacings $1/a\:>\:1.6\,{\rm GeV}$. The \textit{solid, dashed and dashed-dotted blue lines} for the fit parameters denote the central value, statistical and total (statistical plus systematic) error bands, respectively, from our NLO fits, cf.\ Eqs.\,(\ref{eq:result:NLO:chiphys})--(\ref{eq:result:NLO:fratio}). The prior on $\Lambda_{12}$ and its width are indicated by the \textit{shaded gray area} in the respective panel.}
\label{fig:NNLO:priorl12:massrange_fine}
\end{center}
\end{figure}

\begin{figure}
\begin{center}
\includegraphics[width=.45\textwidth]{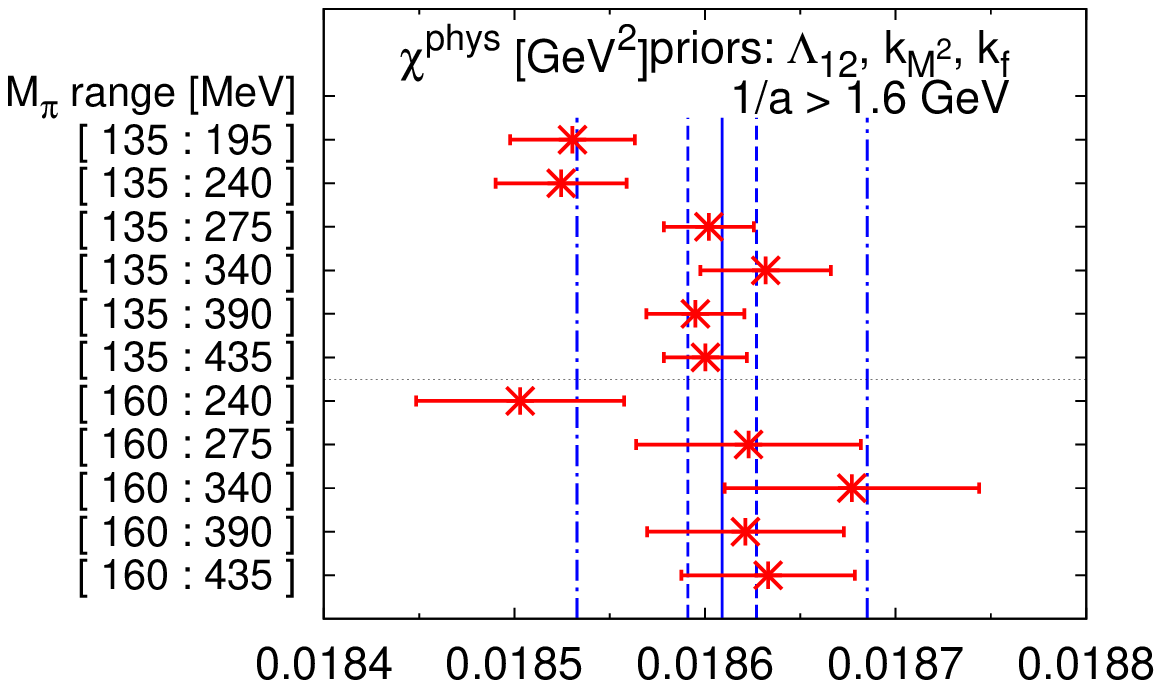}%
\includegraphics[width=.45\textwidth]{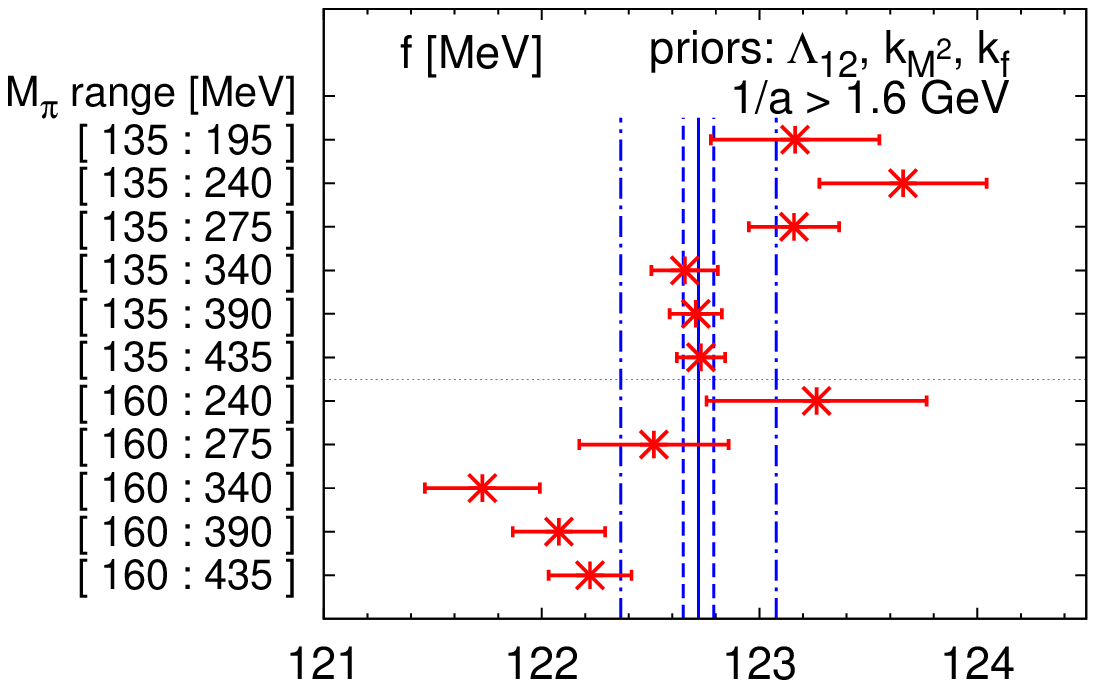}\\
\includegraphics[width=.45\textwidth]{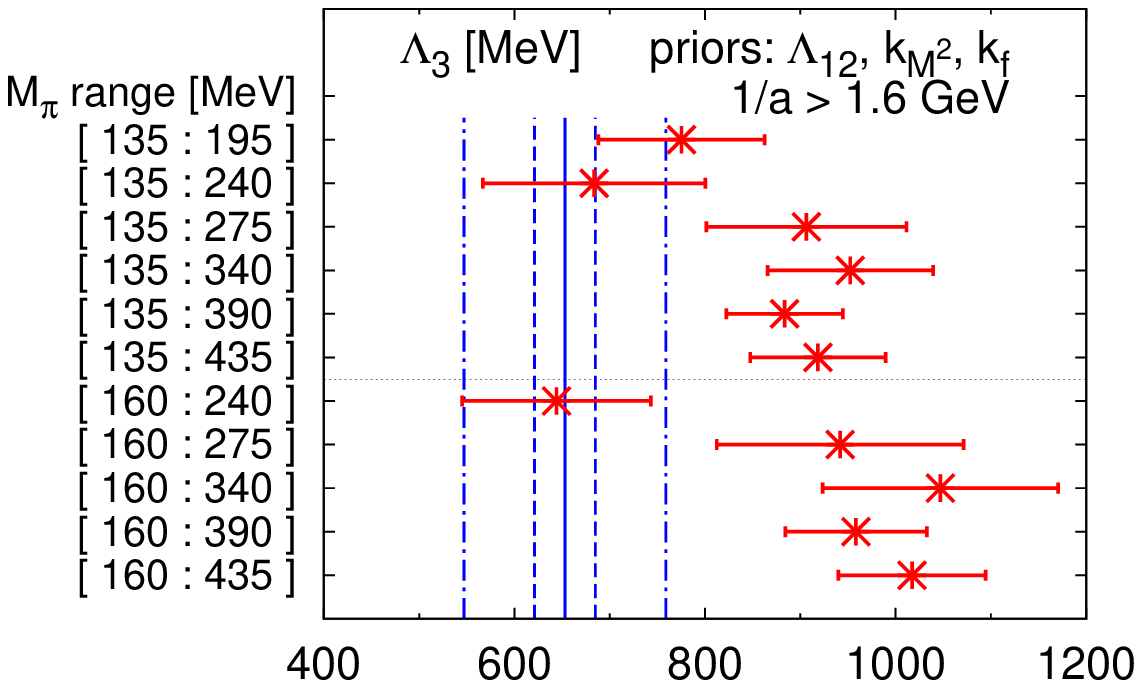}%
\includegraphics[width=.45\textwidth]{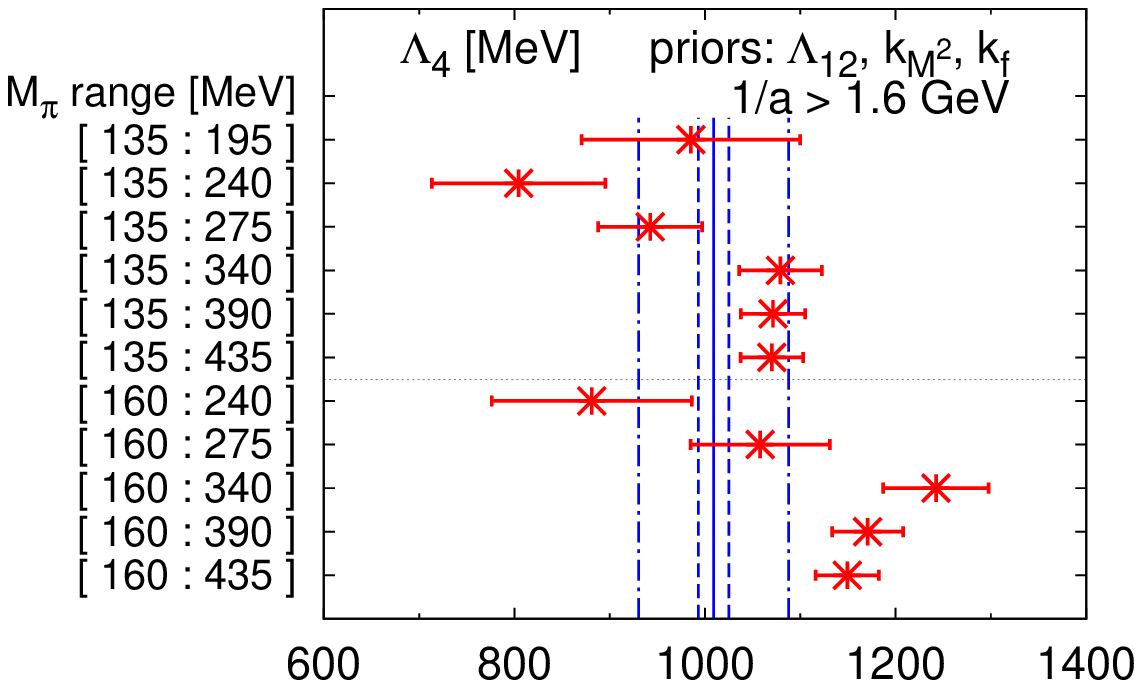}\\
\includegraphics[width=.45\textwidth]{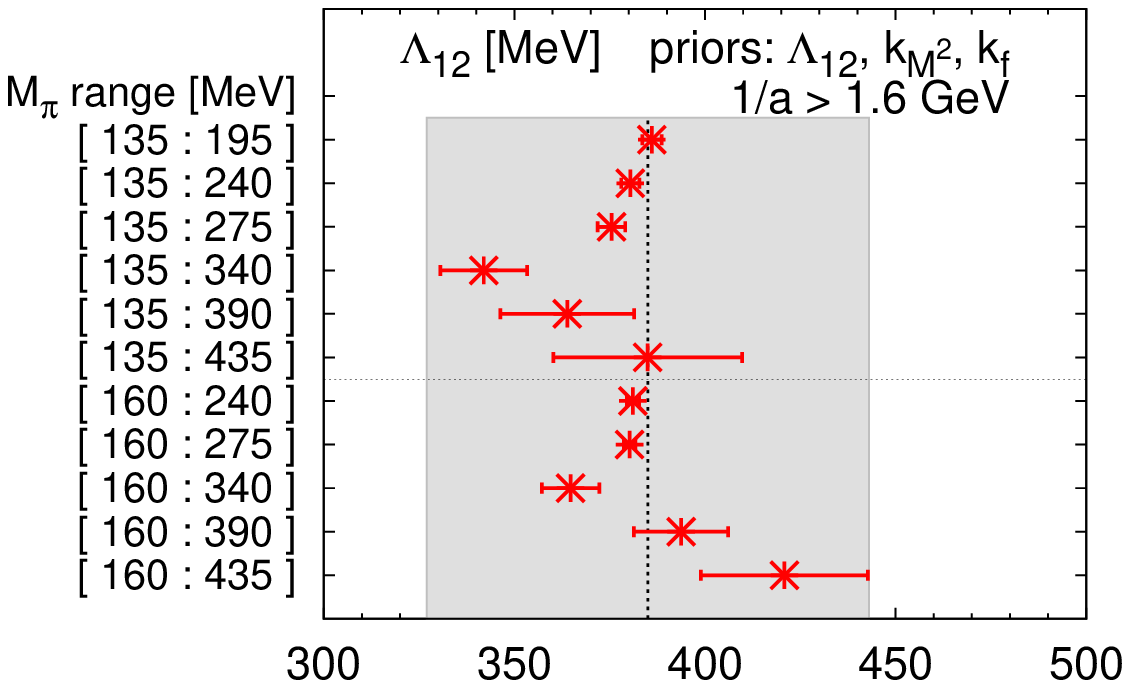}%
\includegraphics[width=.45\textwidth]{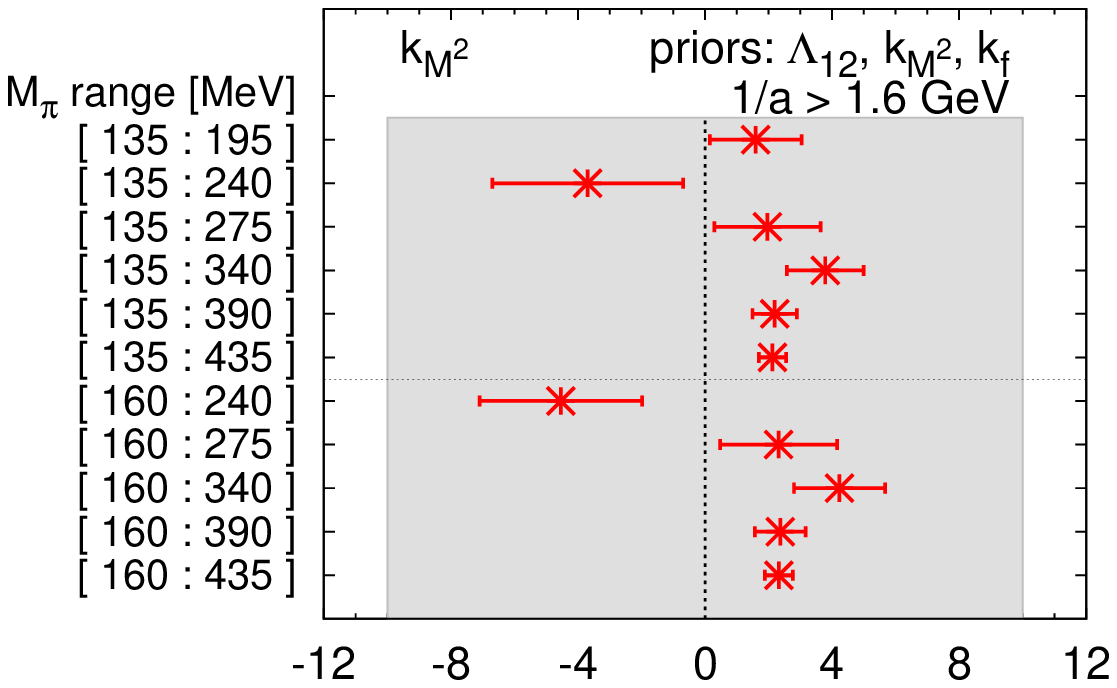}\\
\includegraphics[width=.45\textwidth]{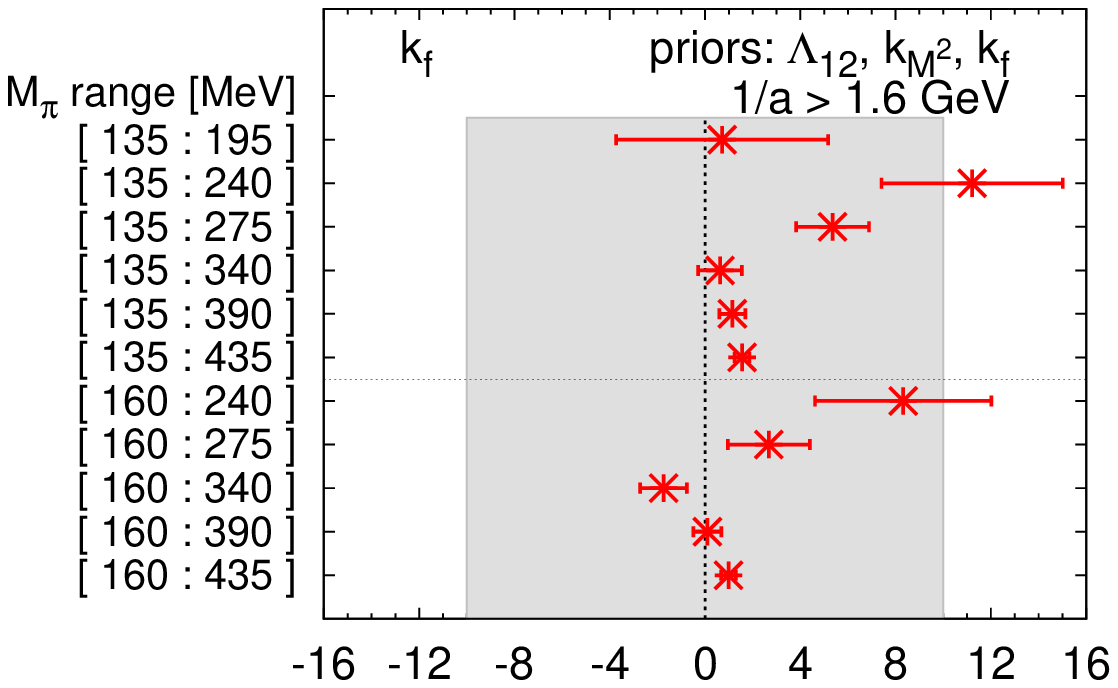}%
\includegraphics[width=.45\textwidth]{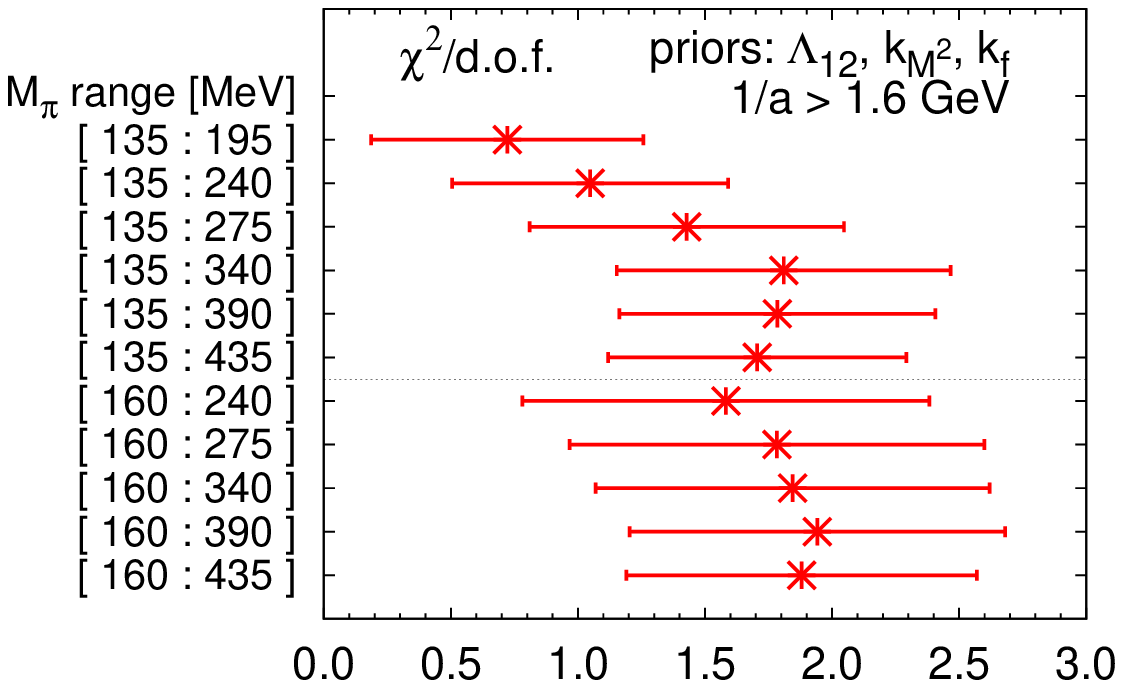}
\caption{Results for the fitted parameters and $\chi^2/\textrm{d.o.f.}$ from NNLO-ChPT fits with priors for $\Lambda_{12}$, $k_{M^2}$, and $k_f$ using different mass ranges and including lattice spacings $1/a>1.6\,{\rm GeV}$. The \textit{solid, dashed and dashed-dotted blue lines} for the fit parameters denote the central value, statistical and total (statistical plus systematic) error bands, respectively, from our NLO fits, cf.\ Eqs.\,(\ref{eq:result:NLO:chiphys})--(\ref{eq:result:NLO:fratio}). The priors and their widths are indicated by the \textit{shaded gray areas} in the respective panels. }
\label{fig:NNLO:priorl12kmkf:massrange_fine}
\end{center}
\end{figure}

\begin{figure}
\begin{center}
\includegraphics[width=.47\textwidth]{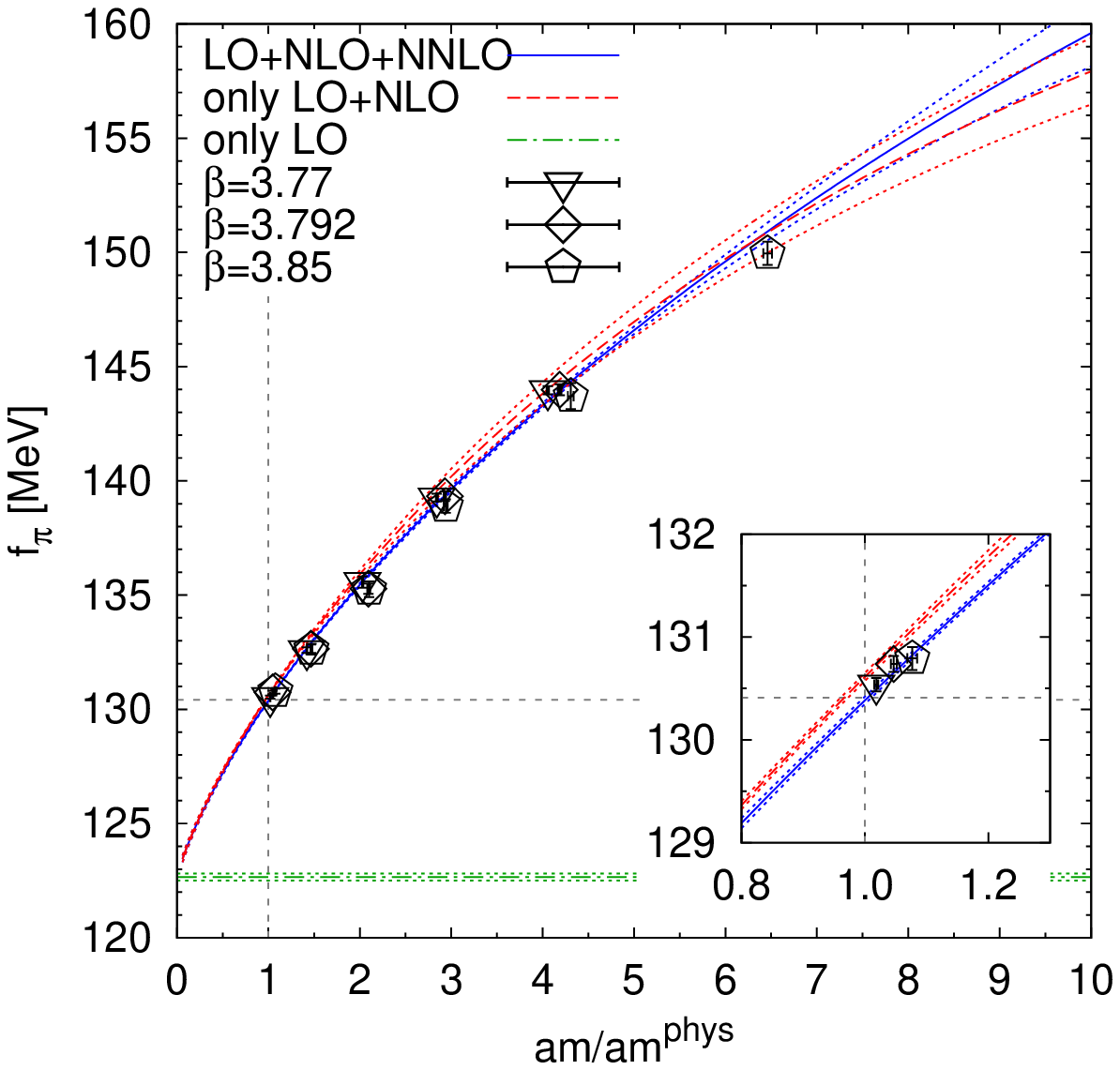}%
\includegraphics[width=.47\textwidth]{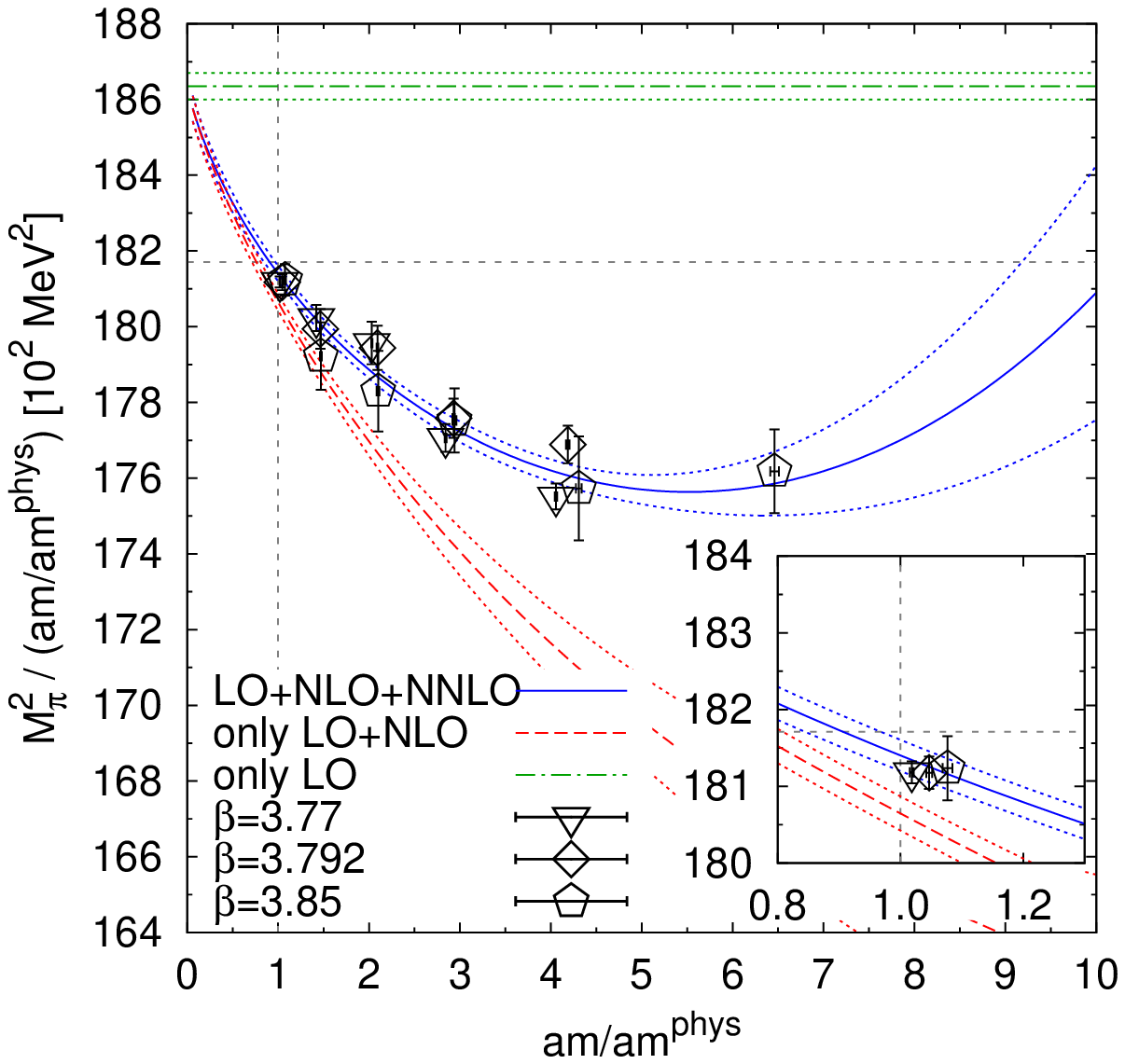}
\caption{Combined NNLO fit for lattice scales $1/a\:>\:1.6\,{\rm GeV}$ and meson masses $135\,{\rm MeV}\:\leq\:M_{\pi}\:\leq\:340\,{\rm MeV}$ using a prior for $\Lambda_{12}$. \textit{Left panel:} meson decay constant, \textit{right panel:} squared meson mass divided by the quark-mass ratio. The \textit{solid blue lines} show the complete (up to NNLO) fit, whereas the \textit{dashed red lines} show the LO+NLO contribution of the full NNLO fit, and the \textit{dash-dotted green lines} show the LO contribution. Only data points included in the fit range are depicted in the plots; the physical values are marked by \textit{dashed gray lines}.}
\label{fig:NNLO:priorl12_fine_135-340}
\end{center}
\end{figure}

\begin{figure}
\begin{center}
\includegraphics[width=.47\textwidth]{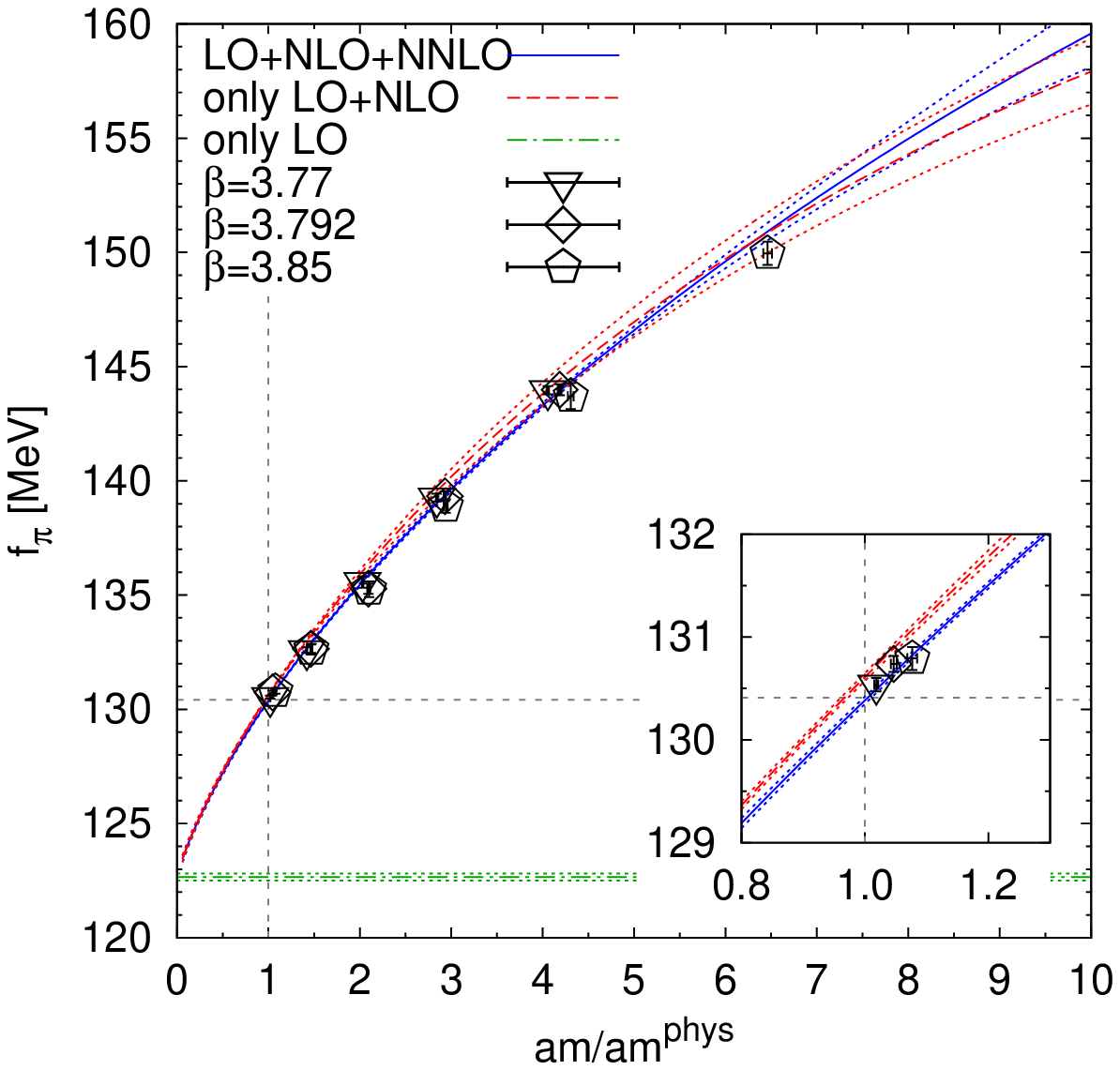}%
\includegraphics[width=.47\textwidth]{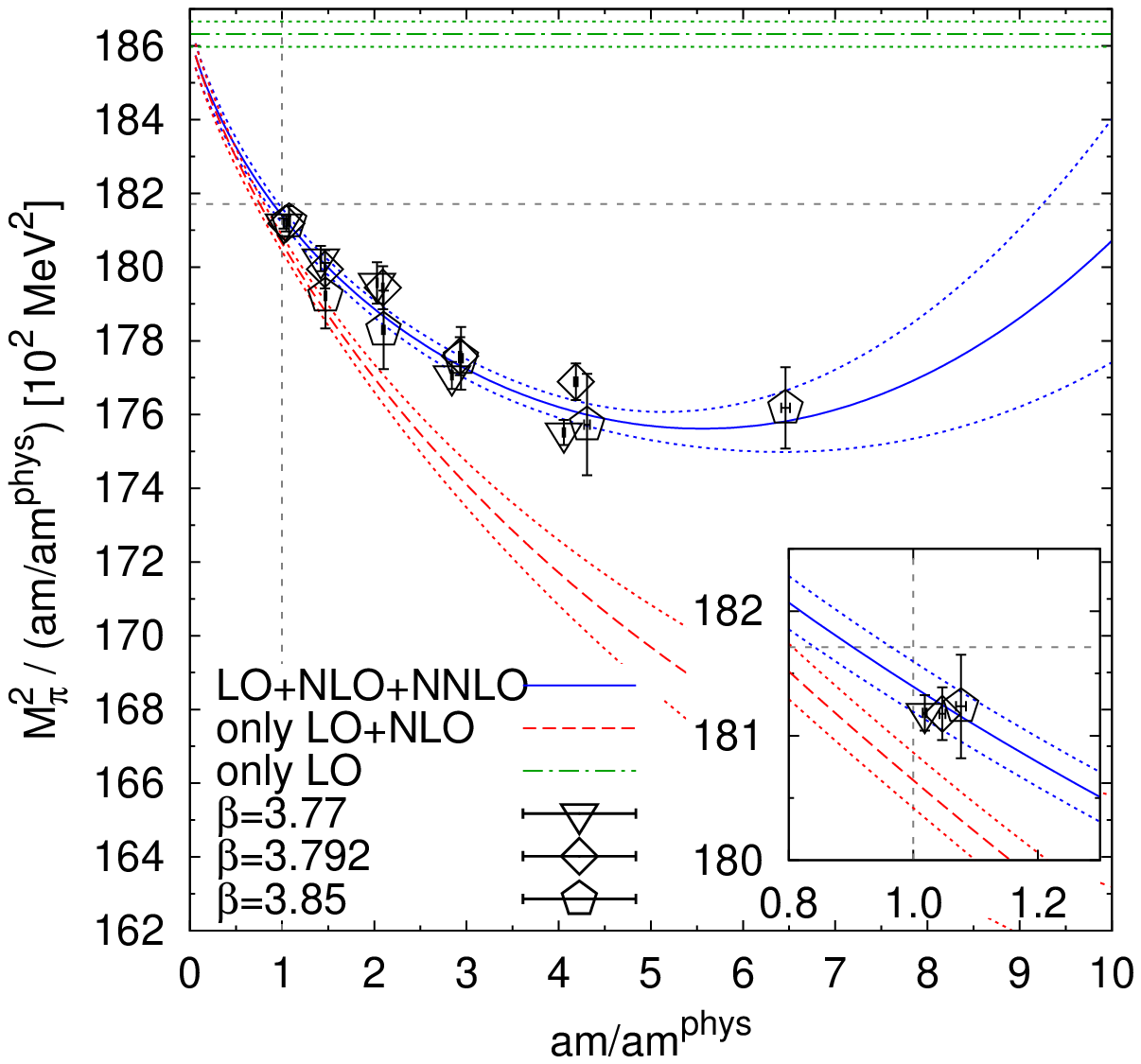}
\caption{Combined NNLO fit for lattice scales $1/a\:>\:1.6\,{\rm GeV}$ and meson masses $135\,{\rm MeV}\:\leq\:M_{\pi}\:\leq\:340\,{\rm MeV}$ using priors for $\Lambda_{12}$, $k_{M^2}$, and $k_f$. \textit{Left panel:} meson decay constant, \textit{right panel:} squared meson mass divided by the quark-mass ratio. The \textit{solid blue lines} show the complete (up to NNLO) fit, whereas the \textit{dashed red lines} show the LO+NLO contribution of the full NNLO fit, and the \textit{dash-dotted green lines} show the LO contribution. Only data points included in the fit range are depicted in the plots; the physical values are marked by \textit{dashed gray lines}.}
\label{fig:NNLO:priorl12kmkf_fine_135-340}
\end{center}
\end{figure}

\begin{figure}
\begin{center}
\includegraphics[width=.45\textwidth]{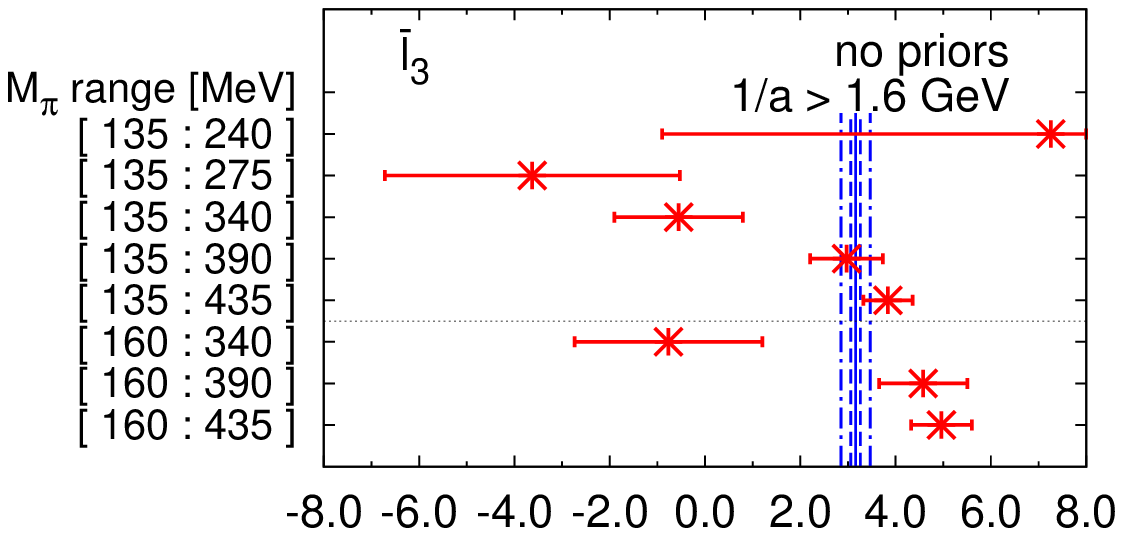}%
\includegraphics[width=.45\textwidth]{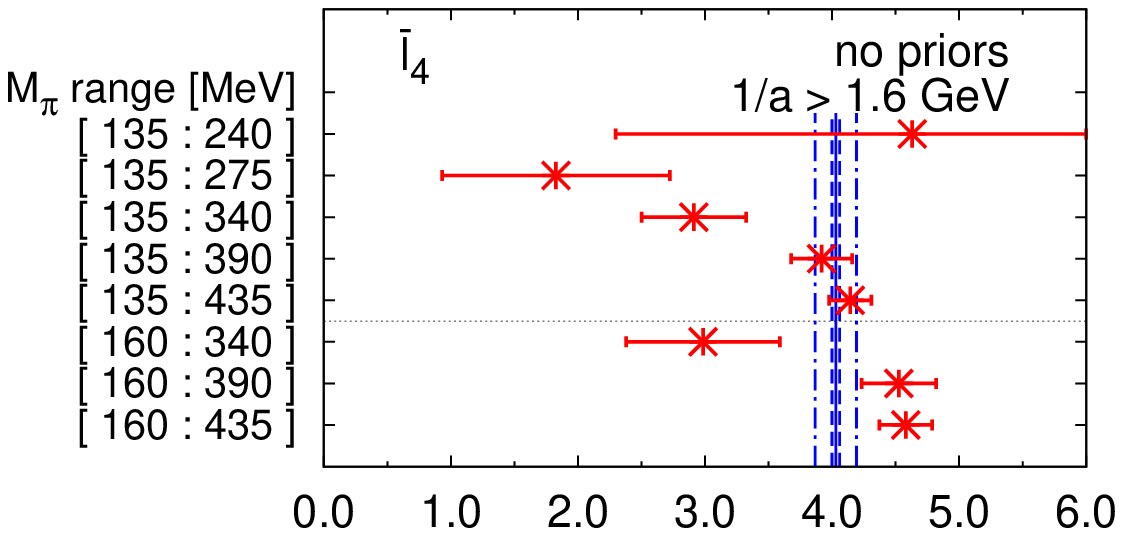}\\
\includegraphics[width=.45\textwidth]{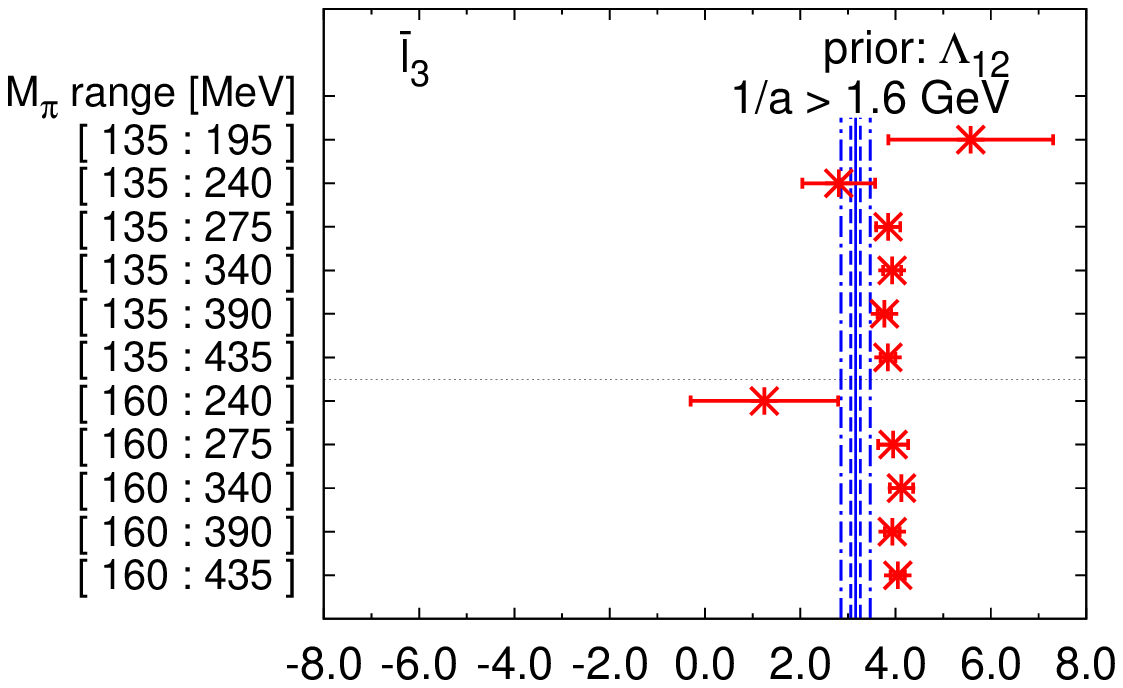}%
\includegraphics[width=.45\textwidth]{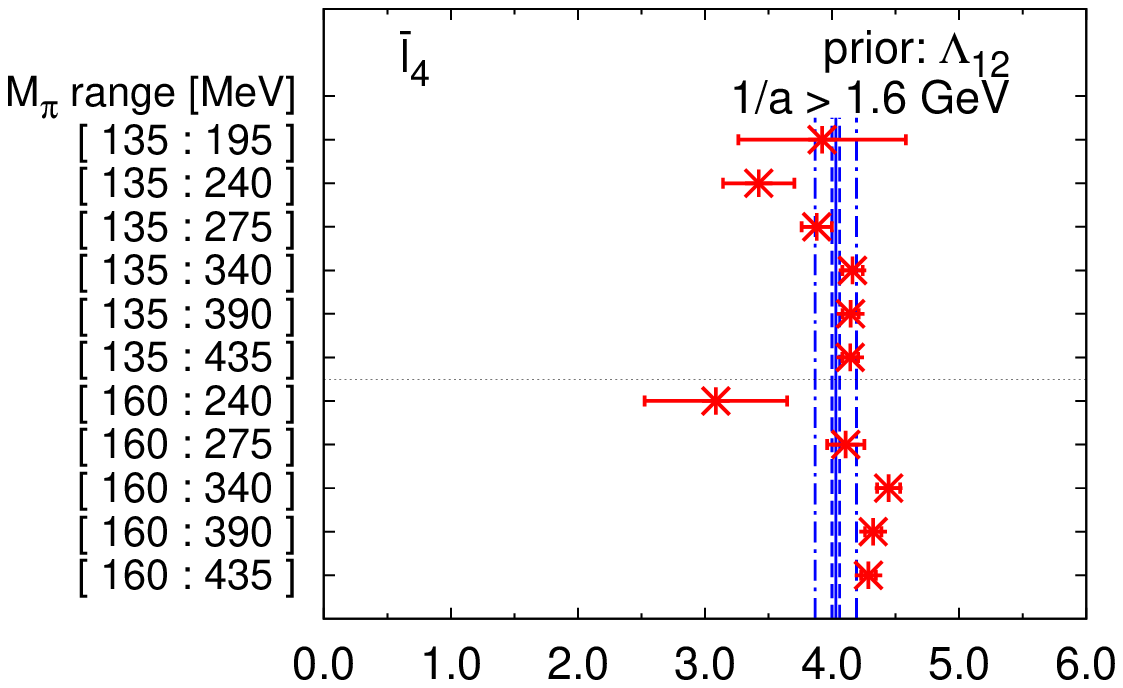}\\
\includegraphics[width=.45\textwidth]{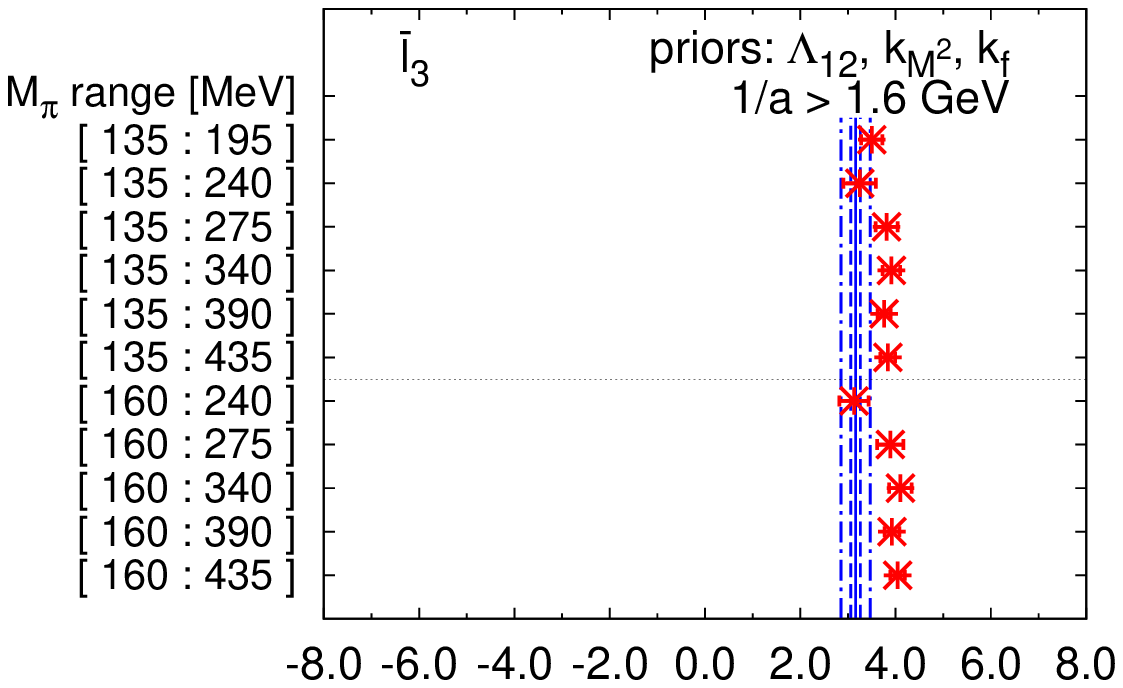}%
\includegraphics[width=.45\textwidth]{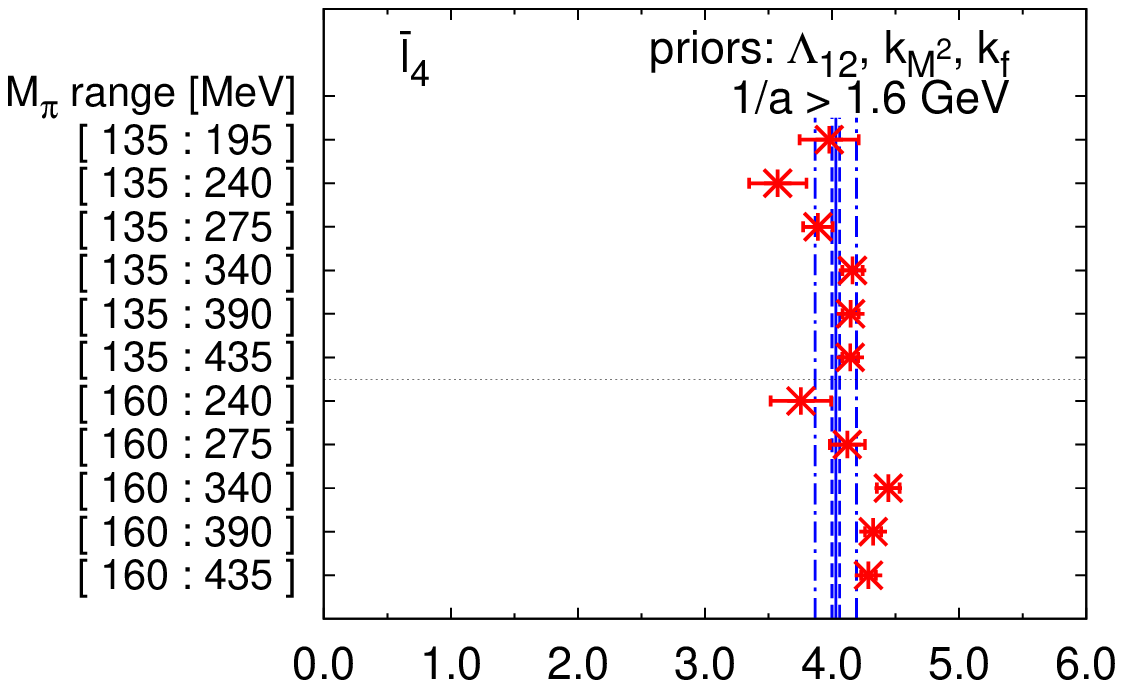}
\caption{Results for the LECs $\bar{\ell}_3$ (\textit{left panels}) and $\bar{\ell}_4$ (\textit{right panels}) from NNLO fits to ensembles with $1/a\,>\,1.6\,{\rm GeV}$ and different mass ranges. The \textit{top}, \textit{middle}, and \textit{bottom panels} show the results using no priors, a prior for $\Lambda_{12}$, and priors for $\Lambda_{12}$, $k_{M^2}$, and $k_f$, respectively. The \textit{solid, dashed, and dashed dotted blue lines} indicate the central value, statistical and total (statistical plus systematic) error bands from our NLO fits, cf.\ Eqs.\,(\ref{eq:result:NLO:chiphys})--(\ref{eq:result:NLO:fratio}).}
\label{fig:NNLO:pheno_lbar34_fine}
\end{center}
\end{figure}

\begin{figure}
\begin{center}
\includegraphics[width=.45\textwidth]{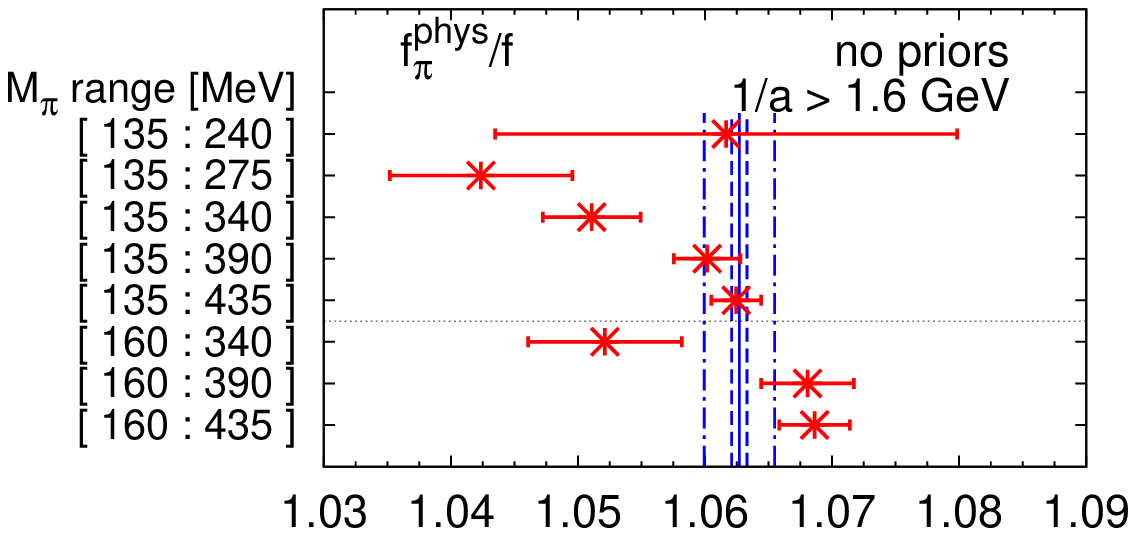}%
\includegraphics[width=.45\textwidth]{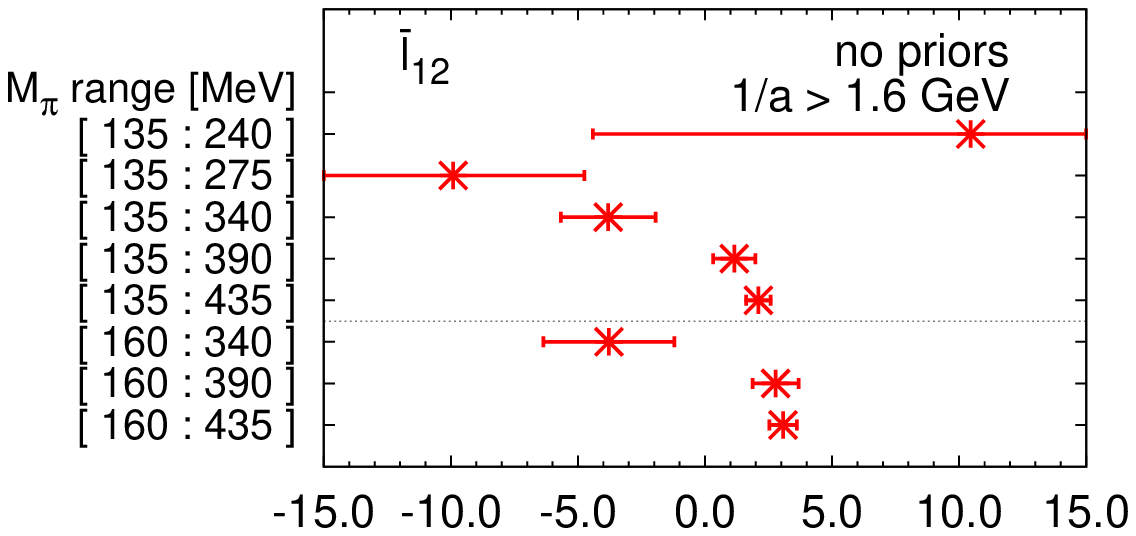}\\
\includegraphics[width=.45\textwidth]{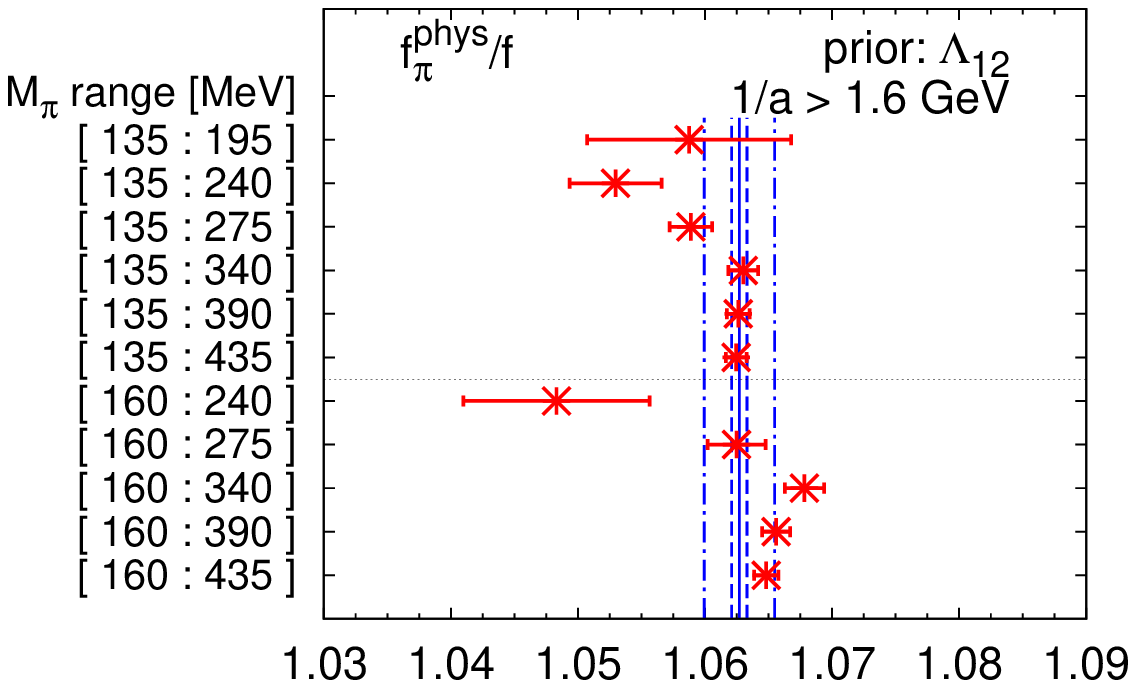}%
\includegraphics[width=.45\textwidth]{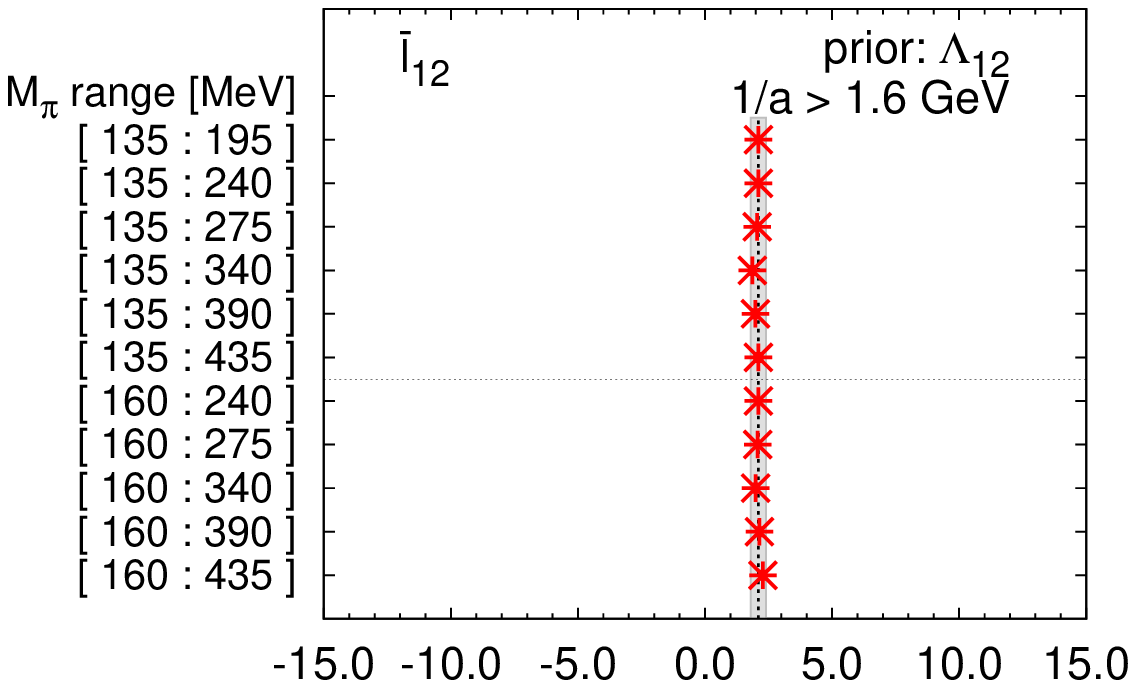}\\
\includegraphics[width=.45\textwidth]{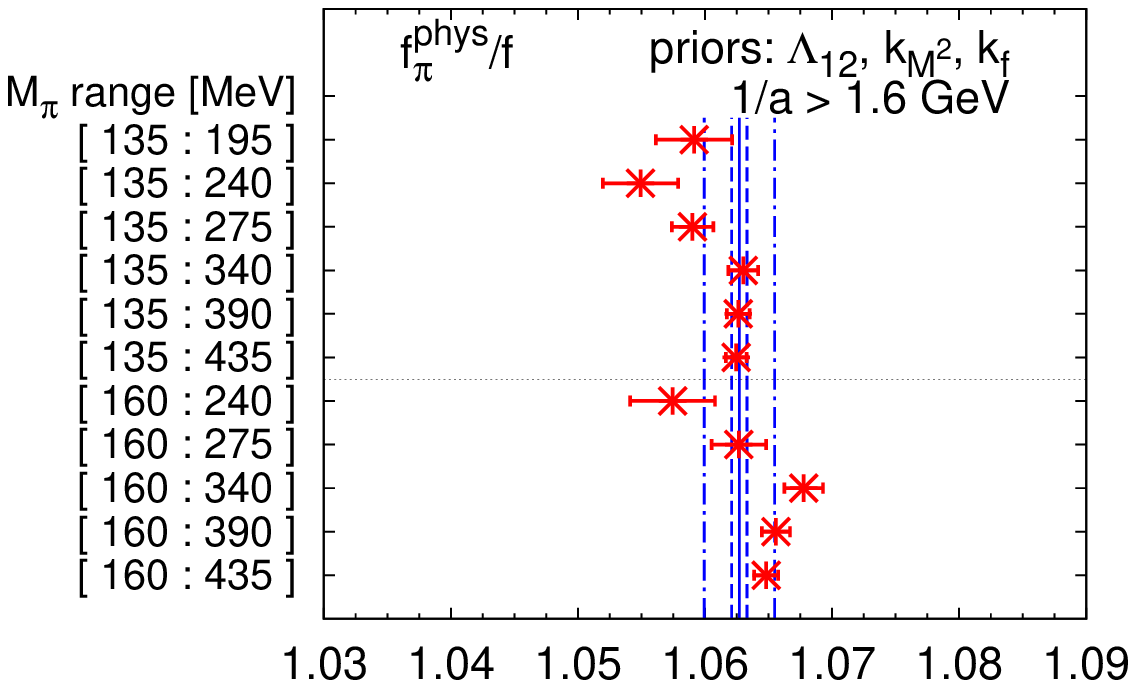}%
\includegraphics[width=.45\textwidth]{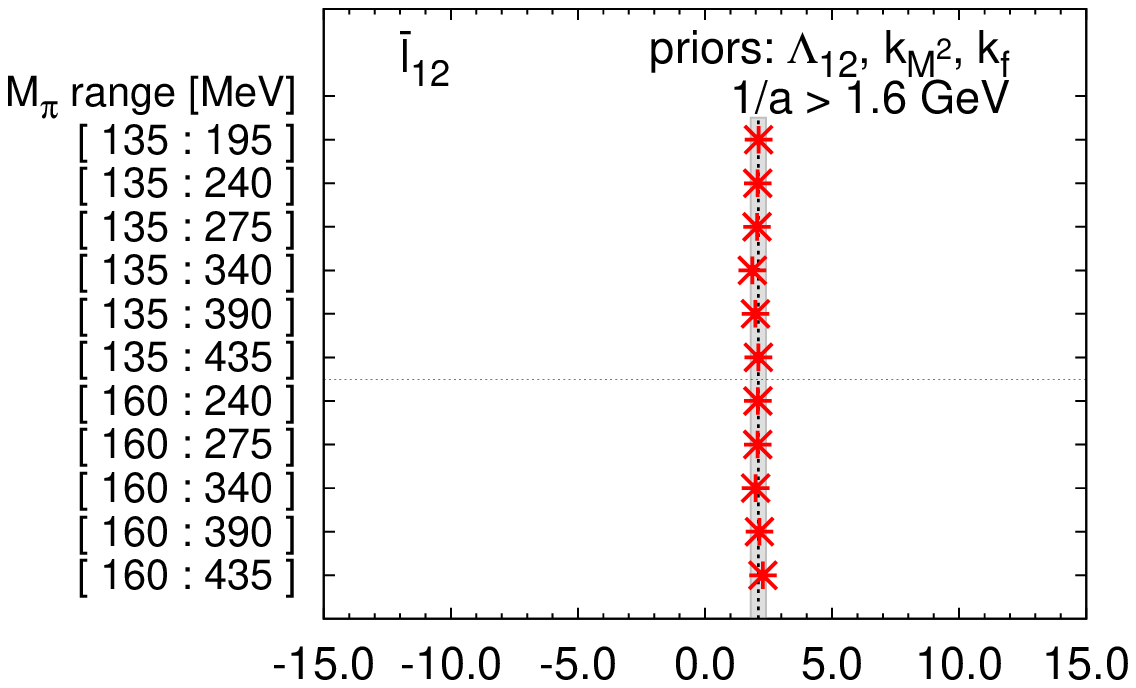}
\caption{Results for the ratio $f_\pi^{\rm phys}/f$ (\textit{left panels}) and the LEC $\bar{\ell}_{12}$ (\textit{right panels}) from NNLO fits to ensembles with $1/a\,>\,1.6\,{\rm GeV}$ and different mass ranges. The \textit{top}, \textit{middle}, and \textit{bottom panels} show the results using no priors, a prior for $\Lambda_{12}$, and priors for $\Lambda_{12}$, $k_{M^2}$, and $k_f$, respectively. The \textit{solid, dashed, and dashed dotted blue lines} indicate the central value, statistical and total (statistical plus systematic) error bands from our NLO fits, cf.\ Eqs.\,(\ref{eq:result:NLO:chiphys})--(\ref{eq:result:NLO:fratio}). The \textit{gray shaded areas} indicate the used prior and its width (where applicable). }
\label{fig:NNLO:pheno_lbar12fratio_fine}
\end{center}
\end{figure}

\subsubsection{NNLO fits using priors}
\label{subsubsec:NNLOfits_priors}

To stabilize the NNLO fits, we examined the effect of using priors for the three additional fit parameters $\Lambda_{12}$, $k_{M^2}$, and $k_f$. Since we are mainly interested to learn how the parameters already appearing at NLO change, when going from NLO to NNLO fits, we did not consider adding priors for these parameters as well. 

The priors were implemented in the fitting procedure via an augmented $\chi^2$. Instead of minimizing the usual $\chi^2$ (sum of deviations of the data points from the fitted function weighted by the uncertainty of that data point) the modified 
\begin{equation}
\label{eq:NNLO:augmentedChisq}
\chi^2_{\rm augmented}\;=\;\chi^2\:+\:\sum_n \frac{(p_n- p^{\rm prior}_n)^2}{(\sigma^{\rm prior}_n)^2}
\end{equation}
is minimized by the fitting routine. Here $n$ runs over all fit parameters $p_n$ to which a prior $p_n^{\rm prior}$ with width $\sigma_n^{\rm prior}$ has been assigned. That is, the fitting routine ``punishes'' a parameter for leaving the prior interval  $p_n^{\rm prior}\,\pm\,\sigma_n^{\rm prior}$.
 
In the following we will use either a prior for $\bar{\ell}_{12}$ alone or together with priors for $k_{M^2}$ and $k_f$. We will assign the following values for the priors to the fit parameters:
%
\begin{eqnarray}
\label{eq:NNLO_priorlbar12}
\bar{\ell}_{12}^{\rm prior} &=& 2.1\:\pm\:0.3\\
\label{eq:NNLO:priorkMsq}
k_{M^2}^{\rm prior} &=& 0\:\pm\:10\\
\label{eq:NNLO:priorkf}
k_{f}^{\rm prior} &=& 0\:\pm\:10\,.
\end{eqnarray}
%
The prior for $\bar{\ell}_{12}$ has been obtained from the estimates for the LECs $\bar{\ell}_1\,=\,-0.4\,\pm\,0.6$ and $\bar{\ell}_2\,=\,4.3\,\pm\,0.1$ extracted from $\pi\pi$-scattering data in Ref.\,\cite{Colangelo:2001df} via
\[ \bar{\ell}_{12}^{\rm prior} \;=\; \frac{7}{15}\bar{\ell}_1^{\rm prior}\:+\:\frac{8}{15}\bar{\ell}_2^{\rm prior}  \] 
cf.\ Eqs.~(\ref{eq:NNLO:Lambda12}) and (\ref{eq:NNLO:lbar:def}). This would translate into a prior for the low-energy scale
\begin{equation}
\label{eq:NNLO:priorLam12}
\Lambda_{12}^{\rm prior} \;=\; 385\,{\rm MeV}\:\pm\:58\,{\rm MeV}\,.
\end{equation}
The priors for the NNLO LECs $k_{M^2}$ and $k_f$ are merely based on assuming a natural order of magnitude for these corrections. We are aware that the latter is a rather weak argument, for that reason we did not use the priors on $k_{M^2}$ and $k_f$ alone, and results from fits using these priors should be taken with some caution. However, these priors are to some extent justified by the NNLO fits where no priors have been used (see Fig.\,\ref{fig:NNLO:massrange_fine}) and also, as we will see, by fits where only the prior on $\bar{\ell}_{12}$ has been used.

The results from the fits to different mass ranges using priors on $\bar{\ell}_{12}$ only or on $k_{M^2}$ and $k_f$ as well are shown in Figs.\,\ref{fig:NNLO:priorl12:massrange_fine} and \ref{fig:NNLO:priorl12kmkf:massrange_fine}, respectively, where only the ensembles with $1/a\,>\,1.6\,{\rm GeV}$ have been used. The used priors are indicated by the gray shaded areas. As an effect, now fits to smaller mass ranges are possible and/or are more stable judging from the uncertainties of the fit parameters. In Figs.\,\ref{fig:NNLO:priorl12_fine_135-340} and \ref{fig:NNLO:priorl12kmkf_fine_135-340}, we show as examples the fits for the mass range $135\,{\rm MeV}\,\leq\,M_\pi\,\leq\,340\,{\rm MeV}$, with a breakup into LO, NLO, and NNLO contributions. Whereas in both cases for the decay constant the NNLO contribution seems to be reasonably small, for the meson mass, the NNLO contribution is more substantial. Nevertheless, the error bands for the LO+NLO contribution are now reasonable as well, showing again the stabilizing effect of using the priors.

In Figs.\,\ref{fig:NNLO:pheno_lbar34_fine} and \ref{fig:NNLO:pheno_lbar12fratio_fine} we compare the LECs and the ratio $f_\pi^{\rm phys}/f$ from the NNLO fits to different mass ranges using either no priors or one of our choices for the set of priors (all for ensembles with $1/a\,>\,1.6\,{\rm GeV}$). Our final results with total error bands from the NLO fits are always indicated by the blue lines. In the case of $\bar{\ell}_3$, the fits using priors shift the value of the LECs a bit upward, whereas $\bar{\ell}_4$ and $f_\pi^{\rm phys}/f$ just fluctuate within the total error band from the NLO fit. The comparison for the LEC $\bar{\ell}_{12}$ again demonstrates the stabilizing effect of the priors or---in other words---the difficulties we encountered in the fits without using priors. In the latter case (top right panel of Fig.\,\ref{fig:NNLO:pheno_lbar12fratio_fine}), the fit result only comes close to the phenomenological estimate, Eq.~(\ref{eq:NNLO_priorlbar12}), for fit ranges including meson masses of 390 MeV or higher.


\section{Conclusions}
\label{sec:Conclusions}

In this paper we presented a determination of the NLO low-energy constants
$\bar\ell_3$ and $\bar\ell_4$ of SU(2) chiral perturbation theory from
2+1 flavor lattice simulations with staggered fermions. 
In addition we gave results for the LO quantities $f$ (or $\fpi^\mr{phys}/f$)
and $B$ (or $\ch^\mr{phys}\!=\!Bm^\mr{phys}$). The quantities
$\bar\ell_{3,4}$ are also expressed in terms of the scales $\Lambda_{3,4}$,
respectively.

Our final results as presented in Eqs.\,(\ref{eq:result:NLO:chiphys})--(\ref{eq:result:NLO:fratio}) stem from fits
which use the NLO functional form.
The result for $\ch^\mr{phys}$ amounts to a condensate parameter
\beq
B(2\GeV)=2.682(36)(39)\GeV
\label{eq:result_B}
\eeq
if one divides out the value of the average light quark mass from
Refs.\,\cite{Durr:2010vn,Durr:2010aw}.
Moreover, after multiplying with $F^2\!=\!f^2/2$ from Eq.~(\ref{eq:result:NLO:f}),
one obtains
\beq
\Sigma(2\GeV)=2.020(27)(31)\,10^{-2}\GeV^3
\qquad\mbox{or}\qquad
\Sigma(2\GeV)^{1/3}=0.2723(12)(14)\GeV
\label{eq:result_Sigma}
\eeq
where all quantities given at the scale $\mu\!=\!2\GeV$ refer to the $\MSbar$
scheme.
These results are reasonably consistent with the high-quality entries in
Table\,10 of Ref.\,\cite{Colangelo:2010et}, in particular with those by the
MILC \cite{Bazavov:2009bb},
RBC/UKQCD \cite{Aoki:2010dy}, and
ETM \cite{Baron:2009wt}
collaborations, to mention some of the most precise determinations.
Similarly, our results 
\[\bar{\ell}_3=3.16(10)(29) \qquad \mbox{and}\qquad  \bar{\ell}_4=4.03(03)(16) \]
 as stated in Eqs.\,(\ref{eq:result:NLO:lbar3}) and
(\ref{eq:result:NLO:lbar4}), respectively,
are in good agreement with the broad majority of the entries in Table\,12 of
Ref.\,\cite{Colangelo:2010et}, in particular with those by the
MILC \cite{Bazavov:2010yq},
RBC/UKQCD \cite{Aoki:2010dy}, and
ETM \cite{Baron:2009wt}
collaborations.
Finally, our result (\ref{eq:result:NLO:fratio}) for the ratio
$\fpi^\mr{phys}/f$ agrees with the entries of Table\,11 in
Ref.\,\cite{Colangelo:2010et}, perhaps with some slight tension when compared
to the recent 2+1+1 flavor result by the ETM Collaboration
\cite{Baron:2010bv}.

We have carefully examined the effect of various cuts on the data, in
particular the effect of requesting $a^{-1}\!>\!1.6\GeV$ and the effect of
limiting the pion mass range that enters the chiral fit.
It turns out that the restriction to fine lattices improves the quality of the
fits, and with this restriction reasonably stable NLO results are obtained for
pion mass windows up to $\sim\!400\MeV$.
Given the fine grained set of pion masses available in our ensemble basis, we
can even explore the effect of dropping some of the lighter data points.
We find that $\bar\ell_3$ is much more robust in this respect than $\bar\ell_4$
(or $\fpi^\mr{phys}/f$), as is evident from Fig.\,\ref{fig:NLO:massrange_fine_pheno}.

Finally, we have explored the effect of adding the NNLO contribution to the
functional Ansatz.
To prevent a dramatic increase in the number of free parameters, we add priors
for the new NLO combinations (in which we are not interested) and the genuine
NNLO coefficients (to which our data show little sensitivity).
We find it very reassuring that these prior-aided NNLO fits remain stable
(for a reasonable range of lattice spacings and pion mass windows) and that
the resulting fits show a very natural ordering between LO, NLO, and NNLO
contributions (out to $\Mpi\!\sim\!400\MeV$).
Moreover, the NLO coefficients $\bar\ell_3$ and $\bar\ell_4$ as determined from
these prior-aided NNLO fits are in good agreement with the results of the
direct NLO fits.
We take this as a sign that our assessment of the systematic uncertainties
of these quantities is true and fair.

The chiral fits presented in Secs.\,\ref{sec:NLOChPT} and \ref{sec:NNLOChPT}
do not include terms designed to absorb cut-off effects in the data.
The purpose of the present work was to explore whether such terms are
mandatory; we find that they are not (with the level of precision of our data),
albeit at the price of pruning the data set to include only lattices with
$a^{-1}>1.6\GeV$.
Still, we did perform an exploratory analysis with such terms included, and it
seems that with our choice of the scaling trajectory (cf.\ discussion in the
beginning of Sec.\,\ref{sec:NLOChPT}), there is hardly any change.

In summary, our results
(\ref{eq:result:NLO:chiphys})--(\ref{eq:result:NLO:fratio}), (\ref{eq:result_B}), and (\ref{eq:result_Sigma}) indicate that SU(2) chiral
low-energy constants can be determined on the lattice with a precision
at the level of a few percent for the LO quantities $B$, $\Sigma$, $f$, and
the level of $O(10\%)$ for the NLO scales $\Lambda_3$, $\Lambda_4$.

\section*{Acknowledgments}

Computations were performed using HPC resources from FZ J\"ulich and on clusters at Wuppertal University. This work is supported in part by EU Grants PITN-GA-2009-238353 (ITN STRONGnet) and PIRG07-GA-2010-268367 and DFG Grants FO 502/2 and SFB-TR~55.

\appendix

\bibliography{references}

\begin{thebibliography}{10}

\bibitem{Fritzsch:1973pi}
H.~Fritzsch, M.~Gell-Mann, and H.~Leutwyler,
\newblock Phys.Lett. {\bf B47}, 365 (1973).

\bibitem{Gasser:1983yg}
J.~Gasser and H.~Leutwyler,
\newblock Annals Phys. {\bf 158}, 142 (1984).

\bibitem{Gasser:1984gg}
J.~Gasser and H.~Leutwyler,
\newblock Nucl.Phys. {\bf B250}, 465 (1985).

\bibitem{Wilson:1974sk}
K.~G. Wilson,
\newblock Phys.Rev. {\bf D10}, 2445 (1974).

\bibitem{Creutz:1980zw}
M.~Creutz,
\newblock Phys.Rev. {\bf D21}, 2308 (1980).

\bibitem{Durr:2008zz}
S.~D{\"u}rr {\em et~al.},
\newblock Science {\bf 322}, 1224 (2008), arXiv:0906.3599.

\bibitem{Scholz:2011rk}
E.~E. Scholz {\em et~al.},
\newblock PoS {\bf LATTICE2011}, 142 (2011), arXiv:1111.3729.

\bibitem{Luscher:1985zq}
M.~L{\"u}scher and P.~Weisz,
\newblock Phys.Lett. {\bf B158}, 250 (1985).

\bibitem{Morningstar:2003gk}
C.~Morningstar and M.~J. Peardon,
\newblock Phys.Rev. {\bf D69}, 054501 (2004), arXiv:hep-lat/0311018.

\bibitem{Aoki:2005vt}
Y.~Aoki {\em et~al.},
\newblock JHEP {\bf 0601}, 089 (2006), arXiv:hep-lat/0510084.

\bibitem{Durr:2008rw}
S.~D{\"u}rr {\em et~al.},
\newblock Phys.Rev. {\bf D79}, 014501 (2009), arXiv:0802.2706.

\bibitem{Aoki:2006br}
Y.~Aoki {\em et~al.},
\newblock Phys.Lett. {\bf B643}, 46 (2006), arXiv:hep-lat/0609068.

\bibitem{Aoki:2006we}
Y.~Aoki {\em et~al.},
\newblock Nature {\bf 443}, 675 (2006), arXiv:hep-lat/0611014.

\bibitem{Aoki:2009sc}
Y.~Aoki {\em et~al.},
\newblock JHEP {\bf 0906}, 088 (2009), arXiv:0903.4155.

\bibitem{Borsanyi:2010bp}
S.~Borsanyi {\em et~al.},
\newblock JHEP {\bf 1009}, 073 (2010), arXiv:1005.3508.

\bibitem{Borsanyi:2010cj}
S.~Borsanyi {\em et~al.},
\newblock JHEP {\bf 1011}, 077 (2010), arXiv:1007.2580.

\bibitem{Bazavov:2009bb}
A.~Bazavov {\em et~al.},
\newblock Rev.Mod.Phys. {\bf 82}, 1349 (2010), arXiv:0903.3598.

\bibitem{Colangelo:2010et}
G.~Colangelo {\em et~al.},
\newblock Eur.Phys.J. {\bf C71}, 1695 (2011), arXiv:1011.4408.

\bibitem{Nakamura:2010zzi}
Particle Data Group, K.~Nakamura {\em et~al.},
\newblock J.Phys.G {\bf G37}, 075021 (2010).

\bibitem{DelDebbio:2007pz}
L.~Del~Debbio, L.~Giusti, M.~\protect{L\"uscher}, R.~Petronzio, and N.~Tantalo,
\newblock JHEP {\bf 0702}, 082 (2007), arXiv:hep-lat/0701009.

\bibitem{Bratt:2010jn}
LHPC Collaboration, J.~Bratt {\em et~al.},
\newblock Phys.Rev. {\bf D82}, 094502 (2010), arXiv:1001.3620.

\bibitem{Colangelo:2005gd}
G.~Colangelo, S.~\protect{D\"urr}, and C.~Haefeli,
\newblock Nucl.Phys. {\bf B721}, 136 (2005), arXiv:hep-lat/0503014.

\bibitem{Aubin:2003mg}
C.~Aubin and C.~Bernard,
\newblock Phys. Rev. {\bf D68}, 034014 (2003), arXiv:hep-lat/0304014.

\bibitem{Aubin:2003uc}
C.~Aubin and C.~Bernard,
\newblock Phys. Rev. {\bf D68}, 074011 (2003), arXiv:hep-lat/0306026.

\bibitem{Blum:1996uf}
T.~Blum {\em et~al.},
\newblock Phys.Rev. {\bf D55}, 1133 (1997), arXiv:hep-lat/9609036.

\bibitem{Orginos:1999cr}
MILC Collaboration, K.~Orginos, D.~Toussaint, and R.~Sugar,
\newblock Phys.Rev. {\bf D60}, 054503 (1999), arXiv:hep-lat/9903032.

\bibitem{Wittig:2012ha}
H.~Wittig,
\newblock PoS {\bf LATTICE2011}, 025 (2012), arXiv:1201.4774.

\bibitem{Colangelo:2001df}
G.~Colangelo, J.~Gasser, and H.~Leutwyler,
\newblock Nucl.Phys. {\bf B603}, 125 (2001), arXiv:hep-ph/0103088.

\bibitem{Durr:2010vn}
S.~D{\"u}rr {\em et~al.},
\newblock Phys.Lett. {\bf B701}, 265 (2011), arXiv:1011.2403.

\bibitem{Durr:2010aw}
S.~D{\"u}rr {\em et~al.},
\newblock JHEP {\bf 1108}, 148 (2011), arXiv:1011.2711.

\bibitem{Aoki:2010dy}
RBC/UKQCD Collaborations, Y.~Aoki {\em et~al.},
\newblock Phys.Rev. {\bf D83}, 074508 (2011), arXiv:1011.0892.

\bibitem{Baron:2009wt}
ETM Collaboration, R.~Baron {\em et~al.},
\newblock JHEP {\bf 1008}, 097 (2010), arXiv:0911.5061.

\bibitem{Bazavov:2010yq}
A.~Bazavov {\em et~al.},
\newblock PoS {\bf LATTICE2010}, 083 (2010), arXiv:1011.1792.

\bibitem{Baron:2010bv}
R.~Baron {\em et~al.},
\newblock JHEP {\bf 1006}, 111 (2010), arXiv:1004.5284.

\end{thebibliography}

\end{document}